\documentclass[preprint,nofootinbib,aps,superscriptaddress]{revtex4}
\usepackage{color,graphicx,shortvrb,epsfig}
\newcommand{\beq}{\begin{equation}}
\newcommand{\eeq}{\end{equation}}
\newcommand{\beqa}{\begin{eqnarray}}
\newcommand{\eeqa}{\end{eqnarray}}
\newcommand{\no}{\nonumber}
\def\lsim{\mathrel{\rlap{\lower4pt\hbox{\hskip1pt$\sim$}}
    \raise1pt\hbox{$<$}}}         
\def\gsim{\mathrel{\rlap{\lower4pt\hbox{\hskip1pt$\sim$}}
    \raise1pt\hbox{$>$}}}         
\newcommand{\re}[1]{\ensuremath{{\cal R}e(#1)}}
\newcommand{\im}[1]{\ensuremath{{\cal I}m(#1)}}

\newcommand{\CP}{CP\ }

\newcommand{\M}{P}
\newcommand{\Mb}{\overline{P}}
\newcommand{\Mz}{P{}^0}
\newcommand{\Mzb}{\overline{P}{}^0}
\newcommand{\Heff}{{\cal H}}
\newcommand{\Meff}{M}
\newcommand{\Geff}{\Gamma}

\newcommand{\fb}{\overline{f}}
\newcommand{\f}{f}

\begin{document}


\vspace*{1cm}

\title{\boldmath CP Violation in Meson Decays\footnote{
Lectures given at the `Third CERN-CLAF School of High Energy Physics' 
Malarg\"ue, Argentina, 27 February - 12 March 2005 and at the Les
Houches Summer School (Session LXXXIV) on
`Particle Physics Beyond the Standard Model,'
Les Houches, France, August 1--26, 2005.}}

\author{Yosef Nir}\email{yosef.nir@weizmann.ac.il}
\affiliation{Department of Particle Physics \\
  Weizmann Institute of Science, Rehovot 76100, Israel}  


\begin{abstract}
This is a written version of a series of lectures aimed at graduate
students in the field of (theoretical and experimental) high energy
physics. The main topics covered are: (i) The flavor sector of the
Standard Model and the Kobayashi-Maskawa mechanism of CP violation;
(ii) Formalism and theoretical interpretation of CP violation in meson
decays; (iii) $K$ decays; (iv) $D$ decays; (v) $B$ decays: $b\to c\bar cs$,
$b\to s\bar ss$, $b\to u\bar ud$ and $b\to c\bar us,u\bar cs$; 
(vi) CP violation as a probe of new physics and, in particular, of
supersymmetry. 
\end{abstract}

\maketitle
\tableofcontents

\section{Introduction}
The Standard Model predicts that the only way that CP is violated is
through the Kobayashi-Maskawa mechanism \cite{Kobayashi:1973fv}. Specifically,
the source of CP violation is a {\it single} phase in the mixing matrix that
describes the charged current weak interactions of quarks.
In the introductory chapter, we briefly review the present evidence that
supports the Kobayashi-Maskawa picture of CP violation, as well as the various 
arguments against this picture.
 
\subsection{Why believe the Kobayashi-Maskawa mechanism?}
Experiments have measured to date nine independent CP violating
observables:\footnote{The list of measured observables in $B$ decays
  is somewhat conservative. I include only observables where the
  combined significance of Babar and Belle measurements (taking an
  inflated error in case of inconsistencies) is above $3\sigma$.}

\begin{enumerate} 
\item Indirect CP violation in $K\to\pi\pi$ decays \cite{Christenson:1964fg}
and in $K\to\pi\ell\nu$ decays is given by
\beq\label{expeps}
\varepsilon_K=(2.28\pm0.02)\times10^{-3}\ e^{i\pi/4}.
\eeq
\item Direct CP violation in $K\to\pi\pi$ decays 
\cite{Burkhardt:1988yh,Fanti:1999nm,Alavi-Harati:1999xp}
is given by
\beq\label{aveepp}
{\varepsilon^\prime/\varepsilon}=(1.72\pm0.18)\times10^{-3}.
\eeq
\item CP violation in the interference of mixing and decay in the
$B\to\psi K_S$ and other, related modes is given by 
\cite{Aubert:2001nu,Abe:2001xe}:
\beq\label{aveapk}
S_{\psi K_S}=+0.69\pm0.03.
\eeq
\item CP violation in the interference of mixing and decay in the
$B\to K^+K^- K_S$ mode is given by
\cite{Abe:2004xp,Aubert:2005ja}
\beq\label{skpkmks}
S_{K^+K^-K_S}=-0.45\pm0.13.
\eeq
\item CP violation in the interference of mixing and decay in the
$B\to D^{*+}D^{*-}$ mode is given by
\cite{Aushev:2004hn,Aubert:2005rn}
\beq\label{sdspdsm}
S_{D^{*+}D^{*-}}=-0.75\pm0.23.
\eeq
\item CP violation in the interference of mixing and decay in the
$B\to\eta^\prime K^0$ modes is given by
\cite{Aubert:2005iy,Abe:2005bt,Aubert:2005vw}
\beq\label{sepks}
S_{\eta^\prime K_S}=+0.50\pm0.09(0.13).
\eeq
\item CP violation in the interference of mixing and decay in the
$B\to f_0 K_S$ mode is given by
\cite{Aubert:2004gk,Abe:2005bt}
\beq\label{sfzks}
S_{f_0 K_S}=-0.75\pm0.24.
\eeq
\item Direct CP violation in the $\overline{B}{}^0\to K^-\pi^+$ mode
is given by 
\cite{Aubert:2004qm,Abe:2005fz}
\beq\label{akpi}
{\cal A}_{K^\mp\pi^\pm}=-0.115\pm0.018.
\eeq
\item Direct CP violation in the $B\to \rho\pi$ mode
is given by
\cite{Wang:2004va,Aubert:2004iu}
\beq\label{arhopi}
{\cal A}^{-+}_{\rho\pi}=-0.48\pm0.14.
\eeq
\end{enumerate}
All nine measurements -- as well as many other, where CP violation is
not (yet) observed at a level higher than $3\sigma$ -- are consistent
with the Kobayashi-Maskawa picture of CP violation. In particular, the
measurement of the phase $\beta$ from the CP asymmetry $B\to\psi K$,
the measurement of the phase $\alpha$ from CP asymmetries and
decay rates in the $B\to\pi\pi,\rho\pi$ and $\rho\rho$ modes, and the
measurement of the phase $\gamma$ from $B\to DK$ decays, have
provided the first three precision tests of CP violation in the Standard
Model. Since the model has passed these tests successfully, we are
able, for the first time, to make the following statement: {\it The
  Kobayashi-Maskawa phase is, very likely, the dominant source of CP
  violation in low-energy flavor-changing processes.} 
 
In contrast, various alternative scenarios of CP violation that have been 
phenomenologically viable for many years are now unambiguously excluded. Two 
important examples are the following:
\begin{itemize}
\item The superweak framework \cite{Wolfenstein:1964ks}, that is, the idea
that CP violation is purely indirect, is excluded by the evidence that
$\varepsilon^\prime/\varepsilon\neq0$.
\item Approximate CP, that is, the idea that all CP violating phases are
small (see, for example, \cite{Eyal:1998bk}), is excluded by the
evidence that $S_{\psi K_S}={\cal O}(1)$.
\end{itemize}
Indeed, I am not aware of any viable, reasonably motivated, scenario
which provides a complete alternative to the KM mechanism, that is of
a framework where the KM phase plays no significant role in the
observed CP violation.

The experimental results from the B-factories, such as those in
Eqs.~(\ref{aveapk}-\ref{arhopi}), and their implications for theory signify
a new era in the study of CP violation. In this series of lectures we  
explain these recent developments and their significance.

\subsection{Why doubt the Kobayashi-Maskawa mechanism?}
\subsubsection{The baryon asymmetry of the Universe}
Baryogenesis is a consequence of CP violating processes
\cite{Sakharov:1967dj}. Therefore the present baryon number, which is 
accurately deduced from nucleosynthesis and CMBR constraints,
\beq\label{barden}
Y_B\equiv \frac{n_B-n_{\overline B}}{s}\simeq9\times10^{-11},
\eeq
is essentially a CP violating observable! It can be added to the list
of known CP violating observables,
Eqs.~(\ref{expeps}-\ref{arhopi}). Within a given model of CP
violation, one can check for consistency between the data from
cosmology, Eq.~(\ref{barden}), and those from laboratory experiments.

The surprising point is that the Kobayashi-Maskawa mechanism for CP violation
fails to account for (\ref{barden}). It predicts present baryon number density
that is many orders of magnitude below the observed value
\cite{Farrar:1994hn,Huet:1995jb,Gavela:1994ds}. This failure is independent
of other aspects of the Standard Model: the suppression of $Y_B$
from CP violation is much too strong, even if the departure from thermal
equilibrium is induced by mechanisms beyond the Standard Model. This situation
allows us to make the following statement: {\it There must exist sources of 
CP violation beyond the Kobayashi-Maskawa phase}.

Two important examples of viable models of baryogenesis are the following:

1. Leptogenesis \cite{Fukugita:1986hr}: a lepton asymmetry is induced
by CP violating decays of heavy fermions that are singlets of the
Standard Model gauge group (sterile neutrinos). Departure from thermal
equilibrium is provided if the lifetime of the heavy neutrino is long
enough that it decays when the temperature is below its
mass. Processes that violate $B+L$ are fast before the electroweak
phase transition and partially convert the lepton asymmetry into a
baryon asymmetry. The CP violating parameters may be related to CP
violation in the mixing matrix for the light neutrinos (but this is a
model dependent issue \cite{Branco:2001pq}).

2. Electroweak baryogenesis (for a review see \cite{Cohen:1993nk}):
the source of the baryon asymmetry is the interactions
of top (anti)quarks with the Higgs field during the electroweak phase
transition. CP violation is induced, for example, by supersymmetric
interactions. Sphaleron configurations provide baryon number violating
interactions. Departure from thermal equilibrium is provided by the
wall between the false vacuum ($\langle\phi\rangle=0$) and the expanding
bubble with the true vacuum, where electroweak symmetry is broken.

\subsubsection{The strong CP problem}
Nonperturbative QCD effects induce an additional term in the SM Lagrangian,
\beq\label{ltheta}
{\cal L}_\theta={\theta_{\rm QCD}\over32\pi^2}\epsilon_{\mu\nu\rho\sigma}
F^{\mu\nu a}F^{\rho\sigma a}.
\eeq
This term violates CP. In particular, it induces an electric dipole moment
(EDM) to the neutron. The leading contribution in the chiral limit is
given by \cite{Crewther:1979pi}
\beq\label{dntheta}
d_N={g_{\pi NN}\bar g_{\pi NN}\over4\pi^2M_N}\ln\frac{M_N}{m_\pi}
\approx5\times10^{-16}\ \theta_{\rm QCD}\ e\ {\rm cm},
\eeq
where $M_N$ is the nucleon mass, and $g_{\pi NN}$ ($\bar g_{\pi NN}$) is the
pseudoscalar coupling (CP-violating scalar coupling) of the pion to the nucleon.
(The leading contribution in the large $N_c$ limit was calculated in the Skyrme
model \cite{Dixon:1991cq} and leads to a similar estimate.) The experimental 
bound on $d_N$ is given by \cite{Harris:1999jx}
\beq\label{dnexp}
d_N\leq 6.3\times10^{-26}\ e\ {\rm cm}.
\eeq
It leads to the following bound on $\theta_{\rm QCD}$:
\beq\label{bouthe}
\theta_{\rm QCD}\lsim10^{-10}.
\eeq

Since $\theta_{\rm QCD}$ arises from nonperturbative QCD effects, it is
impossible to calculate it. Yet, there are good reasons to expect that these
effects should yield $\theta_{\rm QCD}={\cal O}(1)$ (for a review, see
\cite{Dine:2000cj}). Within the SM, a value as small
as in Eq.~(\ref{bouthe}) is unnatural, since setting $\theta_{\rm QCD}$ to zero
does not add symmetry to the model. [In particular, as we will see below, 
CP is violated by $\delta_{\rm KM}={\cal O}(1)$.] Understanding why CP is so
small in the strong interactions is the strong CP problem.

It seems then that the strong CP problem is a clue to new physics. Among
the solutions that have been proposed are a massless $u$-quark (for a review,
see \cite{Banks:1994yg}), the Peccei-Quinn mechanism
\cite{Peccei:1977hh,Peccei:1977ur} and spontaneous CP violation. 

\subsubsection{New Physics}
Almost any extension of the Standard Model provides new sources of CP
violation. For example, in the supersymmetric extension of the
Standard Model (with R-parity), there are 44 independent phases, most
of them in flavor changing couplings. If there is new physics at or
below the TeV scale, it is quite likely that the KM phase is not the
only source of CP violation that is playing a role in meson decays. 

\subsection{Will new CP violation be observed in experiments?}
The SM picture of CP violation is testable because the Kobayashi-Maskawa
mechanism is unique and predictive. These features are mainly related to
the fact that there is a single phase that is responsible to all CP violation.
As a consequence of this situation, one finds two classes of tests:

(i) Correlations: many independent CP violating observables are correlated
within the SM. For example, the SM predicts that the CP asymmetries in 
$B\to\psi K_S$ and in $B\to\phi K_S$, which proceed through different
quark transitions, are equal to each other (to a few percent
accuracy) \cite{Grossman:1997ke,Grossman:1997gr}. Another important
example is the strong SM correlation 
between CP violation in $B\to\psi K_S$ and in $K\to\pi\nu\bar\nu$
\cite{Buchalla:1994tr,Buchalla:1996fp,Bergmann:2000ak}.
It is a significant fact, in this context, that several CP violating
observables can be calculated with very small hadronic uncertainties.
To search for violations of the correlations, precise measurements are
important. 

(ii) Zeros: since the KM phase appears in flavor-changing, weak-interaction
couplings of quarks, and only if all three generations are involved, many
CP violating observables are predicted to be negligibly small. For
example, the transverse lepton polarization in semileptonic meson
decays, CP violation in $t\bar t$ production, tree level $D$ decays,
and (assuming $\theta_{\rm QCD}=0$) the electric dipole moment of the
neutron are all predicted to be orders of magnitude below the (present
and near future) experimental sensitivity. To search for lifted zeros, 
measurements of CP violation in many different systems should be
performed. 

The strongest argument that new sources of CP violation must exist in Nature 
comes from baryogenesis. Whether the CP violation that is responsible for 
baryogenesis would be manifest in measurements of CP asymmetries in $B$ decays 
depends on two issues:

(i) The scale of the new CP violation: if the relevant scale is very high, 
such as in leptogenesis, the effects cannot be signalled in these
measurements. To estimate the limit on the scale, the following three 
facts are relevant: First, the Standard Model contributions to CP
asymmetries in $B$ decays are ${\cal O}(1)$. Second, the expected
experimental accuracy would reach in some cases the few percent
level. Third, the contributions from new physics are expected to be suppressed 
by $(\Lambda_{\rm EW}/\Lambda_{\rm NP})^2$. The conclusion is that, if the new
source of CP violation is related to physics at $\Lambda_{\rm NP}\gg1\ TeV$,
it cannot be signalled in $B$ decays. Only if the true mechanism is
electroweak baryogenesis, it can potentially affect $B$ decays.

(ii) The flavor dependence of the new CP violation: if it is flavor diagonal,
its effects on $B$ decays would be highly suppressed. It can still manifest
itself in other, flavor diagonal  CP violating observables, such as electric 
dipole moments.

We conclude that new measurements of CP asymmetries in meson decays are
particularly sensitive to new sources of CP violation that come from physics
at (or below) the few TeV scale and that are related to flavor changing
couplings. This is, for example, the case, in certain supersymmetric models
of baryogenesis \cite{Worah:1997hk,Worah:1997ni}. The search for electric 
dipole moments can reveal the existence of new flavor diagonal CP violation.

Of course, there could be new flavor physics at the TeV scale that is not 
related to the baryon asymmetry and may give signals in $B$ decays. The best
motivated extension of the SM where this situation is likely is that of
supersymmetry.

Finally, we would like to mention that, in the past, flavor physics
and the physics of CP violation led indeed to the discovery of new
physics or to probing it before it was directly observed in
experiments: 
\begin{itemize}
\item The smallness of $\frac{\Gamma(K_L\to\mu^+\mu^-)}
  {\Gamma(K^+\to\mu^+\nu)}$ led to
    predicting a fourth (the charm) quark;
    \item The size of $\Delta m_K$ led to a successful prediction of
      the charm mass;
      \item The size of $\Delta m_B$ led to a successful prediction of
        the top mass;
        \item The measurement of $\varepsilon_K$ led to predicting the
          third generation.
        \end{itemize}
        
\section{The Kobayashi-Maskawa Mechanism}
\subsection{Yukawa interactions are the source of CP violation}
A model of elementary particles and their interactions is defined
by three ingredients:
\begin{enumerate}
\item The symmetries of the Lagrangian;
\item The representations of fermions and scalars;
\item The pattern of spontaneous symmetry breaking.
\end{enumerate}
The Standard Model (SM) is defined as follows:

1. The gauge symmetry is 
\beq\label{smsym}
G_{\rm SM}=SU(3)_{\rm C}\times SU(2)_{\rm L}\times U(1)_{\rm Y}.
\eeq

2. There are three fermion generations, each consisting of five 
representations of $G_{\rm SM}$:
\beq\label{ferrep}
Q^I_{Li}(3,2)_{+1/6},\ \ U^I_{Ri}(3,1)_{+2/3},\ \ 
D^I_{Ri}(3,1)_{-1/3},\ \ L^I_{Li}(1,2)_{-1/2},\ \ E^I_{Ri}(1,1)_{-1}.
\eeq
Our notations mean that, for example, left-handed quarks, $Q_L^I$, are
triplets of $SU(3)_{\rm C}$, doublets of $SU(2)_{\rm L}$ and carry hypercharge
$Y=+1/6$. The super-index $I$ denotes interaction eigenstates. The sub-index
$i=1,2,3$ is the flavor (or generation) index.
There is a single scalar representation,
\beq\label{scarep}
\phi(1,2)_{+1/2}.
\eeq

3. The scalar $\phi$ assumes a VEV,
\beq\label{phivev}
\langle\phi\rangle=\pmatrix{0\cr {v\over\sqrt2}\cr},
\eeq
so that the gauge group is spontaneously broken,
\beq\label{smssb}
G_{\rm SM}\to SU(3)_{\rm C}\times U(1)_{\rm EM}.
\eeq

The Standard Model Lagrangian, ${\cal L}_{\rm SM}$, is the most general
renormalizable Lagrangian that is consistent with the gauge symmetry 
(\ref{smsym}), the particle content (\ref{ferrep},\ref{scarep}) and
the pattern of spontaneous symmetry breaking (\ref{phivev}). It can be
divided to three parts: 
\beq\label{LagSM}
{\cal L}_{\rm SM}={\cal L}_{\rm kinetic}+{\cal L}_{\rm Higgs}
+{\cal L}_{\rm Yukawa}.
\eeq

As concerns the kinetic terms, to maintain gauge invariance, one has 
to replace the derivative with a covariant derivative:
\beq\label{SMDmu}
D^\mu=\partial^\mu+ig_s G^\mu_a L_a+ig W^\mu_b T_b+ig^\prime B^\mu Y.
\eeq
Here $G^\mu_a$ are the eight gluon fields, $W^\mu_b$ the three
weak interaction bosons and $B^\mu$ the single hypercharge boson.
The $L_a$'s are $SU(3)_{\rm C}$ generators (the $3\times3$
Gell-Mann matrices ${1\over2}\lambda_a$ for triplets, $0$ for singlets),
the $T_b$'s are $SU(2)_{\rm L}$ generators (the $2\times2$
Pauli matrices ${1\over2}\tau_b$ for doublets, $0$ for singlets),
and the $Y$'s are the $U(1)_{\rm Y}$ charges. For example, for the
left-handed quarks $Q_L^I$, we have
\beq\label{DmuQL}
{\cal L}_{\rm kinetic}(Q_L)= i{\overline{Q_{Li}^I}}\gamma_\mu
\left(\partial^\mu+{i\over2}g_s G^\mu_a\lambda_a
+{i\over2}g W^\mu_b\tau_b+{i\over6}g^\prime B^\mu\right)Q_{Li}^I,
\eeq
while for the left-handed leptons $L_L^I$, we have
\beq\label{DmuLL}
{\cal L}_{\rm kinetic}(L_L)=i{\overline{L_{Li}^I}}\gamma_\mu 
\left(\partial^\mu+{i\over2}g W^\mu_b\tau_b-ig^\prime B^\mu\right)L_{Li}^I.
\eeq
These parts of the interaction Lagrangian are always CP conserving.

The Higgs potential, which describes the scalar self interactions, is given by:
\beq\label{HiPo}
{\cal L}_{\rm Higgs}=\mu^2\phi^\dagger\phi-\lambda(\phi^\dagger\phi)^2.
\eeq
For the Standard Model scalar sector, where there is a single doublet,
this part of the Lagrangian is also CP conserving.  For extended scalar
sectors, such as that of a two Higgs doublet model, ${\cal L}_{\rm Higgs}$ can
be CP violating. Even in case that it is CP symmetric, it may lead
to spontaneous CP violation.

The quark Yukawa interactions are given by
\beq\label{Hqint}
-{\cal L}_{\rm Yukawa}^{\rm quarks}=Y^d_{ij}{\overline {Q^I_{Li}}}\phi D^I_{Rj}
+Y^u_{ij}{\overline {Q^I_{Li}}}\tilde\phi U^I_{Rj}+{\rm h.c.}.
\eeq
This part of the Lagrangian is, in general, CP violating.
More precisely, CP is violated if and only if \cite{Jarlskog:1985ht}
\beq\label{JarCon}
\im{\det[Y^d Y^{d\dagger},Y^u Y^{u\dagger}]}\neq0.
\eeq

An intuitive explanation of why CP violation is related to {\it complex} 
Yukawa couplings goes as follows. The hermiticity of the Lagrangian implies
that ${\cal L}_{\rm Yukawa}$ has its terms in pairs of the form
\beq\label{Yukpairs}
Y_{ij}\overline{\psi_{Li}}\phi\psi_{Rj}
+Y_{ij}^*\overline{\psi_{Rj}}\phi^\dagger\psi_{Li}.
\eeq
A CP transformation exchanges the operators 
\beq\label{CPoper}
\overline{\psi_{Li}}\phi\psi_{Rj}\leftrightarrow
\overline{\psi_{Rj}}\phi^\dagger\psi_{Li},
\eeq 
but leaves their coefficients, $Y_{ij}$ and $Y_{ij}^*$, unchanged. This means 
that CP is a symmetry of ${\cal L}_{\rm Yukawa}$ if $Y_{ij}=Y_{ij}^*$.

The lepton Yukawa interactions are given by
\beq\label{Hlint}
-{\cal L}_{\rm Yukawa}^{\rm leptons}=
Y^e_{ij}{\overline {L^I_{Li}}}\phi E^I_{Rj}+{\rm h.c.}.
\eeq
It leads, as we will see in the next section, to charged lepton masses
but predicts massless neutrinos. Recent measurements of the fluxes of 
atmospheric and solar neutrinos provide evidence for neutrino masses
(for a review, see \cite{Gonzalez-Garcia:2002dz}).
That means that ${\cal L}_{\rm SM}$ cannot be a complete description of
Nature. The simplest way to allow for neutrino masses is to add 
dimension-five (and, therefore, non-renormalizable) terms, consistent with the 
SM symmetry and particle content:
\beq\label{Hnint}
-{\cal L}_{\rm Yukawa}^{\rm dim-5}=
{Y_{ij}^\nu\over M}L_iL_j\phi\phi+{\rm h.c.}.
\eeq
The parameter $M$ has dimension of mass. The dimensionless couplings
$Y^\nu_{ij}$ are symmetric ($Y^\nu_{ij}=Y^\nu_{ji}$). We refer to the SM 
extended to include the terms ${\cal L}_{\rm Yukawa}^{\rm dim-5}$ of Eq. 
(\ref{Hnint}) as the ``extended SM" (ESM):
\beq\label{LagESM}
{\cal L}_{\rm ESM}={\cal L}_{\rm kinetic}+{\cal L}_{\rm Higgs}
+{\cal L}_{\rm Yukawa}+{\cal L}_{\rm Yukawa}^{\rm dim-5}.
\eeq
The inclusion of non-renormalizable terms is equivalent to postulating that the 
SM is only a low energy effective theory, and that new physics appears at the 
scale $M$.

How many independent CP violating parameters are there in 
${\cal L}_{\rm Yukawa}^{\rm quarks}$? Each of the two Yukawa matrices $Y^q$ 
($q=u,d$) is $3\times3$ and complex. Consequently, there are 18 real and 18 
imaginary parameters in these matrices. Not all of them are, however, physical.
One can think of the quark Yukawa couplings as spurions that break a global
symmetry,
\beq\label{Gglobq}
U(3)_Q\times U(3)_{D}\times U(3)_{U}\ \to\ U(1)_B.
\eeq
This means that there is freedom to remove 9 real and 17 imaginary parameters 
[the number of parameters in three $3\times3$ unitary matrices minus the phase
related to $U(1)_B$]. We conclude that there are 10 quark flavor parameters: 9 
real ones and a single phase. This single phase is the source of CP violation
in the quark sector.

How many independent CP violating parameters are there in the lepton Yukawa 
interactions? The matrix $Y^e$ is a general complex $3\times3$ matrix and 
depends, therefore, on 9 real and 9 imaginary parameters. The matrix $Y^\nu$ is 
symmetric and depends on 6 real and 6 imaginary parameters. Not all of these 15
real and 15 imaginary parameters are physical. One can think of the lepton 
Yukawa couplings as spurions that break (completely) a global symmetry,
\beq\label{Gglobl}
U(3)_L\times U(3)_{E}.
\eeq
This means that 6 real and 12 imaginary parameters are not physical. We 
conclude that there are 12 lepton flavor parameters: 9 real ones and three 
phases. These three phases induce CP violation in the lepton sector.

\subsection{CKM mixing is the (only!) source of CP violation 
in the quark sector}
Upon the replacement $\re{\phi^0}\to\frac{v+H^0}{\sqrt2}$ [see
Eq.~(\ref{phivev})], the Yukawa interactions (\ref{Hqint}) give rise
to mass terms: 
\beq\label{fermasq}
-{\cal L}_M^q=(M_d)_{ij}{\overline {D^I_{Li}}} D^I_{Rj}
+(M_u)_{ij}{\overline {U^I_{Li}}} U^I_{Rj}+{\rm h.c.},
\eeq
where
\beq\label{YtoMq}
M_q={v\over\sqrt2}Y^q,
\eeq
and we decomposed the $SU(2)_{\rm L}$ quark doublets into their components:
\beq\label{doublets}
Q^I_{Li}=\pmatrix{U^I_{Li}\cr D^I_{Li}\cr}.
\eeq

The mass basis corresponds, by definition, to diagonal mass matrices. We can 
always find unitary matrices $V_{qL}$ and $V_{qR}$ such that
\beq\label{diagMq}
V_{qL}M_q V_{qR}^\dagger=M_q^{\rm diag}\ \ \ (q=u,d),
\eeq
with $M_q^{\rm diag}$ diagonal and real. The quark mass eigenstates
are then identified as
\beq\label{masses}
q_{Li}=(V_{qL})_{ij}q_{Lj}^I,\ \ \ q_{Ri}=(V_{qR})_{ij}q_{Rj}^I\ \ \ (q=u,d).
\eeq

The charged current interactions for quarks [that is the interactions of the 
charged $SU(2)_{\rm L}$ gauge bosons $W^\pm_\mu={1\over\sqrt2}
(W^1_\mu\mp iW_\mu^2)$], which in the interaction basis are described 
by (\ref{DmuQL}), have a complicated form in the mass basis:
\beq\label{Wmasq}
-{\cal L}_{W^\pm}^q={g\over\sqrt2}{\overline {u_{Li}}}\gamma^\mu
(V_{uL}V_{dL}^\dagger)_{ij}d_{Lj} W_\mu^++{\rm h.c.}.
\eeq
The unitary $3\times3$ matrix,
\beq\label{VCKM}
V=V_{uL}V_{dL}^\dagger,\ \ \ 
(VV^\dagger={\bf 1}),
\eeq 
is the Cabibbo-Kobayashi-Maskawa (CKM) {\it mixing matrix} for quarks
\cite{Cabibbo:1963yz,Kobayashi:1973fv}. A unitary $3\times3$ matrix depends on 
nine parameters: three real angles and six phases. 

The form of the matrix is not unique:

$(i)$ There is freedom in defining $V$ in that we can permute between
the various generations. This freedom is fixed by ordering the up quarks and 
the down quarks by their masses, {\it i.e.} $(u_1,u_2,u_3)\to(u,c,t)$ and 
$(d_1,d_2,d_3)\to(d,s,b)$. The elements of $V$ are written as follows:
\beq\label{defVij}
V=\pmatrix{V_{ud}&V_{us}&V_{ub}\cr
V_{cd}&V_{cs}&V_{cb}\cr V_{td}&V_{ts}&V_{tb}\cr}.
\eeq

$(ii)$ There is further freedom in the phase structure of $V$. Let us
define $P_q$ ($q=u,d$) to be diagonal unitary (phase) matrices. Then, if 
instead of using $V_{qL}$ and $V_{qR}$ for the rotation (\ref{masses}) to the 
mass basis we use $\tilde V_{qL}$ and $\tilde V_{qR}$, defined by
$\tilde V_{qL}=P_q V_{qL}$ and $\tilde V_{qR}=P_q V_{qR}$,
we still maintain a legitimate mass basis since $M_q^{\rm diag}$ remains
unchanged by such transformations. However, $V$ does change:
\beq\label{eqphase}
V\to P_u V P_d^*.
\eeq 
This freedom is fixed by demanding that $V$ has the minimal number of
phases. In the three generation case $V$ has a single phase. (There 
are five phase differences between the elements of $P_u$ and $P_d$ and, 
therefore, five of the six phases in the CKM matrix can be removed.) This is 
the Kobayashi-Maskawa phase $\delta_{\rm KM}$ which is the single source of 
CP violation in the quark sector of the Standard Model \cite{Kobayashi:1973fv}. 

As a result of the fact that $V$ is not diagonal, the $W^\pm$ gauge 
bosons couple to quark (mass eigenstates) of different generations. Within the 
Standard Model, this is the only source of {\it flavor changing} quark
interactions. 

\subsection{The three phases in the lepton mixing matrix}
The leptonic Yukawa interactions (\ref{Hlint}) and (\ref{Hnint}) give rise to 
mass terms:
\beq\label{fermasl}
-{\cal L}^\ell_M=(M_e)_{ij}{\overline {e^I_{Li}}}e^I_{Rj}
+(M_\nu)_{ij}\nu^I_{Li}\nu^I_{Lj}+{\rm h.c.},
\eeq
where
\beq\label{YtoMl}
M_e={v\over\sqrt2}Y^e,\ \ \ M_\nu={v^2\over2M}Y^\nu,
\eeq
and we decomposed the $SU(2)_{\rm L}$ lepton doublets into their components:
\beq\label{ldoublets}
L^I_{Li}=\pmatrix{\nu^I_{Li}\cr e^I_{Li}\cr}.
\eeq

We can always find unitary matrices $V_{eL}$ and $V_\nu$ such that
\beq\label{diagMl}
V_{eL}M_eM_e^\dagger V_{eL}^\dagger={\rm diag}(m_e^2,m_\mu^2,m_\tau^2),\ \ \ 
V_\nu M_\nu^\dagger M_\nu V_\nu^\dagger={\rm diag}(m_1^2,m_1^2,m_3^2).
\eeq
The charged current interactions for leptons, which in the interaction basis 
are described by (\ref{DmuLL}), have the following form in the mass basis:
\beq\label{Wmasl}
-{\cal L}_{W^\pm}^\ell={g\over\sqrt2}{\overline {e_{Li}}}\gamma^\mu
(V_{eL}V_{\nu}^\dagger)_{ij}\nu_{Lj} W_\mu^-+{\rm h.c.}.
\eeq
The unitary $3\times3$ matrix,
\beq\label{VMNS}
U=V_{eL}V_{\nu}^\dagger,
\eeq 
is the {\it lepton mixing matrix} \cite{Maki:1962mu}. Similarly to the
CKM matrix, the form of the lepton mixing matrix is 
not unique. But there are differences in choosing conventions:

$(i)$ We can permute between the various generations. This freedom is usually 
fixed in the following way. We order the charged leptons by their masses, 
{\it i.e.} $(e_1,e_2,e_3)\to(e,\mu,\tau)$. As concerns the neutrinos, one takes
into account that the atmospheric and solar neutrino data imply that
$\Delta m^2_{\rm atm}\gg \Delta m^2_{\rm sol}$. It follows that one of
the neutrino mass eigenstates is  
separated in its mass from the other two, which have a smaller mass difference.
The convention is to denote this separated state by $\nu_3$. For the remaining 
two neutrinos, $\nu_1$ and $\nu_2$, the convention is to call the heavier state
$\nu_2$. In other words, the three mass eigenstates are defined by the 
following conventions:
\beq\label{conneu}
|\Delta m^2_{3i}|\gg|\Delta m^2_{21}|,\ \ \ \Delta m^2_{21}>0.
\eeq
Note in particular that $\nu_3$ can be either heavier (`normal
hierarchy') or lighter (`inverted hierarchy') than 
$\nu_{1,2}$. The elements of $U$ are written as  follows:
\beq\label{defVijl}
U=\pmatrix{U_{e1}&U_{e2}&U_{e3}\cr
U_{\mu1}&U_{\mu2}&U_{\mu3}\cr U_{\tau1}&U_{\tau2}&U_{\tau3}\cr}.
\eeq

$(ii)$ There is further freedom in the phase structure of $U$. 
One can change the charged lepton mass basis by the transformation
$e_{(L,R)i}\to e^\prime_{(L,R)i}=(P_e)_{ii} e_{(L,R)i}$, where $P_e$ is a phase 
matrix. There is, however, no similar freedom to redefine the neutrino
mass eigenstates: From Eq. (\ref{fermasl}) one learns that a transformation
$\nu_{L}\to P_\nu\nu_{L}$ will introduce phases into the diagonal mass matrix. 
This is related to the Majorana nature of neutrino masses, assumed in
Eq. (\ref{Hnint}). The allowed transformation modifies $U$:
\beq\label{ephase}
U\to P_e U.
\eeq 
This freedom is fixed by demanding that $U$ will have the minimal 
number of phases. Out of six phases of a generic unitary $3\times3$ matrix,
the multiplication by $P_e$ can be used to remove three. We conclude that the 
three generation $U$ matrix has three
phases. One of these is the analog of the Kobayashi-Maskawa phase. It is 
the only source of CP violation in processes that conserve lepton number,
such as neutrino flavor oscillations. The other two phases
can affect lepton number changing processes.

With $U\neq{\bf 1}$, the $W^\pm$ gauge bosons couple to lepton (mass 
eigenstates) of different generations. Within the ESM, this is the 
only source of {\it flavor changing} lepton interactions. 

\subsection{The flavor parameters}
Examining the quark mass basis, one can easily identify the flavor parameters.
In the quark sector, we have six quark masses and four mixing parameters:
three mixing angles and a single phase. 

The fact that there are only three real and one imaginary physical parameters 
in $V$ can be made manifest by choosing an explicit parameterization.
For example, the standard parameterization \cite{Chau:1984fp}, used by the 
particle data group, is given by 
\beq\label{stapar}
V=\pmatrix{c_{12}c_{13}&s_{12}c_{13}&
s_{13}e^{-i\delta}\cr 
-s_{12}c_{23}-c_{12}s_{23}s_{13}e^{i\delta}&
c_{12}c_{23}-s_{12}s_{23}s_{13}e^{i\delta}&s_{23}c_{13}\cr
s_{12}s_{23}-c_{12}c_{23}s_{13}e^{i\delta}&
-c_{12}s_{23}-s_{12}c_{23}s_{13}e^{i\delta}&c_{23}c_{13}\cr},
\eeq
where $c_{ij}\equiv\cos\theta_{ij}$ and $s_{ij}\equiv\sin\theta_{ij}$. The 
three $\sin\theta_{ij}$ are the three real mixing parameters while $\delta$ is 
the Kobayashi-Maskawa phase. Another, very useful, example is the Wolfenstein 
parametrization, where the four mixing parameters are $(\lambda,A,\rho,\eta)$ 
with $\lambda=|V_{us}|=0.22$ playing the role of an expansion parameter
and $\eta$ representing the CP violating phase
\cite{Wolfenstein:1983yz,Buras:1994ec}: 
\beq\label{wolpar}
V=\pmatrix{
1-\frac12\lambda^2-\frac18\lambda^4 & \lambda & A\lambda^3(\rho-i\eta)\cr
-\lambda +\frac12A^2\lambda^5[1-2(\rho+i\eta)] &
1-\frac12\lambda^2-\frac18\lambda^4(1+4A^2) & A\lambda^2 \cr
A\lambda^3[1-(1-\frac12\lambda^2)(\rho+i\eta)]&-A\lambda^2+\frac12A\lambda^4[1-2(\rho+i\eta)]
& 1-\frac12A^2\lambda^4 \cr}\; .
\eeq

Various parametrizations differ in the way that the freedom of phase rotation, 
Eq. (\ref{eqphase}), is used to leave a single phase in $V$. One can 
define, however, a CP violating quantity in $V_{\rm CKM}$ that is independent 
of the parametrization \cite{Jarlskog:1985ht}. This quantity, $J_{\rm CKM}$, is 
defined through 
\beq\label{defJ}
\im{V_{ij}V_{kl}V_{il}^*V_{kj}^*}=
J_{\rm CKM}\sum_{m,n=1}^3\epsilon_{ikm}\epsilon_{jln},\ \ \ (i,j,k,l=1,2,3).
\eeq
In terms of the explicit parametrizations given above, we have
\beq\label{parJ}
J_{\rm CKM}=c_{12}c_{23}c_{13}^2s_{12}s_{23}s_{13}\sin\delta\simeq \lambda^6 
A^2\eta.
\eeq

It is interesting to translate the condition (\ref{JarCon}) to the language
of the flavor parameters in the mass basis. One finds that the following is
a necessary and sufficient condition for CP violation in the quark sector of
the SM (we define $\Delta m^2_{ij}\equiv m_i^2-m_j^2$):
\beq\label{jarconmas}
\Delta m^2_{tc}\Delta m^2_{tu}\Delta m^2_{cu}\Delta m^2_{bs}\Delta m^2_{bd}
\Delta m^2_{sd}J_{\rm CKM}\neq0.
\eeq
Equation (\ref{jarconmas}) puts the following requirements on the SM in order
that it violates CP:

(i) Within each quark sector, there should be no mass degeneracy;

(ii) None of the three mixing angles should be zero or $\pi/2$;

(iii) The phase should be neither 0 nor $\pi$.

As concerns the lepton sector of the ESM, the flavor parameters are the six
lepton masses, and six mixing parameters: three mixing angles and three phases.
One can parameterize $U$ in a convenient way by factorizing it into 
$U=\hat UP$. Here $P$ is a diagonal unitary matrix that depends on two 
phases, {\it e.g.} $P={\rm diag}(e^{i\phi_1},e^{i\phi_2},1)$, while
$\hat U$ can be parametrized in the same way as (\ref{stapar}). The
advantage of this  
parametrization is that for the purpose of analyzing lepton number conserving
processes and, in particular, neutrino flavor oscillations, the parameters of 
$P$ are usually irrelevant and one can use the same Chau-Keung parametrization 
as is being used for $V$. (An alternative way to understand these
statements is to use a single-phase mixing matrix and put the extra two phases
in the neutrino mass matrix. Then it is obvious that the effects of these
`Majorana-phases' always appear in conjunction with a factor of the
Majorana mass that is lepton number violating parameter.) On the other hand, 
the Wolfenstein parametrization [Eq.~(\ref{wolpar})] is inappropriate for the lepton 
sector: it assumes $|V_{23}|\ll|V_{12}|\ll1$, which does not hold here.

In order that the CP violating phase $\delta$ in $\hat U$ would be physically
meaningful, {\it i.e.} there would be CP violation that is not related to
lepton number violation, a condition similar to Eq.~(\ref{jarconmas}) should hold:
\beq\label{jarconlep}
\Delta m^2_{\tau\mu}\Delta m^2_{\tau e}\Delta m^2_{\mu e}
\Delta m^2_{32}\Delta m^2_{31}\Delta m^2_{21}J_{\ell}\neq0.
\eeq

\subsection{The unitarity triangles}
\begin{figure}[tb]
  \centering
  {\includegraphics[width=0.65\textwidth]{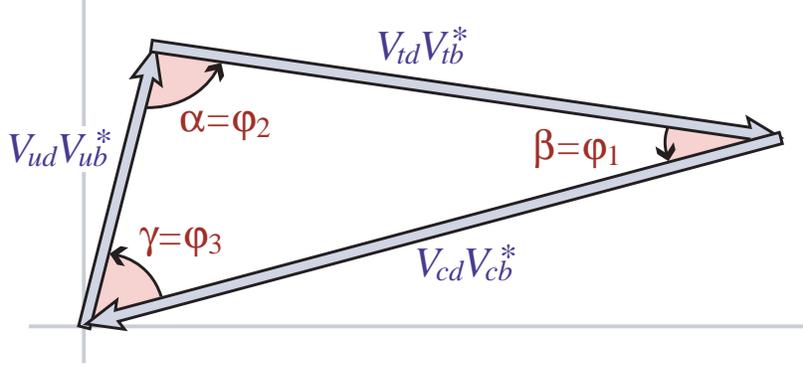}}
  \caption{Graphical representation of the unitarity constraint
  $V_{ud}V_{ub}^*+V_{cd}V_{cb}^*+V_{td}V_{tb}^*=0$ as a triangle in
  the complex plane.}
  \label{fg:tri}
\end{figure}

A very useful concept is that of the {\it unitarity triangles}. We focus
on the quark sector, but analogous triangles can be defined in the lepton
sector. The unitarity of the CKM matrix leads to various relations among
the matrix elements, {\it e.g.}
\beqa\label{Unitds}
V_{ud}V_{us}^*+V_{cd}V_{cs}^*+V_{td}V_{ts}^*=0,\\
\label{Unitsb}
V_{us}V_{ub}^*+V_{cs}V_{cb}^*+V_{ts}V_{tb}^*=0,\\
\label{Unitdb}
V_{ud}V_{ub}^*+V_{cd}V_{cb}^*+V_{td}V_{tb}^*=0.
\eeqa
Each of these three relations requires 
the sum of three complex quantities to vanish and so can be geometrically
represented in the complex plane as a triangle. These are
``the unitarity triangles", though the term ``unitarity triangle"
is usually reserved for the relation (\ref{Unitdb}) only. The
unitarity triangle related to Eq. (\ref{Unitdb}) is depicted in
Fig. \ref{fg:tri}. 

It is a surprising feature of the CKM matrix that all unitarity
triangles are equal in area: the area of each unitarity triangle
equals $|J_{\rm CKM}|/2$ while the sign of $J_{\rm CKM}$ gives the
direction of the complex vectors around the triangles. 

The rescaled unitarity triangle  is derived from (\ref{Unitdb})
by (a) choosing a phase convention such that $(V_{cd}V_{cb}^*)$
is real, and (b) dividing the lengths of all sides by $|V_{cd}V_{cb}^*|$.
Step (a) aligns one side of the triangle with the real axis, and
step (b) makes the length of this side 1. The form of the triangle
is unchanged. Two vertices of the rescaled unitarity triangle are
thus fixed at (0,0) and (1,0). The coordinates of the remaining
vertex correspond to the Wolfenstein parameters $(\rho,\eta)$.
The area of the rescaled unitarity triangle is $|\eta|/2$.

Depicting the rescaled unitarity triangle in the
$(\rho,\eta)$ plane, the lengths of the two complex sides are
\beq\label{RbRt}
R_u\equiv\left|{V_{ud}V_{ub}\over V_{cd}V_{cb}}\right|
=\sqrt{\rho^2+\eta^2},\ \ \
R_t\equiv\left|{V_{td}V_{tb}\over V_{cd}V_{cb}}\right|
=\sqrt{(1-\rho)^2+\eta^2}.
\eeq
The three angles of the unitarity triangle are defined as follows 
\cite{Dib:1989uz,Rosner:1988nx}:
\beq\label{abcangles}
\alpha\equiv\arg\left[-{V_{td}V_{tb}^*\over V_{ud}V_{ub}^*}\right],\ \ \
\beta\equiv\arg\left[-{V_{cd}V_{cb}^*\over V_{td}V_{tb}^*}\right],\ \ \
\gamma\equiv\arg\left[-{V_{ud}V_{ub}^*\over V_{cd}V_{cb}^*}\right].
\eeq
They are physical quantities and can be independently measured by CP
asymmetries in $B$ decays. It is also useful to define the two 
small angles of the unitarity triangles (\ref{Unitsb}) and (\ref{Unitds}):
\beq\label{bbangles}
\beta_s\equiv\arg\left[-{V_{ts}V_{tb}^*\over V_{cs}V_{cb}^*}\right],\ \ \
\beta_K\equiv\arg\left[-{V_{cs}V_{cd}^*\over V_{us}V_{ud}^*}\right].
\eeq

To make predictions for CP violating observables, we 
need to find the allowed ranges for the CKM phases. There are three ways to 
determine the CKM parameters (see {\it e.g.} \cite{Harari:1987ex}):

(i) {\bf Direct measurements} are related to SM tree level  processes. 
At present, we have direct measurements of $|V_{ud}|$, $|V_{us}|$, $|V_{ub}|$,
$|V_{cd}|$, $|V_{cs}|$, $|V_{cb}|$ and $|V_{tb}|$. 

(ii) {\bf CKM Unitarity}  ($V^\dagger V={\bf 1}$) relates 
the various matrix elements. At present, these relations are useful to 
constrain $|V_{td}|$, $|V_{ts}|$, $|V_{tb}|$ and $|V_{cs}|$.
 
(iii) {\bf Indirect measurements} are related to SM loop processes. 
At present, we constrain in this way $|V_{tb}V_{td}|$ (from $\Delta m_B$ 
and  $\Delta m_{B_s}$) and the phase structure of the matrix (for
example, from $\varepsilon_K$ and $S_{\psi K_S}$).

Direct measurements are expected to hold almost model
independently. Most extensions of the SM have a special flavor
structure that suppresses flavor changing couplings and, in addition,
have a mass scale $\Lambda_{\rm NP}$, that is higher than the
electroweak breaking scale. Consequently, new physics contributions to
tree level processes are suppressed, compared to the SM ones, by at
least ${\cal O}(m_Z^2/\Lambda_{\rm NP}^2)\ll1$.

Unitarity holds if the only quarks (that is fermions in color triplets
with electric charge $+2/3$ or $-1/3$) are those of the three
generations of the SM. This is the situation in many extensions of the
SM, including the supersymmetric SM (SSM).

Using tree level constraints and unitarity, the 90\% confidence limits
on the magnitude of the elements are \cite{Eidelman:2004wy}
\beq\label{ckmval}
\pmatrix{0.9739-0.9751&0.221-0.227&0.0029-0.0045\cr
  0.221-0.227&0.9730-0.9744&0.039-0.044\cr
  0.0048-0.014&0.037-0.043&0.9990-0.9992\cr}.
\eeq
Note that $|V_{ub}|$ and $|V_{td}|$ are the only elements with
uncertainties of order one.

Indirect measurements are sensitive to new physics. Take, for example,
the $B^0-\overline{B}{}^0$ mixing amplitude. Within the SM, the leading
contribution comes from an electroweak box diagram and is therefore
${\cal O}(g^4)$ and depends on small mixing angles,
$(V_{td}^*V_{tb})^2$. (It is this dependence on the CKM elements that
makes the relevant indirect measurements, particularly $\Delta m_B$
and $S_{\psi K_S}$, very significant in improving our knowledge of the
CKM matrix.) These suppression factors do not necessarily persist in
extensions of the SM. For example, in the SSM there are
(gluino-mediated) contributions of ${\cal O}(g_s^4)$ and the mixing
angles could be comparable to, or even larger than the SM ones. The
validity of indirect measurements is then model dependent. Conversely,
inconsistencies among indirect measurements (or between indirect and
direct measurements) can give evidence for new physics. 

When all available data are taken into account, one finds \cite{ckmfitter}:
\beqa\label{lacon}
\lambda&=&0.226\pm0.001,\ \ \ A=0.83\pm0.03,\\
\label{recon}
\bar\rho&=&0.21\pm0.04,\ \ \ \bar\eta=0.33\pm0.02,\\
\label{abccon}
\sin2\beta&=&0.720\pm0.025,\ \ \ \alpha=(99\pm7)^o,\ \ \ 
\gamma=(58\pm7)^o,\ \ \ \beta_s=(1.03\pm0.08)^o,\\
R_u&=&0.40\pm0.02,\ \ \ R_t=0.86\pm0.04.
\eeqa

\begin{figure}[tb]
  \centering
  {\includegraphics[width=0.65\textwidth]{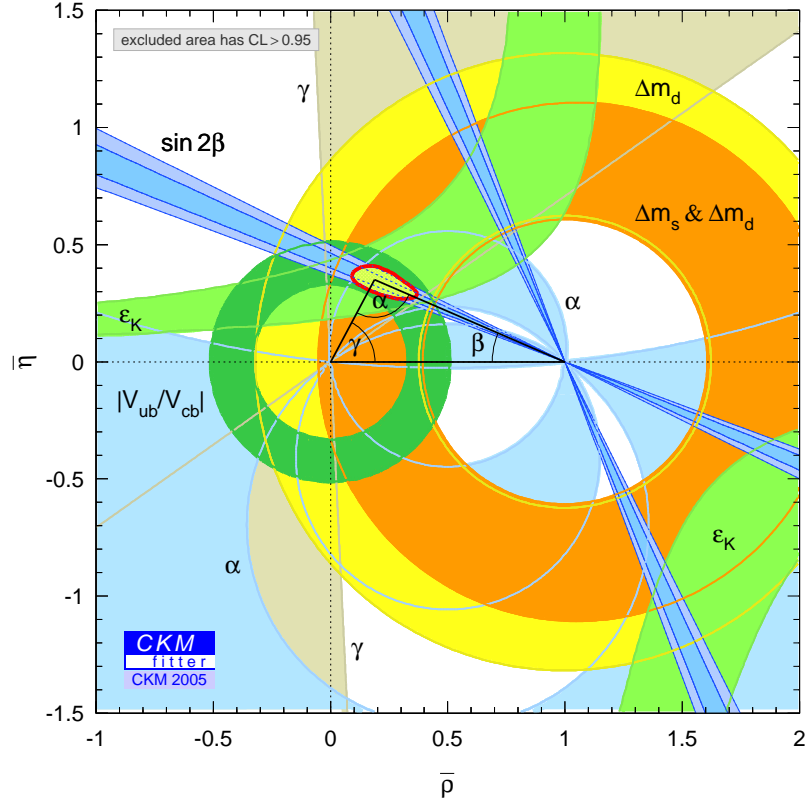}}
  \caption{Allowed region in the $\rho,\eta$ plane. Superimposed are
  the individual constraints from charmless semileptonic $B$ decays
  ($|V_{ub}/V_{cb}|$), mass differences in the $B^0$ ($\Delta m_d$)
  and $B_s$ ($\Delta m_s$) neutral meson systems, and CP violation in
  $K\to\pi\pi$ ($\varepsilon_K$), $B\to\psi K$ ($\sin2\beta$), 
  $B\to\pi\pi,\rho\pi,\rho\rho$ ($\alpha$), and $B\to DK$
  ($\gamma$). Taken from \cite{ckmfitter}.}
  \label{fg:UT}
\end{figure}

Of course, there are correlations between the various parameters.
The present constraints on the shape of the unitarity triangle or,
equivalently, the allowed region in the $\rho-\eta$ plane, are
presented in Fig. \ref{fg:UT}. 

\subsection{The uniqueness of the Standard Model picture of CP violation}
In the previous subsections, we have learnt several features of CP violation 
as explained by the Standard Model. It is important to understand that various 
reasonable (and often well-motivated) extensions of the SM provide examples 
where some or all of these features do not hold. Furthermore, until a few years
ago, none of the special features of the Kobayashi-Maskawa mechanism of CP 
violation has been experimentally tested. This situation has dramatically 
changed recently. Let us survey some of the SM features, how they can be
modified with new physics, and whether experiment has shed light on
these questions.

(i) {\it $\delta_{\rm KM}$ is the only source of CP violation in meson decays.}
This is arguably the most unique feature of the SM and gives the model a strong
predictive power. It is violated in almost any low-energy extension.
For example, in the supersymmetric extension of the SM there are
44 physical CP violating phases, many of which affect meson decays.
The measured value of $S_{\psi K_S}$ is consistent with the correlation 
between $K$ and $B$ decays that is predicted by the SM. The value of
$S_{\phi K_S}$ is equal (within the present experimental accuracy)
with $S_{\psi K_S}$, consistent with the SM correlation between the
asymmetries in $b\to s\bar ss$ and $b\to c\bar cs$ transitions. It is
therefore very likely that $\delta_{\rm KM}$ is indeed the dominant
source of CP violation in meson decays.

(ii) {\it CP violation is small in $K\to\pi\pi$ decays because of flavor
suppression and not because CP is an approximate symmetry.} In many
(though certainly not all) supersymmetric models, the flavor suppression
is too mild, or entirely ineffective, requiring approximate CP to hold.
The measurement of $S_{\psi K_S}={\cal O}(1)$ confirms that not all
CP violating phases are small.
 
(iii) {\it CP violation appears in both $\Delta F=1$ (decay) and
$\Delta F=2$ (mixing) amplitudes.} Superweak models suggest that CP is
violated only in mixing amplitudes. The measurements of non-vanishing
$\varepsilon^\prime/\varepsilon$, ${\cal A}_{K^\mp\pi^\pm}$ and ${\cal
  A}^{-+}_{\rho\pi}$ confirm that there is
CP violation in $\Delta S=1$ and $\Delta B=1$ processes.

(iv) {\it CP is not violated in the lepton sector.} Models that allow for
neutrino masses, such as the ESM framework presented above, predict
CP violation in leptonic charged current interactions. Thus, while
there is no measurement of leptonic CP violation, the data from
neutrino oscillation experiments, which give evidence that neutrinos
are massive and mix, make it very likely that charged current weak
interactions violate CP also in the lepton sector.

(v) {\it CP violation appears only in the charged current weak interactions
and in conjunction with flavor changing processes.} Here both various 
extensions of the SM (such as supersymmetry) and non-perturbative effects 
within the SM ($\theta_{\rm QCD}$) allow for CP violation in other types of 
interactions and in flavor diagonal processes. In particular, it is difficult 
to avoid flavor-diagonal phases in the supersymmetric framework.
The fact that no electric dipole moment has been measured yet poses 
difficulties to many models with diagonal CP violation (and, of course, is
responsible to the strong CP problem within the SM).

(vi) {\it CP is explicitly broken.} In various extensions of the scalar
sector, it is possible to achieve spontaneous CP violation.
It is very difficult to test this question experimentally.

This situation, where the Standard Model has a very unique and predictive
description of CP violation, is the basis for the 
strong interest, experimental and theoretical, in CP violation.

\section{Meson Decays}
\label{sec:formalism}
The phenomenology of CP violation is superficially different in $K$,
$D$, $B$, and $B_s$ decays. This is primarily because each of these
systems is governed by a different balance between decay rates,
oscillations, and lifetime splitting. However, the underlying
mechanisms of CP violation are identical for all pseudoscalar mesons.

In this section we present a general formalism for, and classification
of, CP violation in the decay of a pseudoscalar meson $P$ that might
be a charged or neutral $K$, $D$, $B$, or $B_s$ meson. Subsequent
sections describe the CP-violating phenomenology, approximations, and
alternate formalisms that are specific to each system. We follow here
closely the discussion in \cite{kirkbynir}.

\subsection{Charged and neutral meson decays}
We define decay amplitudes of a pseudoscalar meson $\M$ (which could
be charged or neutral) and its CP conjugate $\Mb$ to a multi-particle
final state $f$ and its CP conjugate $\fb$ as
\beq\label{decamp}
A_{\f}=\langle \f|{\cal
  H}|\M\rangle\quad , \quad \overline{A}_{\f}=\langle \f|{\cal
  H}|\Mb\rangle\quad , \quad A_{\fb}=\langle \fb|{\cal
  H}|\M\rangle\quad , \quad \overline{A}_{\fb}=\langle \fb|{\cal
  H}|\Mb\rangle\; ,
\eeq
where ${\cal H}$ is the Hamiltonian governing
weak interactions.  The action of CP on these states introduces
phases $\xi_\M$ and $\xi_f$ that depend on their flavor content,
according to
\beqa\label{eq:phaseconv}
\CP|\M\rangle &=&
e^{+i\xi_{\M}}\,|\Mb\rangle \quad , \quad \CP|\f\rangle =
e^{+i\xi_{\f}}\,|\fb\rangle \; ,\no\\
\CP|\Mb\rangle& =&
e^{-i\xi_{\M}}\,|\M\rangle \quad , \quad \CP|\fb\rangle =
e^{-i\xi_{\f}}\,|\f\rangle \ ,
\eeqa
so that $(\CP)^2=1$. The phases $\xi_\M$ and $\xi_f$ are arbitrary and
unphysical because of the flavor symmetry of the strong
interaction. If CP is conserved by the dynamics, $[\CP,{\cal H}] =
0$, then $A_f$ and $\overline{A}_{\fb}$ have the same magnitude and an
arbitrary unphysical relative phase 
\beq\label{spupha}
\overline{A}_{\fb} = e^{i(\xi_{\f}-\xi_{\M})}\, A_f\; .
\eeq

\subsection{Neutral meson mixing}
A state that is initially a superposition of $\Mz$ and $\Mzb$, say
\beq
|\psi(0)\rangle = a(0)|\Mz\rangle+b(0)|\Mzb\rangle \; ,
\eeq
will evolve in time acquiring components that describe all possible
decay final states $\{f_1,f_2,\ldots\}$, that is,
\beq
|\psi(t)\rangle =
a(t)|\Mz\rangle+b(t)|\Mzb\rangle+c_1(t)|f_1\rangle+c_2(t)|f_2\rangle+\cdots
\; .
\eeq 
If we are interested in computing only the values of $a(t)$ and $b(t)$
(and not the values of all $c_i(t)$), and if the times $t$ in which we
are interested are much larger than the typical strong interaction
scale, then we can use a much simplified
formalism~\cite{Weisskopf:au}. The simplified time evolution is
determined by a $2\times 2$ effective Hamiltonian $\Heff$ that is
not Hermitian, since otherwise the mesons would only oscillate and not
decay. Any complex matrix, such as $\Heff$, can be written in terms of
Hermitian matrices $\Meff$ and $\Geff$ as
\beq
\Heff = \Meff - \frac{i}{2}\,\Geff \; .
\eeq
$\Meff$ and $\Geff$ are associated with
$(\Mz,\Mzb)\leftrightarrow(\Mz,\Mzb)$ transitions via off-shell
(dispersive) and on-shell (absorptive) intermediate states, respectively.
Diagonal elements of $\Meff$ and $\Geff$ are associated with the
flavor-conserving transitions $\Mz\to\Mz$ and $\Mzb\to\Mzb$ while
off-diagonal elements are associated with flavor-changing transitions
$\Mz\leftrightarrow\Mzb$.

The eigenvectors of $\Heff$ have well defined masses and decay
widths. We introduce complex parameters $p_{L,H}$ and $q_{L,H}$ to
specify the components of the strong interaction eigenstates, $\Mz$ and
$\Mzb$, in the light ($\M_L$) and heavy ($\M_H$) mass eigenstates:
\beq\label{defpq}
|\M_{L,H}\rangle=p_{L,H}|\Mz\rangle\pm q_{L,H}|\Mzb\rangle
\eeq
with the normalization $|p_{L,H}|^2+|q_{L,H}|^2=1$. (Another possible
choice, which is in standard usage for $K$ mesons, defines the mass
eigenstates according to their lifetimes: $K_S$ for the
short-lived and $K_L$ for the long-lived state. The $K_L$ is
experimentally found to be the heavier state.) If either CP or
CPT is a symmetry of $\Heff$ (independently of whether T is conserved or
violated) then $\Meff_{11} = \Meff_{22}$ and $\Geff_{11}=
\Geff_{22}$, and solving the eigenvalue problem for $\Heff$ yields $p_L
= p_H \equiv p$ and $q_L = q_H \equiv q$ with
\beq
\left(\frac{q}{p}\right)^2=\frac{\Meff_{12}^\ast -
    (i/2)\Geff_{12}^\ast}{\Meff_{12}-(i/2)\Geff_{12}}\; .
\eeq
If either CP or T is a symmetry of $\Heff$ (independently of whether
CPT is conserved or violated), then $\Meff_{12}$ and $\Geff_{12}$ are
relatively real, leading to
\beq
\left(\frac{q}{p}\right)^2 = e^{2i\xi_\M} \quad \Rightarrow \quad
\left|\frac{q}{p}\right| = 1 \; ,
\eeq
where $\xi_\M$ is the arbitrary unphysical phase introduced in
Eq.~(\ref{eq:phaseconv}). If, and only if, CP is a symmetry of $\Heff$
(independently of CPT and T) then both of the above conditions hold,
with the result that the mass eigenstates are orthogonal 
\beq
\langle \M_H | \M_L\rangle = |p|^2 - |q|^2 = 0 \; .
\eeq
From now on we assume that CPT is conserved.

The real and imaginary parts of the eigenvalues of $\Heff$
corresponding to $|\M_{L,H}\rangle$ represent their masses and
decay-widths, respectively. The mass difference $\Delta m$ and the
width difference $\Delta\Gamma$ are defined as follows:
\beq\label{DelmG}
\Delta m\equiv M_H-M_L,\ \ \ \Delta\Gamma\equiv\Gamma_H-\Gamma_L.
\eeq
Note that here $\Delta m$ is positive by definition, while the sign of
$\Delta\Gamma$ is to be experimentally determined. (Alternatively, one
can use the states defined by their lifetimes to have
$\Delta\Gamma\equiv\Gamma_S-\Gamma_L$ positive by definition.) 
The average mass and width are given by
\beq\label{aveMG}
m\equiv{M_H+M_L\over2},\ \ \ \Gamma\equiv{\Gamma_H+\Gamma_L\over2}.
\eeq
It is useful to define dimensionless ratios $x$ and $y$:
\beq\label{defxy}
x\equiv{\Delta m\over\Gamma},\ \ \ y\equiv{\Delta\Gamma\over2\Gamma}.
\eeq
Solving the eigenvalue equation gives
\beq\label{eveq}
(\Delta m)^2-{1\over4}(\Delta\Gamma)^2=(4|M_{12}|^2-|\Gamma_{12}|^2),\ \ \ \ 
\Delta m\Delta\Gamma=4\re{M_{12}\Gamma_{12}^*}.
\eeq

\subsection{CP-violating observables}
All CP-violating observables in $\M$ and $\Mb$ decays to final states $f$
and $\fb$ can be expressed in terms of phase-convention-independent
combinations of $A_f$, $\overline{A}_f$, $A_{\overline{f}}$ and
$\overline{A}_{\overline{f}}$, together with, for neutral-meson decays
only, $q/p$. CP violation in charged-meson decays depends only on the
combination $|\overline{A}_{\fb}/A_f|$, while CP violation in
neutral-meson decays is complicated by $\Mz\leftrightarrow\Mzb$
oscillations and depends, additionally, on $|q/p|$ and on $\lambda_f
\equiv (q/p)(\overline{A}_f/A_f)$.

The decay-rates of the two neutral $K$ mass eigenstates,
$K_S$ and $K_L$, are different enough ($\Gamma_S/\Gamma_L\sim500$)
that one can, in most cases, actually study their decays
independently. For neutral $D$, $B$, and $B_s$ mesons, however, values
of $\Delta\Gamma/\Gamma$ are relatively small and so both mass
eigenstates must be considered in their evolution. We denote the state
of an initially pure $|\Mz\rangle$ or $|\Mzb\rangle$ after an elapsed
proper time $t$ as $|\Mz_{\mathrm{phys}}(t)\rangle$ or
$|\Mzb_{\mathrm{phys}}(t)\rangle$,
respectively. Using the effective Hamiltonian approximation, we obtain 
\beqa\label{defphys}
|\Mz_{\rm phys}(t)\rangle&=&g_+(t)\,|\Mz\rangle
- (q/p)\ g_-(t)|\Mzb\rangle,\no\\
|\Mzb_{\rm phys}(t)\rangle&=&g_+(t)\,|\Mzb\rangle
- (p/q)\ g_-(t)|\Mz\rangle \; ,
\eeqa
where 
\beq
g_\pm(t) \equiv \frac12\left(e^{-im_Ht-\frac12\Gamma_Ht}\pm
  e^{-im_Lt-\frac12\Gamma_Lt}\right).
\eeq

One obtains the following time-dependent decay rates:
\beqa
\frac{d\Gamma[\Mz_{\rm phys}(t)\to f]/dt}{e^{-\Gamma t}{\cal N}_f}&=&
\left(|A_f|^2+|(q/p)\overline{A}_f|^2\right)\cosh(y\Gamma t)
  +\left(|A_f|^2-|(q/p)\overline{A}_f|^2\right)\cos(x\Gamma t)\no\\
&+&2\,\re{(q/p)A_f^\ast \overline{A}_f}\sinh(y\Gamma t)
-2\,\im{(q/p)A_f^\ast \overline{A}_f}\sin(x\Gamma t)
\label{decratbt1}\;,\\
\frac{d\Gamma[\Mzb_{\rm phys}(t)\to f]/dt}{e^{-\Gamma t}{\cal N}_f}&=&
\left(|(p/q)A_f|^2+|\overline{A}_f|^2\right)\cosh(y\Gamma t)
  -\left(|(p/q)A_f|^2-|\overline{A}_f|^2\right)\cos(x\Gamma t)\no\\
&+&2\,\re{(p/q)A_f\overline{A}^\ast_f}\sinh(y\Gamma t)
-2\,\im{(p/q)A_f\overline{A}^\ast_f}\sin(x\Gamma t)
\label{decratbt2}\; ,
\eeqa
where ${\cal N}_f$ is a time-independnet normalization factor. Decay rates to
the CP-conjugate final state $\fb$ are obtained analogously, with
${\cal N}_f = {\cal N}_{\fb}$ and the substitutions $A_f\to A_{\fb}$
and $\overline{A}_f\to\overline{A}_{\fb}$ in
Eqs.~(\ref{decratbt1},\ref{decratbt2}). Terms proportional
to $|A_f|^2$ or $|\overline{A}_f|^2 $ are associated with decays that
occur without any net $\M\leftrightarrow\Mb$ oscillation, while terms
proportional to $|(q/p)\overline{A}_f|^2$ or $|(p/q)A_f|^2$ are
associated with decays following a net oscillation. The $\sinh(y\Gamma
t)$ and $\sin(x\Gamma t)$ terms of
Eqs.~(\ref{decratbt1},\ref{decratbt2}) are associated with the
interference between these two cases. Note that, in multi-body decays, 
amplitudes are functions of phase-space variables. Interference may
be present in some regions but not others, and is strongly influenced
by resonant substructure.

\subsection{Classification of CP-violating effects}
We distinguish three types of CP-violating effects in meson decays
\cite{Nir:1992uv}: 

{\bf [I] CP violation in decay} is defined by
\beq\label{cpvdec}
|\overline{A}_{\overline{f}}/A_f|\neq1 \; .
\eeq
In charged meson decays, where mixing effects are absent, this is the only
possible source of CP asymmetries:
\beq\label{asycha}
{\cal A}_{f^\pm}\equiv\frac{\Gamma(P^-\to f^-)-\Gamma(P^+\to f^+)}
{\Gamma(P^-\to f^-)+\Gamma(P^+\to f^+)}=\frac{|\overline{A}_{f^-}/A_{f^+}|^2-1}
{|\overline{A}_{f^-}/A_{f^+}|^2+1}\; .
\eeq

{\bf [II] CP violation in mixing} is defined by
\beq\label{cpvmix}
|q/p|\neq1 \; .
\eeq
In charged-current semileptonic neutral meson decays $\M,\Mb\to
\ell^{\pm} X$ (taking
$|A_{\ell^+ X}|=|\overline{A}_{\ell^- X}|$ and $A_{\ell^-
  X} = \overline{A}_{\ell^+ X} = 0$, as is the case in
the Standard Model, to lowest order in $G_F$, and in most of its
reasonable extensions), this is the only source of CP violation, and
can be measured via the asymmetry of ``wrong-sign'' decays induced by 
oscillations:
\beq\label{asysl}
{\cal A}_{\rm SL}(t)\equiv\frac{d\Gamma/dt[\Mzb_{\rm phys}(t)
    \to\ell^+X]-d\Gamma/dt[\Mz_{\rm phys}(t)\to\ell^-
  X]}{d\Gamma/dt[\Mzb_{\rm phys}(t)\to\ell^+X]+d\Gamma/dt[\Mz_{\rm phys}(t)\to\ell^- X]}
=\frac{1-|q/p|^4}{1+|q/p|^4}.
\eeq
Note that this asymmetry of time-dependent decay rates is actually
time independent.

{\bf [III] CP violation in interference between a decay without mixing,
  $\Mz\to f$, and a decay with mixing, $\Mz\to \Mzb\to f$} (such an
  effect occurs only in decays to final states that are common to $\Mz$
  and $\Mzb$, including all CP eigenstates), is defined by
\beq\label{cpvint}
\im{\lambda_f}\ne 0 \; ,
\eeq
with
\beq\label{deflam}
\lambda_f \equiv \frac{q}{p}\frac{\overline{A}_f}{A_f} \; .
\eeq
This form of CP violation can be observed, for example, using the
asymmetry of neutral meson decays into final CP eigenstates $f_{\CP}$
\beq\label{asyfcp}
{\cal A}_{f_{\CP}}(t)\equiv\frac{d\Gamma/dt[\Mzb_{\rm phys}(t)\to f_{\CP}]-
d\Gamma/dt[\Mz_{\rm phys}(t)\to f_{\CP}]}
{d\Gamma/dt[\Mzb_{\rm phys}(t)\to f_{\CP}]+d\Gamma/dt[\Mz_{\rm phys}(t)\to
  f_{\CP}]}\; .
\eeq
If $\Delta\Gamma = 0$ and $|q/p|=1$, as expected to a good
approximation for $B$ mesons but not for $K$ mesons, then ${\cal
  A}_{f_{\CP}}$ has a particularly simple form
\cite{Dunietz:1986vi,Blinov:ru,Bigi:1986vr}:
\beqa\label{asyfcpb}
{\cal A}_{f}(t)&=&S_f\sin(\Delta mt)-C_f\cos(\Delta mt),\no\\
S_f&\equiv&\frac{2\,\im{\lambda_{f}}}{1+|\lambda_{f}|^2},\ \ \ 
C_f\equiv\frac{1-|\lambda_{f}|^2}{1+|\lambda_{f}|^2}
\; ,
\eeqa
If, in addition, the decay amplitudes
fulfill $|\overline{A}_{f_{\CP}}|=|A_{f_{\CP}}|$, the interference
between decays with and without mixing is the only source of the
asymmetry and
\beq
{\cal A}_{f_{\CP}}(t)=\im{\lambda_{f_{\CP}}}\sin(x\Gamma t).
\eeq

\section{Theoretical Interpretation: General Considerations}
\label{sec:theory}
Consider the $\M\to f$ decay amplitude $A_f$, and the CP conjugate
process, $\Mb\to\fb$, with decay amplitude $\overline{A}_{\fb}$. There
are two types of phases that may appear in these decay amplitudes.
Complex parameters in any Lagrangian term that contributes to the
amplitude will appear in complex conjugate form in the CP-conjugate
amplitude. Thus their phases appear in $A_f$ and
$\overline{A}_{\overline{f}}$ with opposite signs. In the Standard
Model, these phases occur only in the couplings of the $W^\pm$ bosons
and hence are often called ``weak phases''. The weak phase of any
single term is convention dependent. However, the difference between
the weak phases in two different terms in $A_f$ is convention
independent. A second type of phase can appear in scattering or decay
amplitudes even when the Lagrangian is real. Their origin is the
possible contribution from intermediate on-shell states in the decay
process. Since these phases are generated by CP-invariant
interactions, they are the same in $A_f$ and
$\overline{A}_{\overline{f}}$. Usually the dominant rescattering is
due to strong interactions and hence the designation ``strong phases''
for the phase shifts so induced. Again, only the relative strong
phases between different terms in the amplitude are physically
meaningful.

The `weak' and `strong' phases discussed here appear in addition to
the `spurious' CP-transformation phases of Eq.~(\ref{spupha}). Those
spurious phases are due to an arbitrary choice of phase
convention, and do not originate from any dynamics or induce any \CP
violation. For simplicity, we set them to zero from here on.

It is useful to write each contribution $a_i$ to $A_f$ in three parts:
its magnitude $|a_i|$, its weak phase $\phi_i$, and its strong
phase $\delta_i$. If, for example, there are two such
contributions, $A_f = a_1 + a_2$, we have
\beqa\label{weastr}
A_f&=& |a_1|e^{i(\delta_1+\phi_1)}+|a_2|e^{i(\delta_2+\phi_2)},\no\\
\overline{A}_{\overline{f}}&=&
|a_1|e^{i(\delta_1-\phi_1)}+|a_2|e^{i(\delta_2-\phi_2)}.
\eeqa
Similarly, for neutral meson decays, it is useful to write
\beq\label{defmgam}
\Meff_{12} = |\Meff_{12}| e^{i\phi_M} \quad , \quad
\Geff_{12} = |\Geff_{12}| e^{i\phi_\Gamma} \; .
\eeq
Each of the phases appearing in Eqs.~(\ref{weastr},\ref{defmgam}) is
convention dependent, but combinations such as $\delta_1-\delta_2$,
$\phi_1-\phi_2$, $\phi_M-\phi_\Gamma$ and $\phi_M+\phi_1-\overline{\phi}_1$
(where $\overline{\phi}_1$ is a weak phase contributing to $\overline{A}_f$)
are physical. 

It is now straightforward to evaluate the various asymmetries in terms
of the theoretical parameters introduced here. We will do so with
approximations that are often relevant to the most interesting
measured asymmetries.

1. The CP asymmetry in charged meson decays [Eq. (\ref{asycha})] is
given by 
\beq\label{apmth}
{\cal A}_{f^\pm}=-\frac{2|a_1a_2|\sin(\delta_2-\delta_1)
\sin(\phi_2-\phi_1)}{|a_1|^2+|a_2|^2+2|a_1a_2|\cos(\delta_2-\delta_1)
\cos(\phi_2-\phi_1)}.     
\eeq
The quantity of most interest to theory is the weak phase difference
$\phi_2-\phi_1$. Its extraction from the asymmetry requires, however,
that the amplitude ratio and the strong phase are known. Both
quantities depend on non-perturbative hadronic parameters that are
difficult to calculate.

2. In the approximation that $|\Geff_{12}/\Meff_{12}|\ll1$ (valid for
$B$ and $B_s$ mesons), the CP asymmetry in semileptonic neutral-meson
decays [Eq. (\ref{asysl})] is given by
\beq\label{aslth}
{\cal A}_{\rm
  SL}=-\left|\frac{\Geff_{12}}{\Meff_{12}}\right|\sin(\phi_M-\phi_\Gamma). 
\eeq
The quantity of most interest to theory is the weak phase
$\phi_M-\phi_\Gamma$. Its extraction from the asymmetry
requires, however, that $|\Geff_{12}/\Meff_{12}|$ is known. This quantity
depends on long distance physics that is difficult to calculate.

3. In the approximations that only a single weak phase contributes to decay,
$A_f=|a_f|e^{i(\delta_f+\phi_f)}$, and that
$|\Geff_{12}/\Meff_{12}|=0$, we obtain $|\lambda_f|=1$ and
the \CP asymmetries in decays to a final CP
eigenstate $f$ [Eq. (\ref{asyfcp})] with eigenvalue $\eta_f= \pm 1$
are given by
\beq\label{afcth}
{\cal A}_{f_{\CP}}(t) = \im{\lambda_f}\; \sin(\Delta m t) \; \ 
\mathrm{with}\ \   
\im{\lambda_f}=\eta_f\sin(\phi_M+2\phi_f).
\eeq
Note that the phase so measured is purely a weak phase, and no
hadronic parameters are involved in the extraction of its value from
$\im{\lambda_f}$.

The discussion above allows us to introduce another classification:
\begin{enumerate}
\item {\bf Direct CP violation} is one that cannot be accounted for
  by just $\phi_M\neq0$. CP violation in decay (type I) belongs to
  this class. 
\item {\bf Indirect CP violation} is consistent with taking  
  $\phi_M\neq0$ and setting all other CP violating phases to
  zero. CP violation in mixing (type II) belongs to this class.
\end{enumerate}
As concerns type III CP violation, observing  
$\eta_{f_1}\im{\lambda_{f_1}}\neq\eta_{f_2}\im{\lambda_{f_2}}$ (for the 
same decaying meson and two different final CP eigenstates $f_1$ and
$f_2$) would establish direct CP violation. The significance of this
classification is related to theory. In superweak models
\cite{Wolfenstein:1964ks}, CP violation appears only in diagrams that
contribute to $\Meff_{12}$, hence they predict that there is no direct
CP violation. In most models and, in particular, in the Standard
Model, CP violation is both direct and indirect. The experimental
observation of $\epsilon^\prime\neq0$ (see Section \ref{sec:k})
excluded the superweak scenario.

\section{$K$ Decays}
\label{sec:k}
CP violation was discovered in $K\to\pi\pi$ decays in 1964
\cite{Christenson:1964fg}. The same mode provided the first evidence
for direct CP violation
\cite{Burkhardt:1988yh,Fanti:1999nm,Alavi-Harati:1999xp}.

The decay amplitudes actually measured in neutral $K$ decays
refer to the mass eigenstates $K_L$ and $K_S$ rather than to the $K$
and ${\overline{K}}$ states referred to in Eq.~(\ref{decamp}). We define
CP-violating amplitude ratios for two-pion final states,
\beq\label{etapmzz}
\eta_{00}\equiv
\frac{\langle\pi^0\pi^0|{\cal H}|K_L\rangle}{\langle
  \pi^0\pi^0|{\cal H}|K_S\rangle}
\quad,\quad
\eta_{+-}\equiv
\frac{\langle\pi^+\pi^-|{\cal H}|K_L\rangle}{\langle
  \pi^+\pi^-|{\cal H}|K_S\rangle}
\; .
\eeq
Another important observable is the asymmetry of time-integrated
semileptonic decay rates:
\beq
\delta_{L}\equiv \frac
{\Gamma(K_L\to\ell^+\nu_{\ell}\pi^-)-\Gamma(K_L\to\ell^-{\bar\nu}_{\ell}\pi^+)}
{\Gamma(K_L\to\ell^+\nu_{\ell}\pi^-)+\Gamma(K_L\to\ell^-{\bar\nu}_{\ell}\pi^+)}
\; .
\eeq
CP violation has been observed as an appearance of $K_L$ decays to two-pion
final states \cite{Eidelman:2004wy},
\beqa
|\eta_{00}| &=& (2.275\pm 0.017)\times10^{-3},\no\\
|\eta_{+-}| &=& (2.286\pm 0.017)\times10^{-3},\no\\
|\eta_{00}/\eta_{+-}| &=& 0.9950\pm 0.0008,
\eeqa
and, assuming CPT, $\phi_{+-}= \phi_{00}= 43.49^\circ\pm0.07^\circ$
($\phi_{ij}$ is 
the phase of the amplitude ratio $\eta_{ij}$). \CP violation has also
been observed in semileptonic $K_L$ decays~\cite{Eidelman:2004wy}
\beq
\delta_L = (3.27\pm 0.12)\times 10^{-3} \; ,
\eeq
where $\delta_L$ is a weighted average of muon and electron
measurements, as well as in $K_L$ decays to $\pi^+\pi^-\gamma$ and
$\pi^+\pi^-e^+e^-$~\cite{Eidelman:2004wy}. 

Historically, CP violation in neutral $K$ decays has been described
in terms of parameters $\epsilon$ and $\epsilon^\prime$. The
observables $\eta_{00}$, $\eta_{+-}$, and $\delta_L$ are related to
these parameters, and to those of Section~\ref{sec:formalism}, by
\beqa
\eta_{00}& =\frac{1-\lambda_{\pi^0\pi^0}}{1+\lambda_{\pi^0\pi^0}}
\quad=&\ \epsilon - 2\epsilon^\prime\;,\no\\ 
\eta_{+-}& =\frac{1-\lambda_{\pi^+\pi^-}}{1+\lambda_{\pi^+\pi^-}}
\quad=&\ \epsilon + \epsilon^\prime\;,\no\\
\delta_{L}& = \frac{1-|q/p|^2}{1+|q/p|^2}
\quad\ =&\frac{2\re{\epsilon}}
{1+\left|\epsilon\right|^2} \; ,
\eeqa
where, in the last line, we have assumed that
$|A_{\ell^+\nu_{\ell}\pi^-}| =
|{\overline{A}}_{\ell^-{\bar\nu}_{\ell}\pi^+}|$ and
$|A_{\ell^-{\bar\nu}_{\ell}\pi^+}| =
|{\overline{A}}_{\ell^+\nu_{\ell}\pi^-}| = 0$. A fit to the
$K\to\pi\pi$ data yields~\cite{Eidelman:2004wy}
\begin{eqnarray}
|\epsilon|&=&(2.283\pm0.017)\times10^{-3}\; ,\no\\
\re{\epsilon^\prime/\epsilon}&=& (1.67\pm 0.26)\times 10^{-3} \; .
\end{eqnarray}

In discussing two-pion final states, it is useful to express
the amplitudes $A_{\pi^0\pi^0}$ and $A_{\pi^+\pi^-}$ in terms of
their isospin components via
\beqa
A_{\pi^0\pi^0} &=&\sqrt{\frac13}|A_0| e^{i(\delta_0+\phi_0)}
-\sqrt{\frac23} |A_2| e^{i(\delta_2+\phi_2)},\no\\
A_{\pi^+\pi^-} &=& 
\sqrt{\frac23} |A_0| e^{i(\delta_0+\phi_0)}
+\sqrt{\frac13} |A_2| e^{i(\delta_2+\phi_2)}
\; ,
\eeqa
where we parameterize the amplitude $A_I(\overline{A}_I)$ for
$K^0(\overline{K}^0)$ decay into two pions 
with total isospin $I = 0$ or 2 as 
\beq
A_I\equiv \langle(\pi\pi)_I|{\cal H}|K^0\rangle = |A_I|
e^{i(\delta_I+\phi_I)} \; ,\ \ \
\overline{A}_I\equiv \langle(\pi\pi)_I|{\cal H}|\overline{K}{}^0\rangle = |A_I|
e^{i(\delta_I-\phi_I)} \; .
\eeq
The smallness of $|\eta_{00}|$ and $|\eta_{+-}|$ allows us to
approximate
\beq
\epsilon \simeq \frac{1}{2}(1-\lambda_{(\pi\pi)_{I=0}}),
\quad\quad
\epsilon^\prime \simeq \frac{1}{6}
\left(\lambda_{\pi^0\pi^0} - \lambda_{\pi^+\pi^-}\right) \; .
\eeq
The parameter $\epsilon$ represents indirect CP violation,
while $\epsilon^\prime$ parameterizes direct CP violation:
$\re{\epsilon^\prime}$ measures CP violation in
decay (type I), $\re{\epsilon}$ measures CP violation in
mixing (type II), and $\im{\epsilon}$ and $\im{\epsilon^\prime}$
measure the interference 
between decays with and without mixing (type III).

The following expressions for $\epsilon$ and $\epsilon^\prime$ are
useful for theoretical evaluations:
\beq\label{theeps}
\epsilon\simeq \frac{e^{i\pi/4}}{\sqrt2}\frac{\im{\Meff_{12}}}{\Delta
  m},\quad\quad
\epsilon^\prime=\frac{i}{\sqrt2}\left|\frac{A_2}{A_0}\right|e^{i(\delta_2-\delta_0)} 
\sin(\phi_2-\phi_0).
\eeq
The expression for $\epsilon$ is only valid in a phase convention where
$\phi_2=0$, corresponding to a real $V_{ud}^{}V_{us}^*$, and in the
approximation that also $\phi_0 = 0$. The phase of
$\pi/4$ is approximate,  and determined by hadronic parameters,
$\arg\epsilon\approx\arctan(-2\Delta m/\Delta\Gamma)$,
independently of the electroweak model. The calculation of $\epsilon$
benefits from the fact that $\im{\Meff_{12}}$ is
dominated by short distance physics. Consequently, the main
source of uncertainty in theoretical interpretations of $\epsilon$
are the values of matrix elements such as $\langle K^0|(\bar s
d)_{V-A}(\bar sd)_{V-A}|\overline{K}{}^0\rangle$. The expression for  
$\epsilon^\prime$  is valid to first order in $|A_2/A_0|\sim1/20$.
The phase of $\epsilon^\prime$ is experimentally determined,
$\pi/2+\delta_2-\delta_0\approx\pi/4$ and is independent of the
electroweak model. Note that, accidentally, $\epsilon^\prime/\epsilon$
is real to a good approximation.

A future measurement of much interest is that of CP violation in the
rare $K\to\pi\nu\bar\nu$ decays. The signal for CP violation is simply
observing the $K_L\to\pi^0\nu\bar\nu$ decay. The effect here is that
of interference between decays with and without mixing (type III)
\cite{Grossman:1997sk}:
\beq\label{kpinunu}
\frac{\Gamma(K_L\to\pi^0\nu\bar\nu)}{\Gamma(K^+\to\pi^+\nu\bar\nu)}=
\frac12\left[1+|\lambda_{\pi\nu\bar\nu}|^2-
2\,\re{\lambda_{\pi\nu\bar\nu}}\right]
\simeq 1-\re{\lambda_{\pi\nu\bar\nu}},
\eeq
where in the last equation we neglect CP violation in decay and in
mixing (expected, model independently, to be of order $10^{-5}$ and
$10^{-3}$, respectively). Such a measurement
would be experimentally very challenging and theoretically very rewarding
\cite{Littenberg:ix}. Similar to the CP asymmetry in $B\to J/\psi K_S$, the CP
violation in $K\to\pi\nu\bar\nu$ decay is predicted to be large (the
ratio in Eq.~(\ref{kpinunu}) is not CKM suppressed) and
can be very cleanly interpreted. 

Within the Standard Model, the $K_L\to\pi^0\nu\bar\nu$ decay is
dominated by an intermediate top quark contribution and,
consequently, can be cleanly interpreted in terms of CKM
parameters \cite{Buras:1994rj}. (For the charged mode,
$K^+\to\pi^+\nu\bar\nu$, the contribution from an intermediate charm
quark is not negligible and constitutes a source of hadronic
uncertainty.) In particular, ${\cal B}(K_L\to\pi^0\nu\bar\nu)$
provides a theoretically clean way to determine the Wolfenstein
parameter $\eta$~\cite{Buchalla:1993bv}: 
\beq\label{brkpnn}
{\cal B}(K_L\to\pi^0\nu\bar\nu)=\kappa_LX^2(m_t^2/m_W^2)A^4\eta^2,
\eeq
where $\kappa_L=1.80\times10^{-10}$ incorporates the value of the
four-fermion matrix element which is deduced, using isospin relations,
from ${\cal B}(K^+\to\pi^0e^+\nu)$, and $X(m_t^2/m_W^2)$  is a known
function of the top mass. 

\subsection{Implications of $\varepsilon_K$}
The measurement of $\varepsilon_K$ has had (and still has) important
implications. Two implications of historical importance are the
following: 

(i) CP violation was discovered through the measurement of $\varepsilon_K$.
Hence this measurement played a significant role in the history of particle
physics. 

(ii) The observation of $\varepsilon_K\neq0$ led to the prediction
that a third generation must exist, so that CP is violated in the
Standard Model. This provides an excellent example of how precision
measurements at low energy can lead to the discovery of new physics
(even if, at present, this new physics is old...)

Within the Standard Model, $\im{M_{12}}$ is accounted for by box diagrams:
\beq\label{epsCKM}
\varepsilon_K=e^{i\pi/4}C_\varepsilon B_K\im{V_{ts}^*V_{td}}\left\{\re
{V_{cs}^*V_{cd}}[\eta_1 S_0(x_c)-\eta_3 S_0(x_c,x_t)]-\re{V_{ts}^*V_{td}}
\eta_2S_0(x_t)\right\},
\eeq
where $C_\varepsilon\equiv{G_F^2 f_K^2 m_K m_W^2\over6\sqrt{2}\pi^2\Delta m_K}$
is a well known parameter, the $\eta_i$ are QCD correction factors, $S_0$ is a 
kinematic factor, and $B_K$ is the ratio between the matrix element of the four
quark operator and its value in the vacuum insertion
approximation. The measurement of $\varepsilon_K$ has the following
implications within the SM:
\begin{itemize}
\item This measurement allowed one to set the value of $\delta_{\rm KM}$.
Furthermore, by implying that $\delta_{\rm KM}={\cal O}(1)$, it made 
the KM mechanism plausible. Having been the single measured CP violating
parameter it could not, however, serve as a test of the KM
mechanism. More precisely, a value of $|\varepsilon_K|\gg10^{-3}$
would have invalidated the KM mechanism, but any value
$|\varepsilon_K|\lsim10^{-3}$ was acceptable. It is only the
combination of the new measurements in $B$ decays (particularly
$S_{\psi K_S}$) with $\varepsilon_K$ that provides the
first precision test of the KM mechanism.
\item Within the SM, the smallness of $\varepsilon_K$ is not related to 
suppression of CP violation but rather to suppression of flavor violation.
Specifically, it is the smallness of the ratio $|(V_{td}V_{ts})/(V_{ud}V_{us})|
\sim\lambda^4$ that explains $|\varepsilon_K|\sim10^{-3}$.
\item Until recently, the measured value of $\varepsilon_K$ provided a
unique type of information on the KM phase. For example, the 
measurement of $\re{\varepsilon_K}>0$ tells us that $\eta>0$ and
excludes the lower half of the $\rho-\eta$ plane. Such information cannot be
obtained from any CP conserving observable.
\item The $\varepsilon_K$ constraint in Eq.~(\ref{epsCKM}) gives
hyperbole in the $\rho-\eta$ plane. It is shown in
Fig. \ref{fg:UT}. The measured value is consistent with all 
other CKM-related measurements and further narrows the allowed region.
\end{itemize}

Beyond the SM, $\varepsilon_K$ is an extremely powerful probe of new
physics. This aspect will be discussed later.

\section{$D$ Decays}
\label{sec:d}
Unlike the case of neutral $K$, $B$, and $B_s$ mixing,
$D^0-\overline{D}{}^0$ mixing has not yet been observed. 
Long-distance contributions make it difficult to calculate the
Standard Model prediction for the $D^0-\overline{D}{}^0$ mixing
parameters. Therefore, the goal of the search for
$D^0-\overline{D}{}^0$ mixing is not to constrain the CKM parameters
but rather to probe new physics. Here CP violation plays an important
role \cite{Blaylock:1995ay}. Within the Standard Model, the
CP-violating effects are 
predicted to be negligibly small since the mixing and the relevant
decays are described, to an excellent approximation, by physics of the
first two generations. Observation of CP violation in
$D^0-\overline{D}{}^0$ mixing (at a level much higher than ${\cal
  O}(10^{-3})$) will constitute an unambiguous signal of
new physics.\footnote{In contrast, neither $y_D\sim10^{-2}$
\cite{Falk:2001hx}, nor $x_D\sim10^{-2}$ \cite{Falk:2004wg} can be
considered as evidence for new physics.} At present, the most
sensitive searches  involve the $D\to K^+K^-$ and $D\to K^\pm\pi^\mp$
modes. 

The neutral $D$ mesons decay via a singly-Cabibbo-suppressed
transition to the \CP eigenstate $K^+K^-$. Since the decay
proceeds via a Standard-Model tree diagram, it is very likely
unaffected by new physics and, furthermore, dominated by a single weak
phase. It is safe then to assume that direct CP violation plays no
role here \cite{Bergmann:1999pm,D'Ambrosio:2001wg}. In addition, given
the experimental bounds \cite{Eidelman:2004wy},
$x\equiv\Delta m/\Gamma\lsim0.03$ and
$y\equiv\Delta\Gamma/(2\Gamma)=0.008\pm0.005$, 
we can expand the decay rates to first order in these
parameters. Using Eq.~(\ref{decratbt1}) with these assumptions and
approximations yields, for $xt, yt\lsim\Gamma^{-1}$,
\beqa\label{dtokk}
\Gamma[D^0_{\rm phys}(t)\to K^+K^-]&=&e^{-\Gamma
  t}|A_{KK}|^2[1-|q/p|(y\cos\phi_D-x\sin\phi_D)\Gamma t],\no\\
\Gamma[\overline{D}{}^0_{\rm phys}(t)\to K^+K^-]&=&e^{-\Gamma
  t}|A_{KK}|^2[1-|p/q|(y\cos\phi_D+x\sin\phi_D)\Gamma t],
\eeqa
where $\phi_D$ is defined via
$\lambda_{K^+K^-}=-|q/p|e^{i\phi_D}$. (In the limit of \CP
conservation, choosing 
$\phi_D=0$ is equivalent to defining the mass eigenstates by their CP
eigenvalue: $|D_\mp\rangle=p|D^0\rangle\pm q|\overline{D}^0\rangle$,
with $D_-(D_+)$ being the \CP-odd (CP-even) state; that is, the state
that does not (does) decay into $K^+K^-$.) Given the small values of $x$
and $y$, the time dependences of the rates in Eq.~(\ref{dtokk}) can be
recast into purely exponential forms, but with modified decay-rate
parameters~\cite{Bergmann:2000id}:
\beqa\label{gammaom}
\Gamma_{D^0\to K^+K^-}&=&\Gamma\times\left[1+|q/p|(y\cos\phi_D-x\sin\phi_D)\right],\no\\
\Gamma_{\overline{D}{}^0\to
  K^+K^-}&=&\Gamma\times\left[1+|p/q|(y\cos\phi_D+x\sin\phi_D)\right].
\eeqa
One can define CP-conserving and CP-violating combinations of these
two observables (normalized to the true width $\Gamma$):
\beqa\label{gammcv}
Y&\equiv&\frac{\Gamma_{\overline{D}{}^0\to K^+K^-}+\Gamma_{D^0\to
    K^+K^-}}{2\Gamma}-1\no\\ &=&
\frac{|q/p|+|p/q|}{2}\ y\cos\phi_D-\frac{|q/p|-|p/q|}{2}\ x\sin\phi_D,\no\\
\Delta Y&\equiv&\frac{\Gamma_{\overline{D}{}^0\to K^+K^-}-\Gamma_{D^0\to
    K^+K^-}}{2\Gamma}\no\\ &=&
\frac{|q/p|+|p/q|}{2}\ x\sin\phi_D-\frac{|q/p|-|p/q|}{2}\ y\cos\phi_D.
\eeqa
In the limit of \CP conservation (and, in particular, within the
Standard Model), $Y=y$ and $\Delta Y=0$.

The $K^\pm\pi^\mp$ states are not CP eigenstates but they are still common
final states for $D^0$ and $\overline{D}{}^0$ decays. Since
$D^0(\overline{D}{}^0)\to K^-\pi^+$ is a Cabibbo-favored
(doubly-Cabibbo-suppressed) process, these processes are particularly
sensitive to $x$ and/or $y={\cal O}(\lambda^2)$. Taking into account
that $|\lambda_{K^-\pi^+}|,|\lambda_{K^+\pi^-}^{-1}|\ll1$ and
$x,y\ll1$, assuming that there is no direct CP violation (again, these
are Standard Model tree level decays dominated by a single weak phase)
and expanding the time dependent rates for $xt, yt\lsim\Gamma^{-1}$, one
obtains
\beqa\label{dtokpi}
\frac{\Gamma[D^0_{\rm phys}(t)\to
  K^+\pi^-]}{\Gamma[\overline{D}{}^0_{\rm phys}(t)\to
  K^+\pi^-]}&=&
r_d^2+r_d\left|\frac qp\right|(y^\prime\cos\phi_D-x^\prime\sin\phi_D)\Gamma
t+\left|\frac qp\right|^2\frac{y^2+x^2}{4}(\Gamma t)^2,\no\\
\frac{\Gamma[\overline{D}{}^0_{\rm phys}(t)\to
  K^-\pi^+]}{\Gamma[D^0_{\rm phys}(t)\to
  K^-\pi^+]}&=&
r_d^2+r_d\left|\frac pq\right|(y^\prime\cos\phi_D+x^\prime\sin\phi_D)\Gamma
t+\left|\frac pq\right|^2\frac{y^2+x^2}{4}(\Gamma t)^2,
\eeqa
where
\beqa\label{defxypri}
y^\prime&\equiv&y\cos\delta-x\sin\delta,\no\\
x^\prime&\equiv&x\cos\delta+y\sin\delta.
\eeqa
The weak phase $\phi_D$ is the same as that of Eq.~(\ref{dtokk}) (a
consequence of the absence of direct CP violation), $\delta$ is a
strong phase difference for these processes, and $r_d={\cal
  O}(\tan^2\theta_c)$ is the amplitude ratio, $r_d=|\overline{A}_{K^-\pi^+}/
A_{K^-\pi^+}|= |A_{K^+\pi^-}/\overline{A}_{K^+\pi^-}|$, that is,
$\lambda_{K^-\pi^+}=r_d(q/p)e^{-i(\delta-\phi_D)}$ and
$\lambda^{-1}_{K^+\pi^-}=r_d(p/q)e^{-i(\delta+\phi_D)}$. By fitting to
the six coefficients of the various times dependences, one can
extract $r_d$, $|q/p|$, $(x^2+y^2)$,
$y^\prime\cos\phi_D$, and $x^\prime\sin\phi_D$. In particular, finding \CP
violation, that is, $|q/p|\neq1$ and/or $\sin\phi_D\neq0$, would
constitute evidence for new physics.

\section{$B$ Decays}
\label{sec:b}
The upper bound on the CP asymmetry in semileptonic $B$ decays
\cite{Eidelman:2004wy} implies that CP violation in
$B^0-\overline{B}^0$ mixing is a small effect [we use 
${\cal A}_{\rm SL}/2 \approx 1-|q/p|$, see Eq.~(\ref{asysl})]:
\beq\label{expasl}
{\cal A}_{\rm SL}=(0.3\pm1.3)\times10^{-2}\ \ \Longrightarrow\ \
|q/p|=0.998\pm0.007.
\eeq
The Standard Model prediction is
\beq\label{smasl} {\cal A}_{\rm SL}={\cal
  O}\left(\frac{m_c^2}{m_t^2}\sin\beta\right)\lsim 0.001.
\eeq
In models where $\Gamma_{12}/M_{12}$ is approximately real, such as
the Standard Model, an upper bound on $\Delta\Gamma/\Delta
m\approx\re{\Gamma_{12}/M_{12}}$ provides yet another upper bound
on the deviation of $|q/p|$ from one. This constraint does not hold if
$\Gamma_{12}/M_{12}$ is approximately imaginary. 

The small deviation (less than one percent) of $|q/p|$ from $1$
implies that, at the present level of experimental precision, CP
violation in $B$ mixing is a negligible effect.
Thus, for the purpose of analyzing CP asymmetries in hadronic $B$
decays, we can use 
\beq\label{lamhad}
\lambda_f=e^{-i\phi_B}(\overline{A}_f/A_f) \; ,
\eeq
where $\phi_B$ refers to the phase of $M_{12}$ [see
Eq.~(\ref{defmgam})]. 
Within the Standard Model, the corresponding phase factor is given by
\beq\label{phimsm}
e^{-i\phi_B}=(V_{tb}^* V_{td}^{})/(V_{tb}^{}V_{td}^*) \;.
\eeq

\begin{figure}[htb]
\caption{Feynman diagrams for (a) tree and (b) penguin amplitudes
  contributing to $B^0\to f$ or $B_{s}\to f$ via a $\bar b\to\bar q
  q\bar q^\prime$ quark-level process.}
\label{fig:diags}
\begin{center}
\includegraphics[width=2.85in]{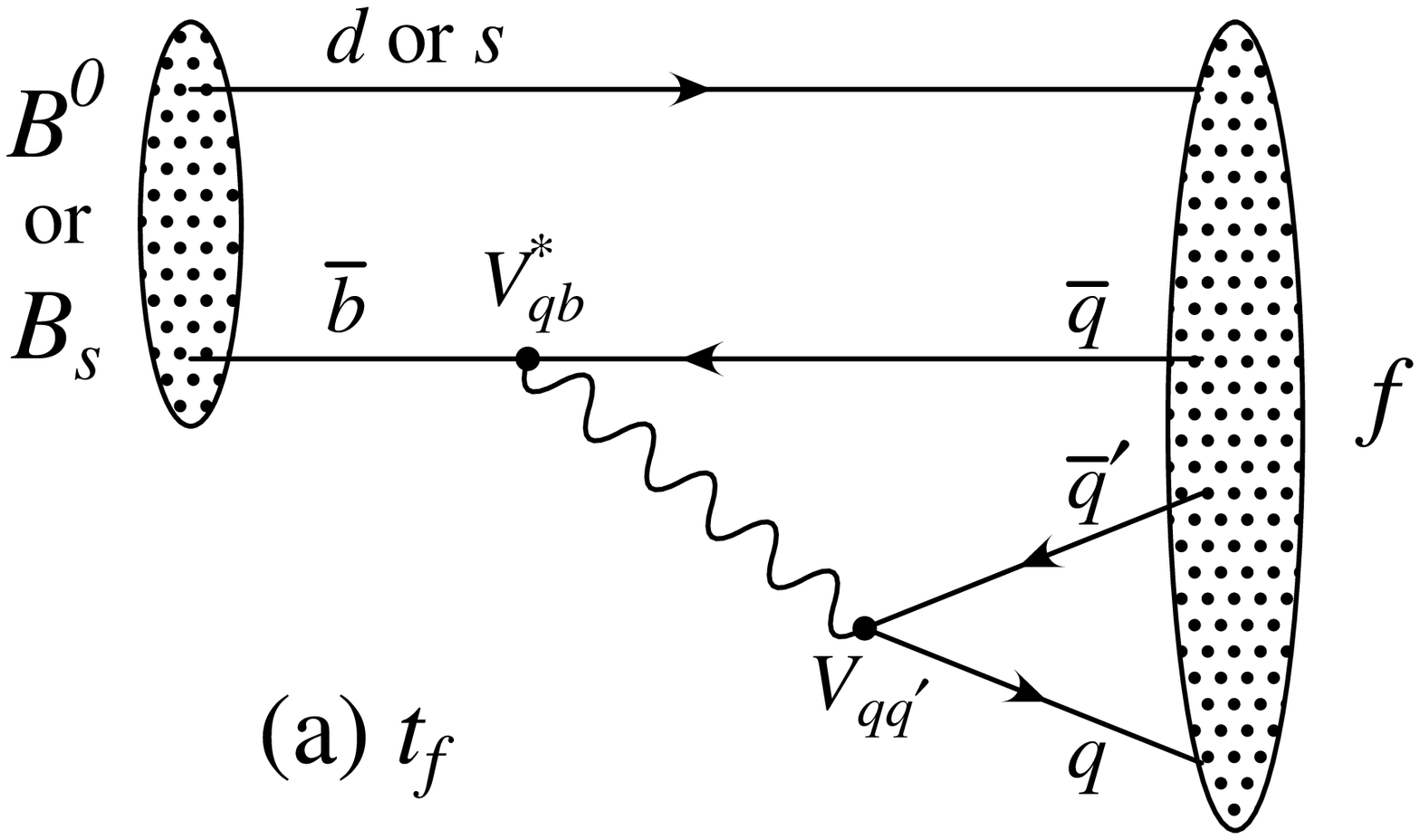}
\hspace{2em}
\includegraphics[width=2.85in]{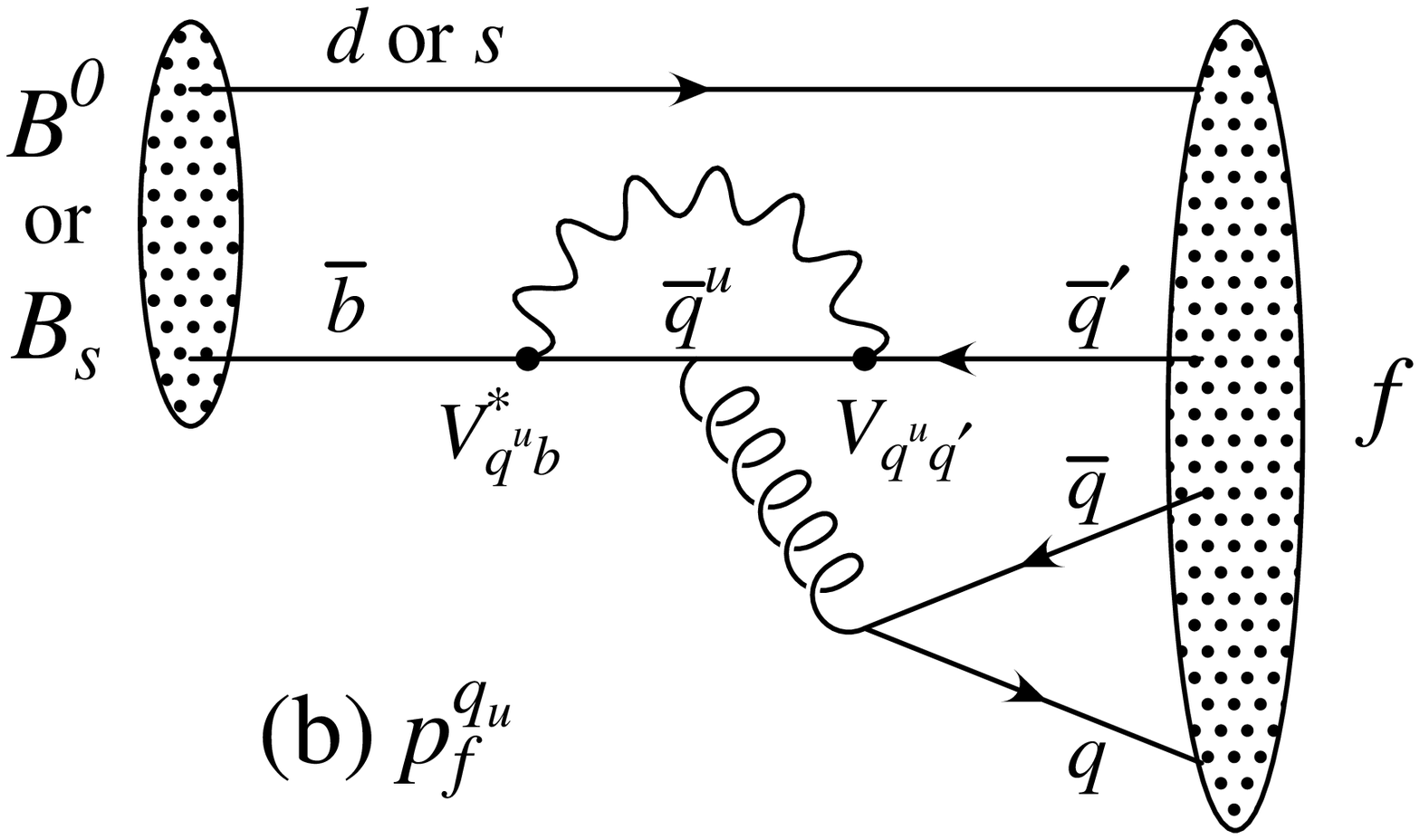}
\end{center}
\end{figure}

Some of the most interesting decays involve final states that are
common to $B^0$ and
$\overline{B}^0$~\cite{Carter:1980hr,Carter:1981tk,Bigi:1981qs}. 
Here Eq. (\ref{asyfcpb}) applies \cite{Dunietz:1986vi,Blinov:ru,Bigi:1986vr}.  
The processes of interest proceed via quark transitions of the form
$\bar b\to\bar q q\bar q^\prime$ with $q^\prime=s$ or $d$. For $q=c$
or $u$, there are contributions from both tree ($t$) and penguin
($p^{q_u}$, where $q_u=u,c,t$ is the quark in the loop) diagrams
(see Fig.~\ref{fig:diags}) which carry different weak phases:
\beq\label{ckmdec}
A_f = \left(V^\ast_{qb} V^{}_{qq'}\right) t_f +
\sum_{q_u= u,c,t}\left(V^\ast_{q_u b} V^{}_{q_u q'}\right) p^{q_u}_f \; .
\eeq
(The distinction between tree and penguin contributions is a heuristic
one, the separation by the operator that enters is more precise. For a
detailed discussion of the more complete operator product approach,
which also includes higher order QCD corrections, see, for example,
ref. \cite{Buchalla:1995vs}.)
Using CKM unitarity, these decay amplitudes can always be written in
terms of just two CKM combinations. For example, for $f=\pi\pi$, which
proceeds via $\bar b\to \bar uu\bar d$ transition, we can write
\beq\label{btouud}
A_{\pi\pi}=\left(V^\ast_{ub} V^{}_{ud}\right)T_{\pi\pi}
+\left(V^\ast_{tb} V^{}_{td}\right)P^t_{\pi\pi}, 
\eeq
where $T_{\pi\pi}=t_{\pi\pi}+p^u_{\pi\pi}-p^c_{\pi\pi}$ and
$P^t_{\pi\pi}=p^t_{\pi\pi}-p^c_{\pi\pi}$. CP violating phases in
Eq.~(\ref{btouud}) appear only in the CKM elements, so that
\beq\label{bbtouud}
\frac{\overline{A}_{\pi\pi}}{A_{\pi\pi}}=
\frac{\left(V_{ub}^{} V^\ast_{ud}\right)T_{\pi\pi}
  +\left(V_{tb}^{} V^\ast_{td}\right)P^t_{\pi\pi}}
{\left(V^\ast_{ub} V^{}_{ud}\right)T_{\pi\pi}
+\left(V^\ast_{tb} V^{}_{td}\right)P^t_{\pi\pi}}. 
\eeq
For $f=J/\psi K$, which proceeds via $\bar b\to \bar cc\bar s$
transition, we can write 
\beq\label{btoccs}
A_{\psi K}=\left(V^\ast_{cb} V^{}_{cs}\right)T_{\psi
  K}+\left(V^\ast_{ub} V^{}_{us}\right)P^u_{\psi K}, 
\eeq
where $T_{\psi K}=t_{\psi K}+p^c_{\psi K}-p^t_{\psi K}$ and
$P^u_{\psi K}=p^u_{\psi K}-p^t_{\psi K}$. A subtlety arises in this
decay that is related to the fact that ${B}^0$ decays into $J/\psi
K^0$ while $\overline{B}^0$ decays into $J/\psi\overline{K}{}^0$. A
common final state, 
e.g. $J/\psi K_S$, is reached only via $K^0-\overline{K}{}^0$ mixing. 
Consequently, the phase factor corresponding to neutral $K$ mixing,
$e^{-i\phi_K}=(V^*_{cd}V^{}_{cs})/(V^{}_{cd}V^*_{cs})$, plays a
role: 
\beq\label{psikmix}
\frac{\overline{A}_{\psi K_S}}{A_{\psi K_S}}
=-\frac{\left(V^{}_{cb} V^\ast_{cs}\right)T_{\psi
    K}+\left(V^{}_{ub} V^\ast_{us}\right)P^u_{\psi K}}
{\left(V^\ast_{cb} V^{}_{cs}\right)T_{\psi
    K}+\left(V^\ast_{ub} V^{}_{us}\right)P^u_{\psi K}}\times
\frac{V_{cd}^\ast V_{cs}^{}}{V_{cd}^{}V_{cs}^\ast}.
\eeq
For $q=s$ or $d$, there are only penguin contributions
to $A_f$, that is, $t_f=0$ in Eq. (\ref{ckmdec}). (The tree $\bar
b\to\bar uu\bar q^\prime$ transition followed by $\bar uu\to\bar qq$
rescattering is included below in the $P^u$ terms.) Again, CKM
unitarity allows us to write $A_f$ in terms of two CKM
combinations. For example, for $f=\phi K_S$,
which proceeds via $\bar b\to \bar ss\bar s$ transition, we can write
\beq\label{btosss}
\frac{\overline{A}_{\phi K_S}}{A_{\phi K_S}}=
-\frac{\left(V^{}_{cb} V^\ast_{cs}\right)P^c_{\phi
    K}+\left(V^{}_{ub} V^\ast_{us}\right)P^u_{\phi K}}
{\left(V^\ast_{cb} V^{}_{cs}\right)P^c_{\phi
  K}+\left(V^\ast_{ub} V^{}_{us}\right)P^u_{\phi K}}\times
\frac{V_{cd}^\ast V_{cs}^{}}{V_{cd}^{}V_{cs}^\ast}, 
\eeq
where $P^c_{\phi K}=p^c_{\phi K}-p^t_{\phi K}$ and
$P^u_{\phi K}=p^u_{\phi K}-p^t_{\phi K}$.

Since the amplitude $A_f$ involves two different weak phases, the
corresponding decays can exhibit both CP violation in the
interference of decays with and without mixing, $S_f\neq0$, and CP
violation in decays, $C_f\neq0$. [At the present level of experimental
precision, the contribution to $C_f$ from CP violation in mixing is
negligible, see Eq. (\ref{expasl}).]  If the contribution from a
second weak phase is suppressed, then the interpretation of $S_f$ in
terms of Lagrangian CP-violating parameters is clean, while $C_f$ is
small. If such a second contribution is not suppressed, $S_f$ depends
on hadronic parameters and, if the relevant strong phase is large,
$C_f$ is large.

A summary of $\bar b\to\bar q q\bar q^\prime$ modes with $q^\prime=s$
or $d$ is given in Table~\ref{tab:bqqq}.  The $\bar b\to\bar dd\bar q$
transitions lead to final states that are similar to the $\bar
b\to\bar uu\bar q$ transitions and have similar phase dependence.
Final states that consist of two vector-mesons ($\psi\phi$ and
$\phi\phi$) are not CP eigenstates, and angular analysis is
needed to separate the CP-even from the CP-odd contributions.

\begin{table}[t]
\caption{Summary of $\bar b\to\bar q q\bar q^\prime$ modes with
  $q^\prime=s$ or $d$. The second and third columns give examples of
  final hadronic states. The fourth column gives the CKM dependence of
  the amplitude $A_f$, using the notation of
  Eqs. (\ref{btouud},\ref{btoccs},\ref{btosss}), with the dominant
  term first and the sub-dominant second. The suppression factor of the
  second term compared to the first is given in the last column.
  ``Loop'' refers to a penguin versus tree suppression factor (it is
  mode-dependent and roughly ${\cal O}(0.2-0.3)$) and $\lambda = 0.22$
  is the expansion parameter of Eq.~(\ref{wolpar}).}
\label{tab:bqqq}
\begin{center}
\begin{tabular}{lcccc} \hline
\rule{0pt}{1.2em}%
$\bar b\to q\bar q\bar q^\prime$ & $B^0\to f$ & $B_s\to f$ & CKM
dependence of $A_f$ &  Suppression \\[2pt] \hline
\rule{0pt}{1.2em}%
$\bar b\to\bar cc\bar s$ & $\psi K_S$ & $\psi\phi$
& $(V^\ast_{cb} V^{}_{cs})T+(V^\ast_{ub} V^{}_{us})P^u$ & ${\rm loop}\times\lambda^2$ \\
$\bar b\to\bar ss\bar s$ & $\phi K_S$ & $\phi\phi$ &
$(V^\ast_{cb} V^{}_{cs})P^c+(V^\ast_{ub} V^{}_{us})P^u$ & $\lambda^2$ \\
$\bar b\to\bar uu\bar s$ & $\pi^0 K_S$ & $K^+K^-$ &
$(V^\ast_{cb} V^{}_{cs})P^c+(V^\ast_{ub} V^{}_{us})T$ & $\lambda^2/{\rm loop}$ \\
$\bar b\to\bar cc\bar d$ & $D^+D^-$ & $\psi K_S$ &
$(V^\ast_{cb} V^{}_{cd})T+(V^\ast_{tb} V^{}_{td})P^t$ & ${\rm loop}$ \\
$\bar b\to\bar ss\bar d$ & $\phi\pi$ & $\phi K_S$ &
$(V^\ast_{tb} V^{}_{td})P^t+(V^\ast_{cb} V^{}_{cd})P^c$ & $\lsim 1$ \\
\rule[-0.6em]{0pt}{1em}%
$\bar b\to\bar uu\bar d$ & $\pi^+\pi^-$ & $\pi^0 K_S$ &
$(V^\ast_{ub} V^{}_{ud})T+(V^\ast_{tb} V^{}_{td})P^t$ & ${\rm loop}$\\
\hline
\end{tabular}
\end{center}
\end{table}

The cleanliness of the theoretical interpretation of $S_f$ can be
assessed from the information in the last column of
Table~\ref{tab:bqqq}.  In case of small uncertainties, the expression
for $S_f$ in terms of CKM phases can be deduced from the fourth column
of Table~\ref{tab:bqqq} in combination with Eq. (\ref{phimsm}) (and,
for $b\to q\bar qs$ decays, the example in Eq.~(\ref{psikmix})). In
the next three sections, we consider three interesting classes.

For $B_s$ decays, one has to replace Eq.~(\ref{phimsm}) with
\beq
e^{-i\phi_{B_s}}=(V_{tb}^* V_{ts}^{})/(V_{tb}^{}V_{ts}^*).
\eeq
Note that one expects $\Delta\Gamma_s/\Gamma_s={\cal O}(0.1)$, and
therefore $y_{B_s}$ should not be put to zero in the expressions for
the time dependent decay rates, but $|q/p|=1$ is expected to hold to
an even better approximation than for $B$ mesons. The CP asymmetry in
$B_s\to D_s^+D_s^-$ (or in $B_s\to\psi\phi$ with angular analysis to
disentangle the CP-even and CP-odd components of the final state) will
determine $\sin2\beta_s$, where $\beta_s$ is defined in
Eq.~(\ref{bbangles}). Since the SM prediction is that this asymmetry
is small [see Eq.~(\ref{abccon})], $\sin2\beta_s\sim0.036$, an
observation of a $S_{B_s\to D_s^+D_s^-}\gg0.04$ will provide evidence
for new physics.

\section{$b\to c\bar cs$ transitions}
\label{sec:bccs}
For $B\to J/\psi K_S$ and other $\bar b\to\bar cc\bar s$ processes, we
can neglect the $P^u$ contribution to $A_{\psi K}$, in the SM, to
an approximation that is better than one percent:
\beq\label{btopsik}
\lambda_{\psi K_S}=-e^{-2i\beta}\ \Rightarrow\
S_{\psi K_S}=\sin2\beta,\ \ \ C_{\psi K_S}=0 \; .
\eeq
(Below the percent level, several effects modify this equation
\cite{Grossman:2002bu,Boos:2004xp}.) The experimental measurements
give the following ranges \cite{ichep:2005-abe}:
\beq\label{scpkexp}
S_{\psi K_S}=0.69\pm0.03,\ \ \ C_{\psi K_S}=0.02\pm0.05 \; .
\eeq

The consistency of the experimental results (\ref{scpkexp}) with the
SM predictions means that the KM mechanism of CP violation has
successfully passed its first precision test. For the first time, we
can make the following statement based on experimental evidence:\\
{\bf Very likely, the Kobayashi-Maskawa mechanism is the dominant
  source of CP violation in flavor changing processes.}

There are three qualifications implicit in this statement, and we now
explain them in little more detail \cite{Nir:2002gu}.
\begin{itemize}
\item {\it `Very likely'}: It could be that the success is
  accidental. It could happen, for example, that $\sin2\beta$ is
  significantly different from the SM value and that, at the same
  time, there is a significant CP violating contribution to the
  $B^0-\overline{B}{}^0$ mixing amplitude, and the sum of $M_{12}^{\rm
    SM}+M_{12}^{\rm NP}$ accidentally carries the same phase as the
  one predicted by the SM alone. It could also happen that the size of
  NP contributions to $b\to d$ transitions is small, or that its phase
  is similar to the SM one, but that in $b\to s$ transitions the
  deviation is significant.
\item {\it `Dominant'}: While $S_{\psi K}$ is measured with an
  accuracy of order 0.04, the accuracy of the SM prediction for
  $\sin2\beta$ is only at the level of 0.2. Thus, it is quite possible
  that there is a new physics contribution at the level of
  $|M_{12}^{\rm NP}/M_{12}^{\rm SM}|\lsim{\cal O}(0.2)$. 
\item {\it `Flavor changing'}: It may well happen that the KM phase,
  which is closely related to flavor violation through the CKM matrix,
  dominates meson decays while new, flavor diagonal phases (such as
  the two unavoidable phases in the universal version of the MSSM)
  dominate observables such as electric dipole moments by many orders
  of magnitude.
\end{itemize}

\begin{figure}[htb]
\caption{Constraints in the $\rho-\eta$ plane from
  (a) CP conserving or (b) CP violating loop processes.}
\label{fig:cpccpv}
\begin{center}
\includegraphics[width=2.85in]{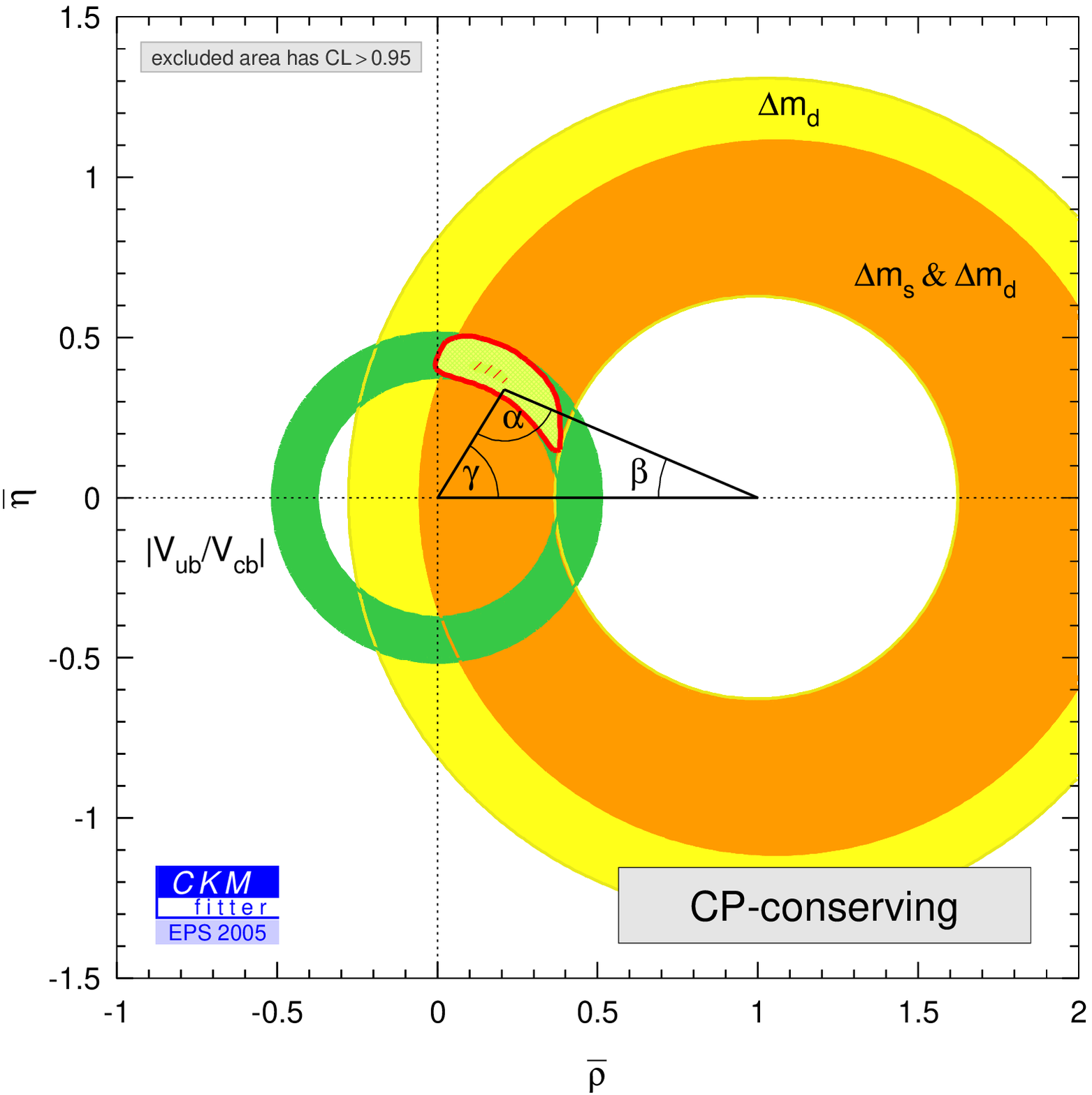}
\hspace{2em}
\includegraphics[width=2.85in]{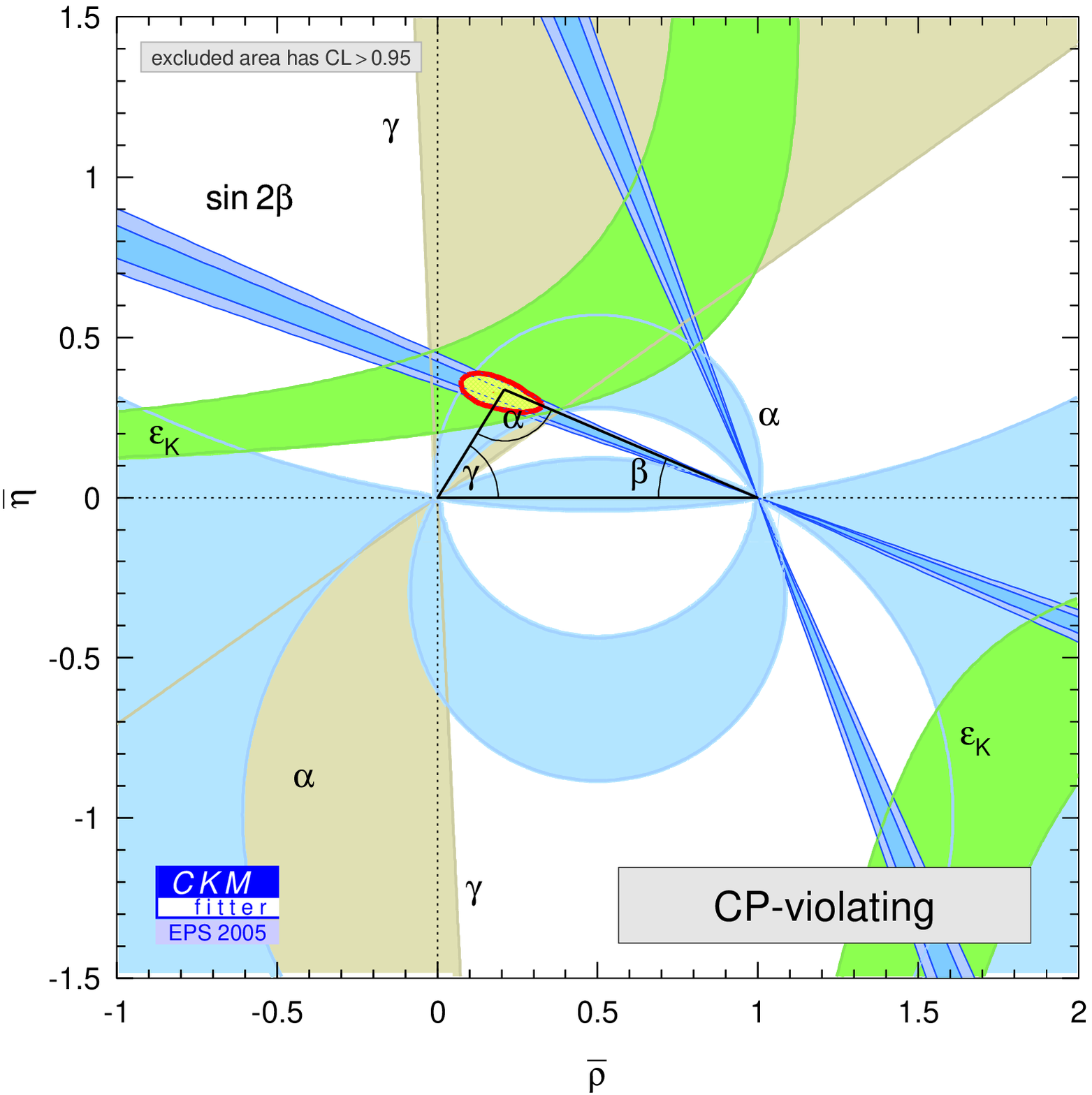}
\end{center}
\end{figure}

The measurement of $S_{\psi K}$ provides a significant constraint on
the unitarity triangle. In the $\rho-\eta$ plane, it reads:
\beq
\sin2\beta=\frac{2\eta(1-\rho)}{\eta^2+(1-\rho)^2}=0.69\pm0.03.
\eeq
One can get an impression of the impact of this constraint by looking
at Fig. \ref{fg:UT}, where the blue region represents
$\sin2\beta=0.69\pm0.03$. An impression of the KM test can be achieved
by observing that the blue region has an excellent
overlap with the region allowed by all other measurements. 
A comparison between the constraints in the $\rho-\eta$ plane from CP
conserving and CP violating processes is provided in
Fig. \ref{fig:cpccpv}. The impressive consistency between the two
allowed regions is the basis for our statement that the KM mechanism
has passed its first precision tests. The fact that the allowed region
from the CP violating processes is more strongly constrained is
related to the fact that CP is a good symmetry of the strong
interactions and that, therefore, various CP violating observables --
in particular $S_{\psi K}$ -- can be cleanly interpreted.

\section{Penguin dominated $b\to s$ transitions}
\label{sec:bsss}
\subsection{General considerations}
The present experimental situation concerning CP asymmetries in decays
to final CP eigenstates dominated by $b\to s$ penguins is summarized
in Table \ref{tab:bsss}. 

\begin{table}[t]
\caption{CP asymmetries in $b\to s$ penguin dominated modes.}
\label{tab:bsss}
\begin{center}
     \begin{tabular}{ccc} \hline
       $f_{\rm CP}$ &
        $-\eta_{f_{\rm CP}}S_{f_{\rm CP}}$ & $C_{f_{\rm CP}}$ \\ \hline
$\phi K_S$  & $+0.47\pm0.19$ &   $-0.09\pm0.15$ \\
$\eta^\prime K_S$ &  $+0.50\pm0.09(0.13)$ & $-0.07\pm0.07(0.10)$ \\
$f_0 K_S$ & $+0.75\pm0.24$ & $+0.06\pm0.21(0.23)$ \\
$\pi^0 K_S$ & $+0.31\pm0.26$ & $-0.02\pm0.13$ \\
$\omega K_S$ & $+0.63\pm0.30$ & $-0.44\pm0.24$ \\
$K_SK_SK_S$ & $+0.61\pm0.23$ & $-0.31\pm0.17(0.20)$ \\
    \hline
  \end{tabular}
\end{center}
\end{table}

For $B\to\phi K_S$ and other $\bar b\to\bar ss\bar s$ processes, we
can neglect the $P^u$ contribution to $A_f$, in the Standard Model, to
an approximation that is good to order of a few percent:
\beq\label{btophik}
\lambda_{\phi K_S}\approx-e^{-2i\beta}\ \Rightarrow\ S_{\phi
  K_S}\approx\sin2\beta,\ \ \ C_{\phi K_S}\approx0 \; .
\eeq
In the presence of new physics, both $A_f$ and $M_{12}$ can get
contributions that are comparable in size to those of the Standard
Model and carry new weak phases \cite{Grossman:1997ke}. Such a
situation gives several interesting consequences for $\bar b\to \bar s
s \bar s$ decays:
\begin{enumerate}
\item A new CP violating phase in the $b\to s$ decay amplitude will
  lead to a deviation of $-\eta_f S_f$ from $S_{\psi K}$.
\item The $S_f$'s will be different, in general, among
  the various $f$'s. Only if the new physics contribution to $A_f$
  dominates over the SM we should expect a universal $S_f$.
\item A new CP violating phase in the $b\to s$ decay amplitude in
  combination with a strong phase will lead to $C_f\neq0$.
\end{enumerate}

\subsection{Calculating the deviations from $S_f=S_{\psi K}$}
It is important to understand how large a deviation from the
approximate equalities in Eq.~(\ref{btophik}) is expected within the
SM. The SM contribution to the decay amplitudes, related to $\bar
b\to\bar qq\bar s$ transitions, can always be written as a sum of two
terms, $A_f^{\rm SM}=A_f^c+A_f^u$, with $A_f^c\propto V_{cb}^*V_{cs}$
and $A_f^u\propto V_{ub}^*V_{us}$. Defining the ratio $a_f^u\equiv
e^{-i\gamma}(A_f^u/A_f^c)$, we have
\beq\label{defafu}
A_f^{\rm SM}=A_f^c(1+a_f^ue^{i\gamma}).
\eeq
The size of the deviations from Eq.~(\ref{btophik}) is set by
$a_f^u$. For $|a_f^u|\ll1$, we obtain
\beqa
-\eta_f S_f&\simeq&\sin2\beta+2\cos2\beta\ {\cal R}e(a_f^u)\sin\gamma,\no\\ 
C_f&\simeq&-2{\cal I}m(a_f^u)\sin\gamma. 
\eeqa
For charmless modes, the effects of the $a_f^u$ terms (often called
`the SM pollution') are at least of order
$|(V_{ub}^*V_{us})/(V_{cb}^*V_{cs})|\sim$ a few percent.

To calculate them explicitly, we use the operator product expansion
(OPE). We follow the notations of ref. \cite{Beneke:2000ry}. We
consider the following  effective Hamiltonian for $\Delta B=\pm1$
decays: 
\beq\label{heffbone}
{\cal H}_{\rm eff}={G_F\over\sqrt2}\sum_{p=u,c}
  V_{ps}^*V_{pb}\left(C_1O_1^p+C_2O_2^p+ 
  \sum_{i=3}^{10}C_iO_i+C_{7\gamma}O_{7\gamma}+C_{8g}O_{8g}\right)+{\rm h.c.},
\eeq
with
\beqa\label{curcur}
O_1^p=(\bar pb)_{V-A}(\bar sp)_{V-A},&\ \ \ \ \ &
O_2^p=(\bar p_{\beta}b_{\alpha})_{V-A}(\bar s_{\alpha}p_{\beta})_{V-A},\no\\
O_3=(\bar sb)_{V-A}\sum_{q}(\bar qq)_{V-A},&\ \ \ \ \ & 
O_4=(\bar s_\alpha b_{\beta})_{V-A}\sum_{q}(\bar q_{\beta}q_{\alpha})_{V-A},\nonumber\\
O_5=(\bar sb)_{V-A}\sum_{q}(\bar qq)_{V+A},&\ \ \ \ \ &
O_6=(\bar s_{\alpha}b_{\beta})_{V-A}\sum_{q}(\bar q_{\beta}q_{\alpha})_{V+A},\no\\
O_7=\frac{3}{2}(\bar sb)_{V-A}\sum_{q}e_q(\bar qq)_{V+A},&\ \ \ \ \ &
O_8=\frac{3}{2}(\bar s_{\alpha}b_{\beta})_{V-A}\sum_{q}e_q(\bar
q_{\beta}q_{\alpha})_{V+A},\nonumber\\
O_9=\frac{3}{2}(\bar sb)_{V-A}\sum_{q}e_q(\bar qq)_{V-A},&\ \ \ \ \ &
O_{10}=\frac{3}{2}(\bar s_{\alpha}b_{\beta})_{V-A}\sum_{q}e_q(\bar q_{\beta}q_{\alpha})_{V-A},\no\\
O_{7\gamma}=-\frac{em_b}{8\pi^2}\bar s\sigma^{\mu\nu}(1+\gamma_5) F_{\mu\nu}b,&\ \ \ \ \ & 
O_{8g}=-\frac{g_sm_b}{8\pi^2}\bar s\sigma^{\mu\nu}(1+\gamma_5)G_{\mu\nu}b,
\eeqa
where $(\bar q_1 q_2)_{V\pm A}=\bar q_1\gamma_\mu(1\pm\gamma_5)q_2$,
the sum is over active quarks, with $e_q$ denoting their electric
charge in fractions of $|e|$ and $\alpha,\beta$ are color indices. The
decay amplitudes can be calculated from this effective Hamiltonian:
\beq
A_f=\langle f|{\cal H}_{\rm eff}|B^0\rangle,\ \ \
\overline{A}_f=\langle f|{\cal H}_{\rm eff}|\overline{B}^0\rangle.
\eeq
The electroweak model determines the Wilson coefficients    
while QCD (or, more practically, a calculational method such as QCD
factorization) determines the matrix elements $\langle
f|O_i|B^0(\overline{B}^0)\rangle$.

Take, for example, the ${B}^0\to {K}^0\pi^0$ decay amplitude. It can
be written as follows (for simplicity, we omit the contributions from
$O_{7-10}$): 
\beqa
A_{{K}^0\pi^0}^c&\approx&i V_{cb}^*V_{cs}\frac{G_F}{2}
f_KF^{B\to\pi}(m_K^2)(m_B^2-m_\pi^2)\left(
a_4+r_\chi a_6\right),\\
A_{{K}^0\pi^0}^u&\approx&i V_{ub}^*V_{us}\frac{G_F}{2}\left[
f_KF^{B\to\pi}(m_K^2)(m_B^2-m_\pi^2)\left(
a_4+r_\chi a_6\right)-
f_\pi F^{B\to K}(m_\pi^2)(m_B^2-m_K^2)a_2\right],\no
\eeqa
where $r_\chi=2m_K^2/[m_b(m_s+m_d)]$. The $a_i$ parameters are related
to the Wilson coefficients as follows:
\beq
a_i\equiv C_i+\frac{1}{N_c}C_{i\pm1}\ {\rm for}\
i={\rm odd,even}.
\eeq
Within the SM, at leading order,
\beq
C_1(m_W)=1,\ \ \ C_{i\neq1}(m_W)=0.
\eeq
(Strictly speaking, $C_{7\gamma}(m_W)$ and $C_{8g}(m_W)$ are also
different from zero. Their contributions to the decay processes of
interest occur, however, at next-to-leading order which we neglect
here for simplicity.) To run the Wilson coefficients from the weak
scale $m_W$ to the low scale of order $m_b$, we use
\beq
\vec C(\mu)=[\alpha_s(m_W)/\alpha_s(\mu)]^{\gamma/2\beta_0},
\eeq
where $\beta_0=(33-2f)/3$, with $f=5$ for $m_b\leq\mu\leq m_W$, and  
$\gamma$ is the 12-dimensional leading-log anomalous dimension matrix 
which can be found, for example, in ref. \cite{Buchalla:2005us}. The
bottom line is the following set of values for the relevant $a_i$
parameters at the scale $\mu=m_b$:
\beq
a_1=1.028,\ \ a_2=0.105,\ \ a_4=-0.0233,\ \ a_6=-0.0314.
\eeq
We use the following values for the relevant hadronic parameters:
\beq
f_\pi=131\ MeV,\ \ f_K=160\ MeV,\ \ F^{B\to\pi}(0)=0.28,\ \ \ F^{B\to
  K}(0)=0.34,\ \ r_\chi(m_b)=1.170.
\eeq
Thus we can estimate $a_{\pi K}^u$:
\beq
a_{\pi K}^u\approx\lambda^2 R_u\left(1-\frac{f_\pi}{f_K}
  \frac{F^{B\to K}}{F^{B\to\pi}}\frac{a_2}{a_4+r_\chi
    a_6}\right)\approx 2.75\lambda^2 R_u\approx0.052.
\eeq
We learn that the SM and factorization predict that $-S_{\pi^0
  K_S}-S_{\psi K_S}\approx+0.05$. 

In Table \ref{tab:afu} we give the values of the $a_f^u$ parameter
(obtained in ref. \cite{Buchalla:2005us} by using factorization
\cite{Beneke:2000ry,Ali:1998eb,Beneke:2002jn}) for
all relevant modes.

\begin{table}[t]
  \caption{The $a_f^u$ parameters, calculated in QCD factorization at
    leading log and to zeroth order in $\Lambda/m_b$
    (except for chirally enhanced corrections), and 
    the SM values of $S_f$ for $\mu=m_b$ and in parentheses the
    respective values for $\mu=2 m_b$ (first) and $\mu= m_b/2$
    (second) if different from the central one. In the last column,
    the results of ref. \cite{Beneke:2005pu}, using QCD factorization
    at NLO, are given. Taken from \cite{Buchalla:2005us}.
\label{tab:afu}}
\vspace{0.4cm}
\begin{center}
  \begin{tabular}{|c|c|c|c|}
  \hline
$f$               & $a_f^u$ \cite{Buchalla:2005us}  & $-\eta_{\rm
  CP}S_f$ \cite{Buchalla:2005us} & $-\eta_{\rm
  CP}S_f$ \cite{Beneke:2005pu} \\ \hline 
$\psi K_S$        & $0$      & $0.69$ & $0.69$ \\ \hline
$\phi K_S$        & $0.019$  & $0.71$  &  $0.71\pm0.01$ \\
$\pi^0 K_S$       & $0.052 \, \left[0.094, 0.021\right]$  & $0.75 \,
\left[0.79, 0.72\right]$ & $0.76^{+0.05}_{-0.04}$  \\
$\eta K_S$        & $0.08 \, \left[0.16, 0.02\right] $   & $0.78 \,
\left[0.84, 0.72\right]$ & $0.79^{+0.11}_{-0.07}$  \\
$\eta^\prime K_S$ & $0.007 \, \left[-0.006, 0.019\right] $ & $0.70 \,
\left[0.68, 0.71\right]$ & $0.70\pm0.01$  \\
$\omega K_S$      & $0.22 \, \left[0.37, 0.04\right]$   & $0.88 \,
\left[0.94, 0.74\right]$ & $0.82\pm0.08$  \\
$\rho^0 K_S$      & $-0.16 \, \left[-0.32, 0.005\right]$  & $0.45 \,
\left[0.15, 0.70\right]$ & $0.61^{+0.08}_{-0.12}$ \\ 
\hline
\end{tabular}
\end{center}
\end{table}

An examination of Table \ref{tab:afu} shows that the SM pollution is
small (that is, at the naively expected level of
$|(V_{ub}V_{us}^*)/(V_{cb}V_{cs}^*)|\sim$ a few percent) for $f=\phi K_S,\
\eta^\prime K_S$ and $\pi^0 K_S$. It is larger for $f=\eta
K_S,\ \omega K_S$ and $\rho^0 K_S$. In these modes, $a_f^u$ is
enhanced because, within the QCD factorization approach, there is an
accidental cancellation between the leading contributions to $A_f^c$.
The reason for the suppression of the leading $A^c_f$ piece in 
$f=\rho K,\ \omega K$ versus $f=\pi^0 K$ is that the dominant QCD-penguin
coefficients $a_4$ and $a_6$ appear in $A_{(\rho,\omega)K}^c$ as
$(a_4 -r_\chi a_6)$ and in $A^c_{\pi^0 K}$ as
$(a_4 +r_\chi a_6)$. Since $r_\chi \simeq 1$ and,
within the Standard Model, $a_4\sim a_6$, there is a cancellation in
$A_{(\rho,\omega)K}^c$ while there isn't one in $A^c_{\pi^0 K}$. 
The suppression for $A^c_{\eta K}$ with respect to $A^c_{\eta^\prime
  K}$ has a different reason: it is due to the octet-singlet mixing,
which causes destructive (constructive) interference in the
$\eta(\eta^\prime)K$ penguin amplitude \cite{Lipkin:1980tk}. 

\section{$b\to u\bar ud$ transitions}
\label{sec:buud}
The present experimental situation concerning CP asymmetries in decays
to final CP eigenstates via $b\to d$ transitions is summarized
in Table \ref{tab:buud}. 

\begin{table}[t]
  \caption{CP asymmetries in $b\to c\bar cd$ (above line) or $b\to
u\bar ud$ (below line) modes.}
\label{tab:buud}
\begin{center}
\begin{tabular}{ccc} \hline
$f_{\rm CP}$ & $-\eta_{f_{\rm CP}}S_{f_{\rm CP}}$ & $C_{f_{\rm CP}}$ \\ \hline
$\psi\pi^0$  & $+0.69\pm0.25$ &   $-0.11\pm0.20$ \\
$D^+D^-$ &  $+0.29\pm0.63$ & $+0.11\pm0.35$ \\
$D^{*+}D^{*-}$ & $+0.75\pm0.23$ & $-0.04\pm0.14$ \\ \hline
$\pi^+\pi^-$ & $+0.50\pm0.12(0.18)$ & $-0.37\pm0.10(0.23)$ \\
$\pi^0\pi^0$ & & $-0.28\pm0.39$ \\
$\rho^+\rho^-$ & $+0.22\pm0.22$ & $-0.02\pm0.17$ \\
    \hline
  \end{tabular}
\end{center}
\end{table}

For $B\to\pi\pi$ and other $\bar b\to\bar uu\bar d$ processes,
the penguin-to-tree ratio can be estimated using SU(3)
relations and experimental data on related $B\to K\pi$ decays. The
result is that the suppression is of order $0.2-0.3$ and so cannot be
neglected. The expressions for $S_{\pi\pi}$ and $C_{\pi\pi}$ to
leading order in
$R_{PT}\equiv(|V_{tb}V_{td}|P^t_{\pi\pi})/(|V_{ub}V_{ud}|T_{\pi\pi})$
are: 
\beqa\label{btopipi}
\lambda_{\pi\pi}&=&e^{2i\alpha}\left[(1-
    R_{PT}e^{-i\alpha})/(1-R_{PT}e^{+i\alpha})\right]\ \Rightarrow\no\\
S_{\pi\pi}&\approx&\sin2\alpha+2\,
\re{R_{PT}}\cos2\alpha\sin\alpha,\ \ \ 
C_{\pi\pi}\approx2\,\im{R_{PT}}\sin\alpha.
\eeqa
$R_{PT}$ is mode-dependent and, in particular, could be
different for $\pi^+\pi^-$ and $\pi^0\pi^0$.
If strong phases can be neglected then $R_{PT}$ is real,
resulting in $C_{\pi\pi}=0$. As concerns
$S_{\pi\pi}$, it is clear from (\ref{btopipi}) that the relative size
and strong phase of the penguin contribution must be known to extract
$\alpha$. (Only one of the two is required if both $C_{\pi\pi}$ and
$S_{\pi\pi}$ are measured.) This is the problem of penguin pollution. 

The cleanest solution involves isospin relations among the 
$B\to\pi\pi$ amplitudes. Let us derive this relation step by step. The
$SU(2)$-isospin representations of the $\pi\pi$ states are as follows:
\beqa\label{isopp}
\langle \pi^+\pi^-|&=&\tiny{\sqrt{\frac12}}\langle(1,+1)(1,-1)+(1,-1)(1,+1)|
=\sqrt{1\over3}\ \langle2,0|+\sqrt{2\over3}\ \langle0,0|,\no\\
\langle\pi^0\pi^0|&=&\langle(1,0)(1,0)|=
\sqrt{\frac23}\ \langle2,0|-\sqrt{\frac13}\ \langle0,0|,\no\\
\langle\pi^+\pi^0|&=&\sqrt{\frac12}\langle(1,+1)(1,0)+(1,0)(1,+1)|=\langle 2,+1|. 
\eeqa
The Hamiltonian, with its four quark operators, has two features that
are important for our purposes:
\begin{enumerate}
  \item There are $\Delta I=1/2$ and $\Delta I=3/2$ pieces, but no
    $\Delta I=5/2$ one. The absence of the latter gives isospin
    relations among the $B\to\pi\pi$ amplitdues.
    \item The penguin operatores are purely $\Delta I=1/2$. Thus we
      will find that they do not contribute to the $\pi^\pm\pi^0$
      modes.
    \end{enumerate}
We contract the Hamiltonian with with the $(B^+,B^0)=(1/2,\pm1/2)$ states:
\beqa\label{isohb}
H_{3/2,+1/2}|1/2,-1/2\rangle&\propto&\sqrt{\frac12}\ |2,0\rangle+\sqrt{\frac12}\ |1,0\rangle,\no\\
H_{3/2,+1/2}|1/2,+1/2\rangle&\propto&\sqrt{\frac34}\ |2,1\rangle-\sqrt{\frac14}\ |1,1\rangle,\no\\
H_{1/2,+1/2}|1/2,-1/2\rangle&\propto&\sqrt{\frac12}\ |1,0\rangle-\sqrt{\frac12}\ |0,0\rangle,\no\\
H_{1/2,+1/2}|1/2,+1/2\rangle&\propto&|1,0\rangle.
\eeqa
Combining (\ref{isopp}) and (\ref{isohb}), we obtain:
\beqa
A_{\pi^+\pi^-}&=&\sqrt{1/6}\ A_{3/2}-\sqrt{1/3}\ A_{1/2},\no\\
A_{\pi^0\pi^0}&=&\sqrt{1/3}\ A_{3/2}+\sqrt{1/6}\ A_{1/2},\no\\
A_{\pi^+\pi^0}&=&\sqrt{3/4}\ A_{3/2}.
\eeqa
Analogous relation hold for the CP-conjugate amplitudes,
$\overline{A}_{\pi^i\pi^j}$. These isospin decompositions lead to the
Gronau-London triangle relations \cite{Gronau:1990ka}: 
\beqa\label{grolon}
\frac{1}{\sqrt2}A_{\pi^+\pi^-}+A_{\pi^0\pi^0}&=&A_{\pi^+\pi^0},\no\\
\frac{1}{\sqrt2}\overline{A}_{\pi^+\pi^-}+\overline{A}_{\pi^0\pi^0}&=&\overline{A}_{\pi^-\pi^0}. 
\eeqa
The method further exploits the fact that the penguin contribution
to $P_{\pi\pi}$ is pure $\Delta I=\frac12$ (this is not true for the
electroweak penguins which, however, are expected to be small), while
the tree contribution  to $T_{\pi\pi}$ contains pieces which are both
$\Delta I=\frac12$ and $\Delta I=\frac32$. A simple geometric
construction then allows one to find $R_{PT}$ and extract $\alpha$
cleanly from $S_{\pi^+\pi^-}$. Explicitly, one notes that, since
$A_{3/2}$ comes purely from tree contributions, we have
\beq
\frac qp \frac{\overline{A}_{3/2}}{A_{3/2}}=-e^{2i\alpha}.
\eeq
The branching ratios of the various modes determine $|A_{\pi^i\pi^j}|$
and $|\overline{A}_{\pi^i\pi^j}|$ (with
$|A_{\pi^+\pi^0}|=|\overline{A}_{\pi^-\pi^0}|$). This would determine
the shape of each of the triangles (\ref{grolon}). Defining
\beq
A_0\equiv(1/\sqrt{6})\ A_{1/2},\ \ \ \
A_2\equiv(1/\sqrt{12})\ A_{3/2},
\eeq
we can obtain $A_2=(1/3)A_{\pi^+\pi^0}$ and
$A_0=(1/\sqrt{2})A_{\pi^+\pi^-}-A_2$. Similarly, we can obtain
$\overline{A}_2$ and $\overline{A}_0$. Next, we define (and obtain)
\beq
\theta\equiv\arg(A_0A_2^*),\ \ \ \
\overline{\theta}\equiv\arg(\overline{A}_0\overline{A}_2^*).
\eeq
Then we have
\beq
{\cal I}m{\lambda_{\pi^+\pi^-}}={\cal I}m\left(-e^{-2i\alpha}
  \frac{|\overline{A}_2|-|\overline{A}_0|e^{i\overline{\theta}}}
  {|A_2|-|A_0|e^{i\theta}}\right).
\eeq
On the other hand, we can use the experimentally measured quantities
to extract ${\cal I}m{\lambda_{\pi^+\pi^-}}$:
\beq
{\cal I}m{\lambda_{\pi^+\pi^-}}=\frac{S_{\pi^+\pi^-}}{1+C_{\pi^+\pi^-}}.
\eeq

The key experimental difficulty is
that one must measure accurately the separate rates for
$B^0,\overline{B}^0\to\pi^0\pi^0$. It has been noted that an upper bound on the
average rate allows one to put a useful upper bound on the deviation of
$S_{\pi^+\pi^-}$ from $\sin2\alpha$
\cite{Grossman:1998jr,Charles:1998qx,Gronau:2001ff}. Parametrizing the
asymmetry by
$S_{\pi^+\pi^-}/\sqrt{1-(C_{\pi^+\pi^-})^2}=\sin(2\alpha_{\rm eff})$, the
bound reads
\beq\label{grqu}
\cos(2\alpha_{\rm
  eff}-2\alpha)\geq\frac{1}{\sqrt{1-(C_{\pi^+\pi^-})^2}}
\left[1-\frac{2{\cal B}_{00}}{{\cal B}_{+0}}
  +\frac{({\cal B}_{+-}-2{\cal B}_{+0}+2{\cal B}_{00})^2}{4{\cal
    B}_{+-}{\cal B}_{+0}}\right]\,,
\eeq
where ${\cal B}_{ij}$ are the averages over CP-conjugate branching
ratios; {\it e.g.},
${\cal B}_{00}=\frac12[{\cal B}(B^0\to\pi^0\pi^0)+{\cal
  B}(\overline{B}^0\to\pi^0\pi^0)]$. CP asymmetries in $B\to\rho\pi$ and, in
particular, in $B\to\rho\rho$ can also be used to determine
$\alpha$
\cite{Lipkin:1991st,Gronau:1991dq,Snyder:1993mx,Quinn:2000by,Falk:2003uq}.
At present, the constraints read \cite{ckmfitter}
\beqa
|\alpha_{\rm eff}^{\pi^+\pi^-}-\alpha|&<38^o,\ \ \
R_{PT}^{\pi^+\pi^-}&=0.37\pm0.17,\no\\
|\alpha_{\rm eff}^{\rho^+\rho^-}-\alpha|&<14^o,\ \ \
R_{PT}^{\rho^+\rho^-}&=0.07^{+0.14}_{-0.07}.
\eeqa

Using isospin analyses for all three systems ($\pi\pi$, $\rho\pi$ and
$\rho\rho$), one obtains \cite{ckmfitter}
\beq\label{alpbfa}
\alpha(\pi\pi,\pi\rho,\rho\rho)=\left[101^{+16}_{-9}\right]^o,
\eeq
to be compared with the result of the CKM fit,
\beq\label{alpckm}
\alpha({\rm CKM\ fit})=96\pm16^o.
\eeq
We would like to emphasize the following points:
\begin{itemize}
\item The consistency of (\ref{alpbfa}) with (\ref{alpckm}) means that
  {\bf the KM mechanism of CP violation has successfully
  passed a second precision test.}
\item The $\alpha$ measurement via the $b\to u\bar ud$ transitions
  provides a significant constraint on the unitarity triangle.
  \item The isospin analysis determines the relative phase
    between the $B^0-\overline{B}^0$ mixing amplitude and the tree
    decay amplitude $A_{3/2}$, independent of the electroweak
    model. The tree decay amplitude is unliley to bne significantly
    affected by new 
    physics. Any new physics modification of the mixing amplitude is
    measured by $S_{\psi k}$. Thus, the combination of $S_{\psi K}$
    and the isospin analysis of $S_{\pi\pi,\rho\pi,\rho\rho}$
    constrains $\alpha$ even in the presence of new physics in
    $B^0-\overline{B}^0$ mixing.
  \end{itemize}

\section{$b\to c\bar us,u\bar c s$ transitions}
An interesting set of measurements is that of $B\to DK$ which proceed
via the quark transitions $\bar b\to\bar cu\bar s$ or $\bar b\to\bar
uc\bar s$ (and their CP conjugates). Given the quark processes, it is
clear that there is no penguin contribution here. Thus, the quark
transitions are purely tree processes. The interference between the two
quark transitions (if they lead to the same final states -- see below)
is sensitive to $\arg[(V_{ub}^*V_{us})/(V_{cb}^*V_{cs})]\approx\gamma$.

There are three variants on this method: GLW
\cite{Gronau:1990ra,Gronau:1991dp}, ADS \cite{Atwood:1996ci} and GGSZ 
\cite{Giri:2003ty}. The simplest one to explain involves branching
ratios of charged $B$ decays, and thus $B^0-\overline{B}^0$
mixing plays no role. Consider the decay $B^\pm\to D_1^0 K^\pm$, where
$D_{1,2}^0=\frac{1}{\sqrt2}(D^0\pm \overline{D}^0)$ are the CP
eigenstates. Taking into account that
\beqa
A(B^+\to D^0 K^+)\times A(D^0\to D_1^0)&\propto&
(V_{ub}^*V_{cs})\times(V_{cs}^*V_{us}),\no\\
A(B^+\to \overline{D}^0 K^+)\times A(\overline{D}^0\to D_1^0)&\propto&
(V_{cb}^*V_{us})\times(V_{us}^*V_{cs}),
\eeqa
we can write the relevant decay amplitudes as follows:
\beqa\label{bdktri}
\sqrt{2}A_{D_1^0 K^+}&=&|A_{D^0 K^+}|e^{i(\delta+\gamma)}+|A_{\overline{D}^0 K^+}|=A_{D^0
  K^+}+A_{\overline{D}^0 K^+},\no\\
\sqrt{2}A_{D_1^0 K^-}&=&|A_{D^0
  K^-}|e^{i(\delta-\gamma)}+|A_{\overline{D}^0 K^-}|=A_{\overline{D}^0
  K^-}+A_{{D}^0 K^-}.
\eeqa
Measuring the rates for the six relevant decay modes ($D_1^0 K^+$,
$D^0 K^+$, $\overline{D}^0 K^+$ and the CP conjugate modes), one can
construct an amplitude triangle for each of the two relations in
Eq.~(\ref{bdktri}). We can choose a phase convention where
$A_{\overline{D}^0 K^+}=A_{{D}^0 K^-}$. Then, the relative angle
between $A_{D^0 K^+}$ and $A_{\overline{D}^0 K^-}$ is $2\gamma$. 

The method of \cite{Giri:2003ty} gives, at present, the most
significant constraints. It allows one to determine the amplitude
ratios, $r(DK)=0.12^{+0.03}_{-0.04}$ and $r(D^*K)=0.09^{+0.03}_{-0.04}$,
and the weak phase $\gamma$ \cite{ckmfitter}:
\beq\label{gambfa}
\gamma(DK)=(63^{+15}_{-13})^o.
\eeq
This range is to be compared with the range of $\gamma$ derived from
the CKM fit (not including the direct $\gamma$ measurements):
\beq\label{gamckm}
\gamma({\rm CKM\ fit})=(57^{+7}_{-14})^o.
\eeq
We would like to emphasize the following points:
\begin{itemize}
\item The consistency of (\ref{gambfa}) with (\ref{gamckm}) means that
  {\bf the KM mechanism of CP violation has successfully
  passed a third precision test.}
\item The $\gamma$ measurement via the $b\to c\bar us,u\bar cs$
  transitions provides yet another constraint on the unitarity
  triangle. The constraint will become more significant when the
  experimental precision improves. 
  \item The determination of $\gamma$ here relies on tree decay
    amplitudes. Thus, the analysis of $B\to DK$ decays constrains
    $\gamma$ even in the presence of new physics in loop processes. 
  \end{itemize}

\section{CP Violation as a Probe of New Physics}
We have argued that the Standard Model picture of CP violation is unique
and highly predictive. We have also stated that reasonable extensions of the 
Standard Model have a very different picture of CP violation. Experimental 
results are now starting to decide between the various possibilities. Our 
discussion of CP violation in the presence of new physics is aimed to 
demonstrate that, indeed, models of new physics can significantly modify the 
Standard Model predictions and that present and near future
measurements have therefore a strong impact on the theoretical
understanding of CP violation. 

To understand how the Standard Model predictions could be modified by New 
Physics, we focus on CP violation in the interference between decays with and 
without mixing. As explained above, this type of CP violation may give, due to 
its theoretical cleanliness, unambiguous evidence for New Physics most easily.
We now demonstrate what type of questions can be (or have already
been) answered when these observables are measured.

{\bf I.} Consider $S_{\psi K_S}$, the CP asymmetry in
$B\rightarrow\psi K_S$. This  
measurement cleanly determines the relative phase between the
$B^0-\overline{B}^0$ mixing amplitude and the $b\to c\bar cs$ decay
amplitude ($\sin2\beta$ in the SM). The $b\to c\bar cs$ decay has
Standard Model tree contributions and therefore is very unlikely to be
significantly affected by new physics. On the other hand, the mixing
amplitude can be easily modified by new physics. We parametrize such a
modification as follows:
\beq\label{derthed}
r_d^2\ e^{2i\theta_d}=\frac{M_{12}}{M_{12}^{\rm SM}}.
\eeq
Then the following observables provide constraints on $r_d^2$ and
$2\theta_d$: 
\beqa\label{apksNP}
S_{\psi K_S}&=&\sin(2\beta+2\theta_d),\no\\
\Delta m_{B}&=&r_d^2(\Delta m_B)^{\rm SM},\no\\
{\cal A}_{\rm SL}&=&-{\cal
    R}e\left(\frac{\Gamma_{12}}{M_{12}}\right)^{\rm
    SM}\frac{\sin2\theta_d}{r_d^2}
  +{\cal I}m\left(\frac{\Gamma_{12}}{M_{12}}\right)^{\rm
    SM}\frac{\cos2\theta_d}{r_d^2}.
\eeqa
Examining whether $S_{\psi K_S}$, $\Delta m_B$ and ${\cal A}_{\rm SL}$
fit the SM prediction, that is, whether $\theta_d\neq0$ and/or
$r_d^2\neq1$, we can answer the following question 
(see {\it e.g.} \cite{Grossman:1997dd}):

(i) {\it Is there new physics in $B^0-\overline{B}^0$ mixing}?

\begin{figure}[htb]
\caption{Constraints in the (a) $\rho-\eta$ plane (b)
  $r_d^2-2\theta_d$ plane, assuming that NP contributions to tree
  level processes are negligible \cite{ckmfitter}.}
\label{fig:rdtd}
\begin{center}
\includegraphics[width=2.85in]{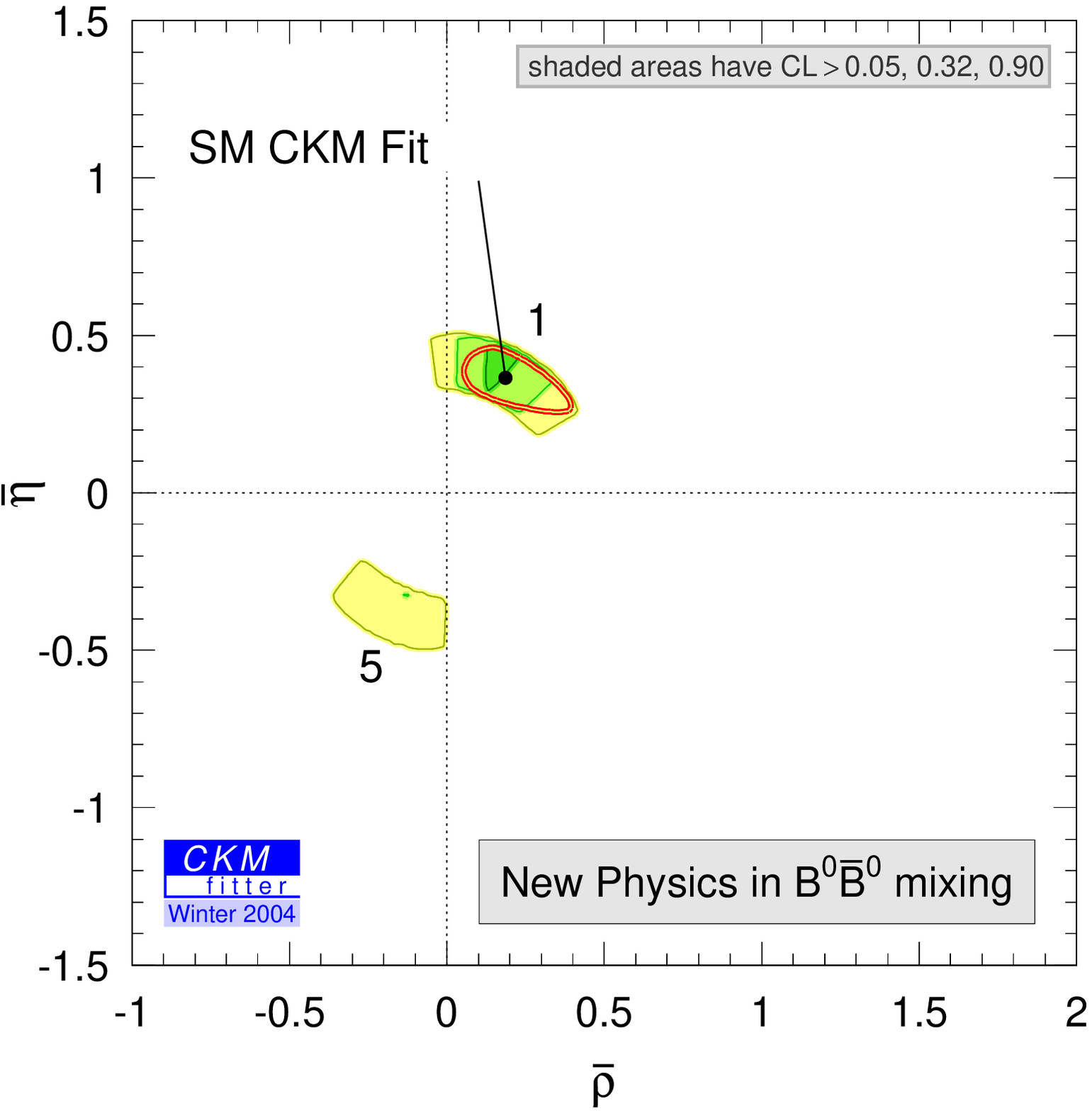}
\hspace{2em}
\includegraphics[width=2.85in]{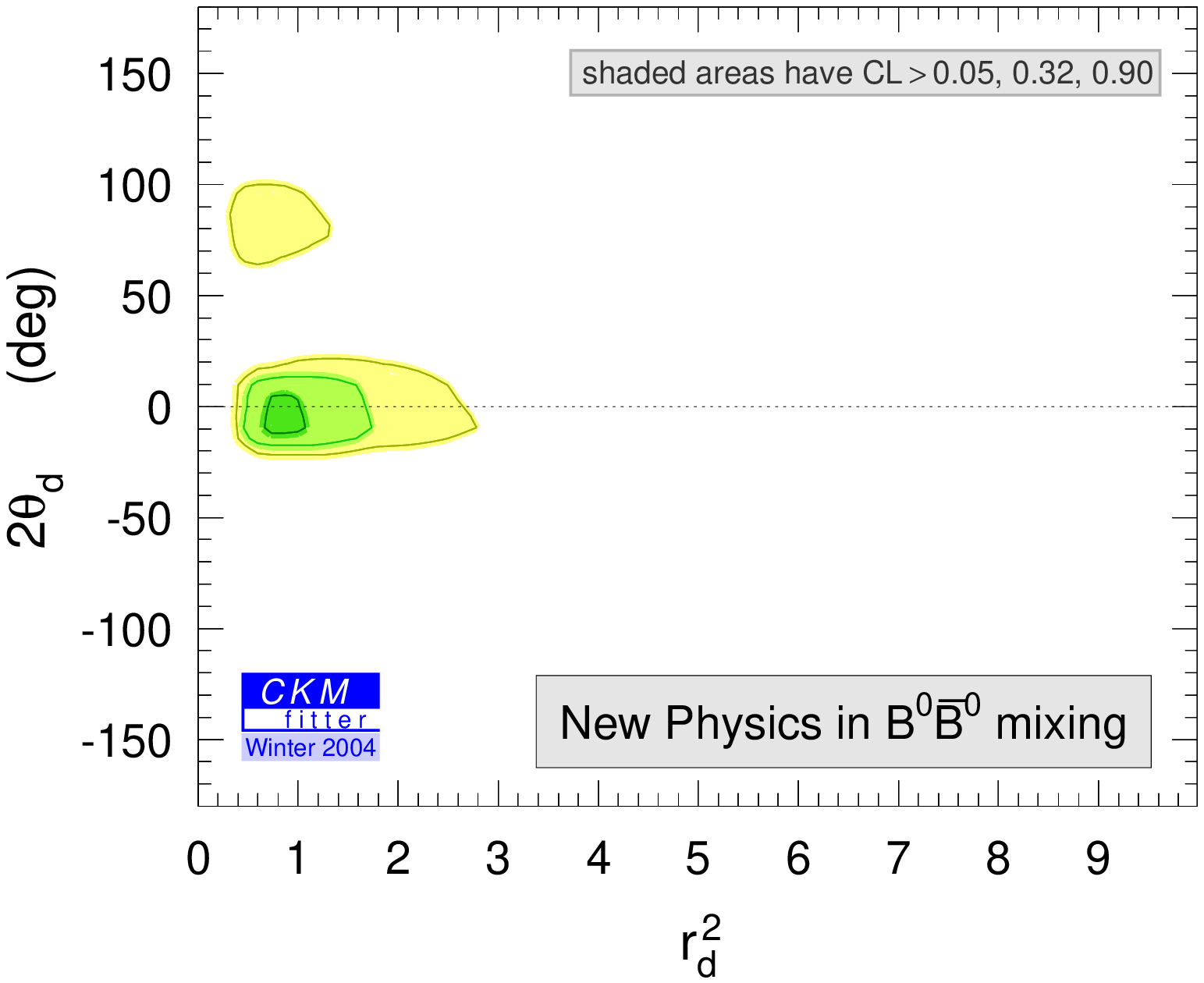}
\end{center}
\end{figure}

Thanks to the fact that quite a few observables that are related to
SM tree level processes have already been measured, we are able to
refer to this question in a quantitative way. The tree level processes
are insensitive to new physics and can be used to constrain $\rho$ and
$\eta$ even in the presence of new physics contributions to loop
processes, such as $\Delta m_B$. Among these observables we have
$|V_{cb}|$ and $|V_{ub}|$ from semileptonic $B$ decays, the phase
$\gamma$ from $B\to DK$ decays, and the phase $\alpha$ from
$B\to\rho\rho$ decays (in combination with $S_{\psi K}$). One can fit
these observables, and the ones in Eq.~(\ref{apksNP}) to the four
parameters $\rho,\eta,r_d^2$ and $2\theta_d$. The resulting
constraints are shown in Fig.~\ref{fig:rdtd}. 

A long list of models that require a significant modification of the
$B^0-\overline{B}^0$ mixing amplitude are excluded. We can further
conclude from Fig.~\ref{fig:rdtd} that a new physics contribution to
the $B^0-\overline{B}^0$ mixing amplitude at a level higher than about
30\%
is now disfavored. Yet, it is still possible that $\rho$ and $\eta$ are
well outside their SM range and that NP gives $2\theta_d$ very
different from zero and/or $r_d^2$ very different from one. In this
case, the SM and the NP `conspire' to mimic the SM values of the
observables (\ref{apksNP}). This is what we meant concretely in our 
statement that the KM dominance of the observed CP violation is now
very likely but not guaranteed.

{\bf II.} Consider $S_{\phi K_S}$, the CP asymmetry in $B\to\phi K_S$. This 
measurement is sensitive to the relative phase between the $B-\bar B$ 
mixing amplitude and the $b\to s\bar ss$ decay amplitude ($\sin2\beta$ in the 
SM). The $b\to s\bar ss$ decay has only Standard Model penguin contributions 
and therefore is sensitive to new physics. We parametrize the size and
phase of a NP contribution as follows (for simplicity, we neglect here the
$a_f^u$ terms of Eq.~(\ref{defafu})):
\beq\label{dertheA}
A_f=A_f^c\left(1+b_f\ e^{i\phi_{bs}}\right).
\eeq
Here $b_f$ is complex only if it carries a strong phase. The effects
of this new physics contribution are simple to understand in two limits:
\begin{enumerate}
\item The new physics contribution is dominant, $|b_f|\gg1$. The shift
  in all modes where this condition is valid is
universal and depends only on $\phi_{bs}$:
\beqa\label{unishi}
-\eta_f
S_f&\simeq&\sin(2\beta+2\theta_d)\cos2\phi_{bs}+\cos(2\beta+2\theta_d)
\sin2\phi_{bs},\no\\
C_f&\simeq&0.
\eeqa
\item The new physics contribution is small. Explicitly,
$|b_{f}|\ll1$. The shift is mode dependent and depends on both $b_f$
and $\sin\phi_{bs}$:
\beqa\label{smallnp}
-\eta_f S_f&\simeq&\sin(2\beta+2\theta_d)+2\cos(2\beta+2\theta_d){\cal 
  R}e(b_{f})\sin\phi_{bs},\no\\ 
C_f&\simeq&-2{\cal I}m(b_{f}^c)\sin\phi_{bs}. 
\eeqa
\end{enumerate}
Note that the effect of the NP is similar to that of the SM $a_f^u$
terms (with $b_f\leftrightarrow a_f^u$ and
$\phi_{bs}\leftrightarrow\gamma$), so that the latter have to be known
in order to probe the $b_f$ terms. Once that is done, the value of
$S_{\psi K}$ determines $2\beta+2\theta_d$ and one can examine whether
$\phi_{bs}\neq0$  and answer the following questions: 

(ii) {\it Is there new physics in $b\to s$ transitions?}

\begin{figure}[htb]
\caption{Constraints in the $\rho-\eta$ plane from tree processes and
  (a) $b\to d$ or (b) $b\to s$ loop processes.}
\label{fig:bdbs}
\begin{center}
\includegraphics[width=2.85in]{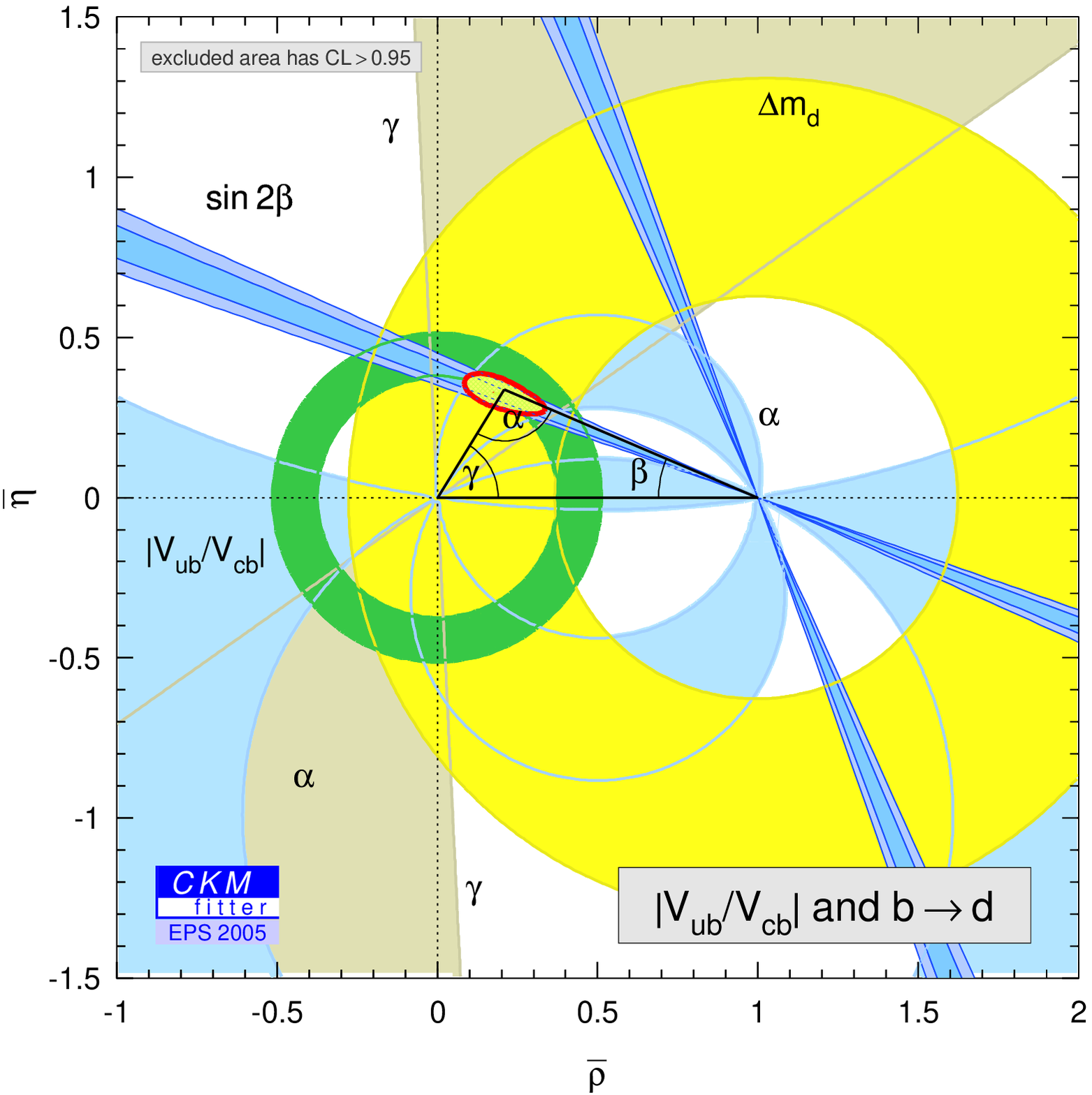}
\hspace{2em}
\includegraphics[width=2.85in]{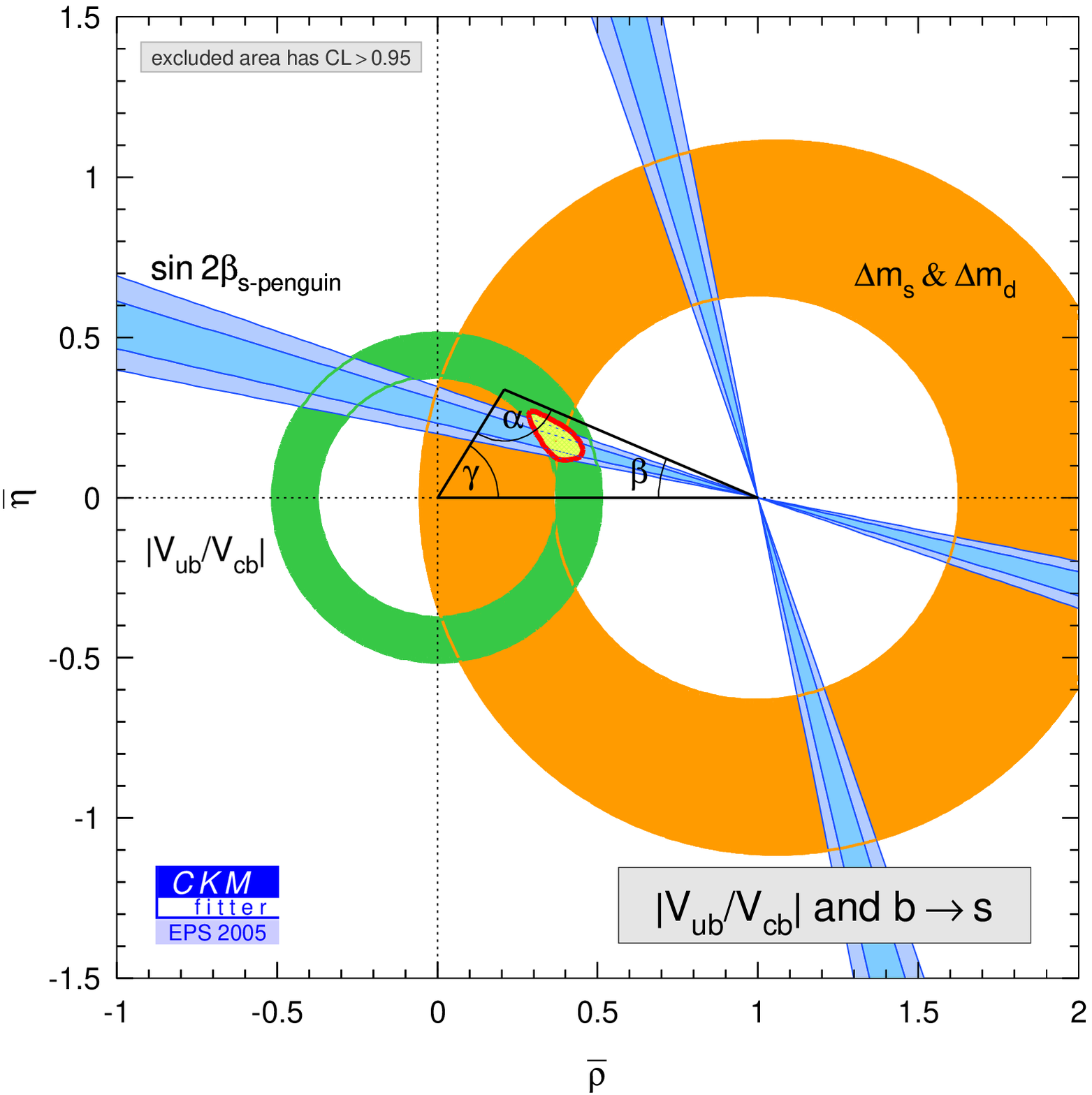}
\end{center}
\end{figure}

So far, the experimental data -- see Table \ref{tab:bsss} -- do not
provide any evidence for $\phi_{bs}\neq0$. Yet, the experimental
accuracy is still not sufficient to make qualitative statements such
as we made for $b\to d$ transitions ($B^0-\overline{B}^0$
mixing). To see this, we compare the constraints in the $\rho-\eta$
plane that arise from tree plus $b\to d$ loops ($\Delta m_B$, $S_{\psi
  K_S}$, $S_{\rho\rho}$, etc.) to those from tree plus $b\to s$ loops
($S_{\phi K_S}$, $S_{\eta^\prime K_S}$, $\Delta m_s$). This is done in
Fig.~\ref{fig:bdbs}. 

{\bf III.} Together with a future measurement of $B_s-\overline{B}_s$ 
mixing, we may also try to answer the following question:

(iii) {\it Is there new physics in $\Delta B=1$ processes? in $\Delta B=2$? 
in both?}

{\bf IV.} Consider $a_{\pi\nu\bar\nu}\equiv
{\Gamma_{K_L\to\pi^0\nu\bar\nu}}/{\Gamma_{K^+\to\pi^+\nu\bar\nu}}$,
see Eq.~(\ref{kpinunu}). This measurement will cleanly determine the
relative phase between the $K^0-\overline{K}^0$ mixing amplitude and the 
$s\to d\nu\bar\nu$ decay amplitude (of order $\sin^2\beta$ in the SM). The
experimentally measured small value of $\varepsilon_K$ requires that the phase 
of the $K^0-\overline{K}^0$ mixing amplitude is not modified from the
SM prediction. (More precisely, it requires that the phase of the mixing amplitude
is very close to twice the phase of the $s\to d\bar uu$ decay amplitude 
\cite{Nir:1990hj}.) On the other hand, the decay, which in the SM is a loop 
process with small mixing angles, can be easily modified by new physics.
Examining whether the SM correlation between $a_{\pi\nu\bar\nu}$ and 
$S_{\psi K_S}$ is fulfilled, we can answer the following  question:

(iv) {\it Is there new physics related solely to the third generation?
  to all generations?}

\begin{figure}[htb]
\caption{Constraints in the $\rho-\eta$ plane from tree processes and
(a) $s\to d$ or (b) $b\to d$ and $b\to s$ loop processes.}
\label{fig:sd}
\begin{center}
\includegraphics[width=2.85in]{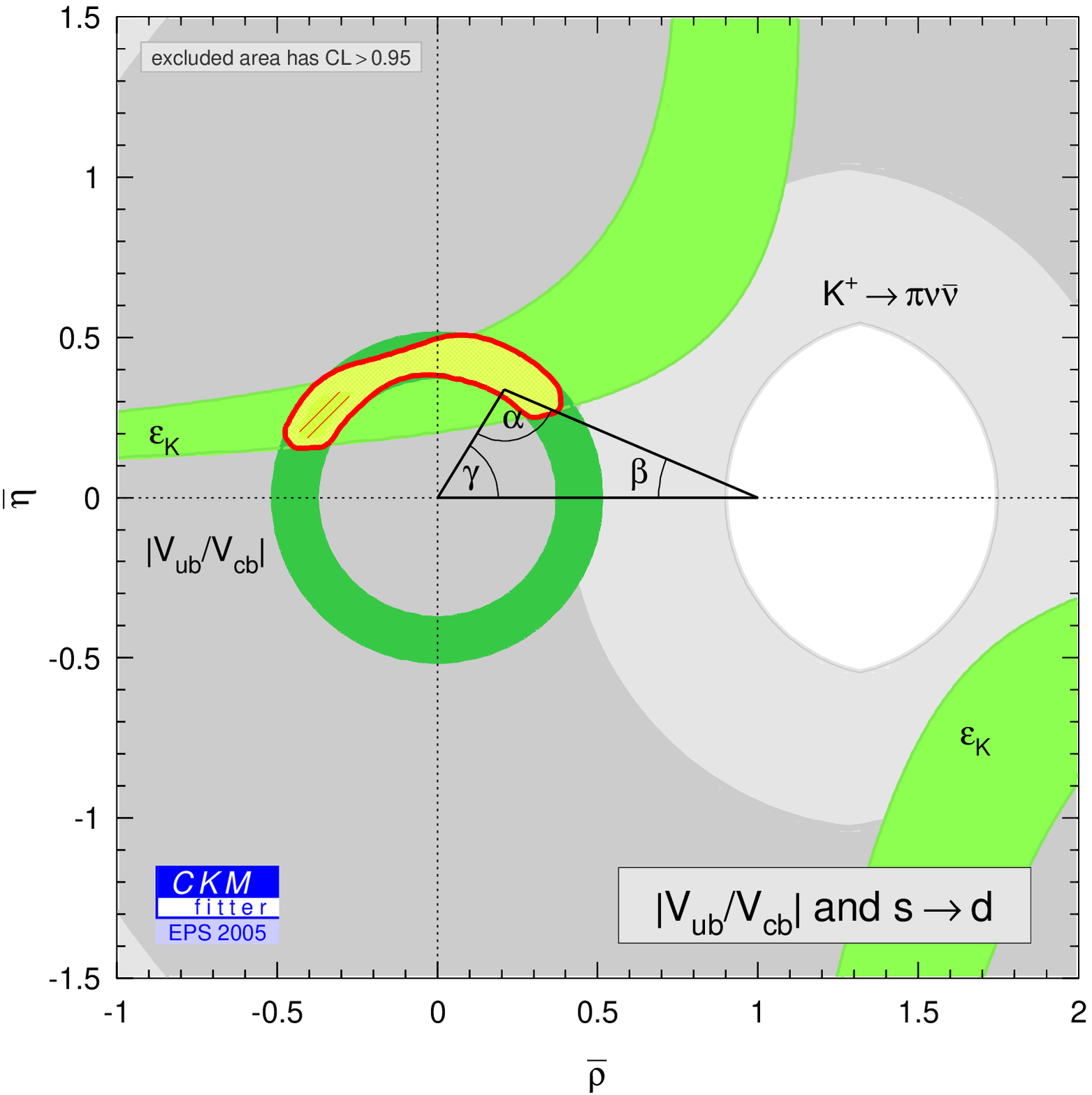}
\hspace{2em}
\includegraphics[width=2.85in]{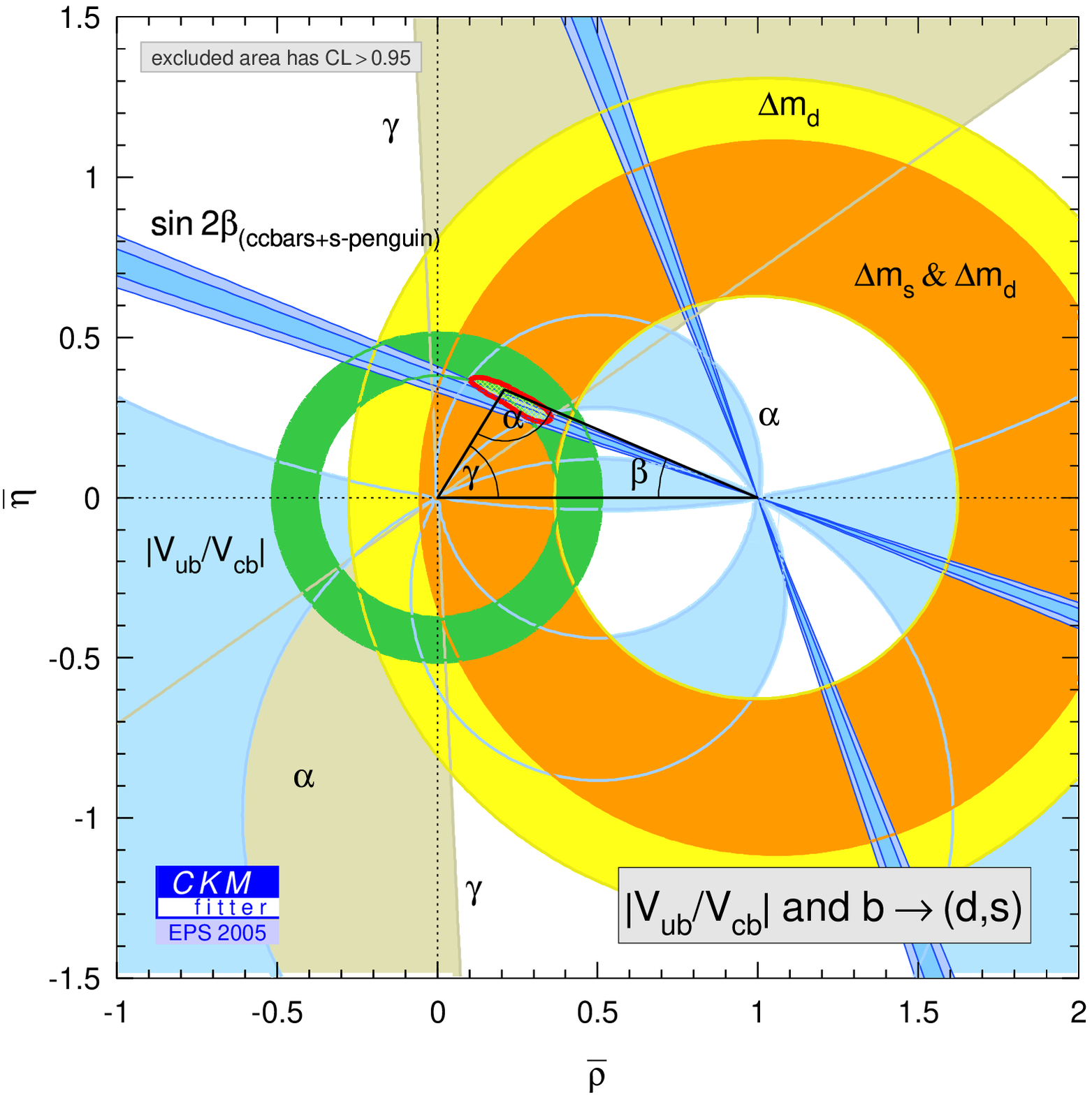}
\end{center}
\end{figure}

To understand the present situation, we present in Fig.~\ref{fig:sd}
the constraints in the $\rho-\eta$ plane from tree plus loop processes
that do not involve external third generation quarks, namely $s\to d$
transitions only ($\epsilon$ and ${\cal B}(K^+\to\pi^+\nu\bar\nu$)).
This can be compared with the constraints from tree plus loop
processes that do involve the third generation, namely $b\to d$ and
$b\to s$ transitions. Again, one can see that there is a lot to be
learnt from future measurements. (For a recent, comprehensive analysis
of this question, see ref. \cite{Agashe:2005hk}.)

{\bf V.} Consider $\phi_D$, defined in Eq.~(\ref{dtokpi}), which is
the relative phase between the $D^0-\overline{D}^0$ mixing amplitude and
the $c\to d\bar su$ and $c\to s\bar du$ decay amplitudes. Within the
Standard Model, the two decay channels are tree level. It is unlikely
that they are affected by new physics. On the other hand, the mixing
amplitude can be easily modified by new physics.
Examining whether $\phi_D\neq0$, we can answer the following question:

(v) {\it Is there new physics in the down sector? in the up sector? in
  both?}

{\bf VI.} Consider $d_N$, the electric dipole moment of the neutron. We did not discuss 
this quantity so far because, unlike CP violation in meson decays, flavor 
changing couplings are not necessary for $d_N$. In other words, the CP 
violation that induces $d_N$ is {\it flavor diagonal}. It does in general get 
contributions from flavor changing physics, but it could be induced by sectors 
that are flavor blind. Within the SM (and ignoring $\theta_{\rm QCD}$), the 
contribution from $\delta_{\rm KM}$ arises at the three loop level and is at 
least six orders of magnitude below the experimental bound (\ref{dnexp}).
If the bound is further improved (or a signal observed), we can answer
the following question:

(vi) {\it Are there new sources of CP violation that are flavor
  changing? flavor diagonal? both?}

It is no wonder then that with such rich information, flavor and CP violation 
provide an excellent probe of new physics. We next demonstrate this 
situation more concretely by discussing CP violation in supersymmetry.

\section{Supersymmetric CP Violation}
Supersymmetry solves the fine-tuning problem of the Standard Model and has
many other virtues. But at the same time, it leads to new problems:
baryon number violation, lepton number violation, large flavor changing
neutral current processes and large CP violation. The first two problems
can be solved by imposing $R$-parity on supersymmetric models. There is no
such simple, symmetry-related solution to the problems of flavor and CP 
violation. Instead, suppression of the relevant couplings can be achieved
by demanding very constrained structures of the soft supersymmetry breaking
terms. There are two important questions here: First, can theories of
dynamical supersymmetry breaking naturally induce such structures? 
Second, can measurements of flavor changing and/or CP violating processes
shed light on the structure of the soft supersymmetry breaking terms?
Since the answer to both questions is in the affirmative, we conclude that
flavor changing neutral current processes and, in particular, CP violating
observables will provide clues to the crucial question of how supersymmetry
breaks.

\subsection{CP violating parameters}
A generic supersymmetric extension of the Standard Model contains a host of new
flavor and CP violating parameters. (For a review of CP violation in 
supersymmetry see \cite{Grossman:1997pa,Dine:2001ne}.) It is an amusing 
exercise to count the number of parameters \cite{Haber:1998if}. The 
supersymmetric part of the Lagrangian depends, in addition to the three gauge 
couplings of $G_{\rm SM}$, on the parameters of the superpotential $W$:
\beq\label{superp}
W=\sum_{i,j}\left(Y^u_{ij}H_u Q_{Li} \overline{U}_{Lj}
+Y^d_{ij}H_d Q_{Li} \overline{D}_{Lj}
+Y^\ell_{ij}H_d L_{Li} \overline{E}_{Lj}\right)+\mu H_u H_d.
\eeq
In addition, we have to add soft supersymmetry breaking terms:
\beqa\label{Lsoft}
{\cal L}_{\rm soft}=&-&\left(A^u_{ij}H_u\tilde Q_{Li}\tilde{\overline{U}}_{Lj}
+A^d_{ij}H_d\tilde Q_{Li}\tilde{\overline{D}}_{Lj}+A^\ell_{ij}H_d\tilde L_{Li}
\tilde{\overline{E}}_{Lj}+B H_u H_d+{\rm h.c.}\right)\no\\
&-&\sum_{\rm all\ scalars}(m^2_S)_{ij}A_i\bar A_j-{1\over2}\sum_{(a)=1}^3
\left(\tilde m_{(a)}(\lambda\lambda)_{(a)}+{\rm h.c.}\right).
\eeqa
where $S=Q_L,\overline{D}_L,\overline{U}_L,L_L,\overline{E}_L$. The three 
Yukawa matrices $Y^f$ depend on 27 real and 27 imaginary parameters. Similarly,
the three $A^f$-matrices depend on 27 real and 27 imaginary parameters. The 
five $m^2_S$ hermitian $3\times3$ mass-squared matrices for sfermions have 30 
real parameters and 15 phases. The gauge and Higgs sectors depend on 
\beq\label{GaHi}
\theta_{\rm QCD},\tilde m_{(1)},\tilde m_{(2)},\tilde m_{(3)},
g_1,g_2,g_3,\mu,B,m_{h_u}^2,m_{h_d}^2,
\eeq
that is 11 real and 5 imaginary parameters. Summing over all sectors, we get
95 real and 74 imaginary parameters. The various couplings (other than the
gauge couplings) can be thought of as spurions that break a global symmetry,
\beq\label{MSBro}
U(3)^5\times U(1)_{\rm PQ}\times U(1)_R\ \to\ U(1)_B\times U(1)_L.
\eeq
The $U(1)_{\rm PQ}\times U(1)_R$ charge assignments are:
\beq\label{PQRcharges}
\matrix{&H_u&H_d&Q\overline{U}&Q\overline{D}&L\overline{E}\cr
U(1)_{\rm PQ}&1&1&-1&-1&-1\cr U(1)_{\rm R}&1&1&1&1&1\cr}.
\eeq
Consequently, we can remove 15 real and 30 imaginary parameters, which leaves
\beq\label{MSpar}
124\ =\ \cases{80&real\cr 44&imaginary}\ {\rm physical\ parameters}.
\eeq 
In particular, there are 43 new CP violating phases! In addition to the single
Kobayashi-Maskawa of the SM, we can put 3 phases in $M_1,M_2,\mu$ (we used the 
$U(1)_{\rm PQ}$ and $U(1)_R$ to remove the phases from $\mu B^*$ and $M_3$, 
respectively) and the other 40 phases appear in the mixing matrices of the 
fermion-sfermion-gaugino couplings. (Of the 80 real parameters, there are 11 
absolute values of the parameters in (\ref{GaHi}), 9 fermion masses, 21 
sfermion masses, 3 CKM angles and 36 SCKM angles.) Supersymmetry provides a 
nice example to our statement that reasonable extensions of the Standard Model 
may have more than one source of CP violation.

The requirement of consistency with experimental data provides strong 
constraints on many of these parameters. For this reason, the physics of flavor
and CP violation has had a profound impact on supersymmetric model building. 
A discussion of CP violation in this context can hardly avoid addressing the 
flavor problem itself.  Indeed, many of the supersymmetric models that we
analyze below were originally aimed at solving flavor problems.
 
As concerns CP violation, one can distinguish two classes of experimental
constraints. First, bounds on nuclear and atomic electric dipole moments
determine what is usually called the {\it supersymmetric CP problem}. Second, 
the physics of neutral mesons and, most importantly, the small experimental 
value of $\varepsilon_K$ pose the {\it supersymmetric $\varepsilon_K$ problem}.
In the next two subsections we describe the two problems.
 
\subsection{The Supersymmetric CP problem}
One aspect of supersymmetric CP violation involves effects that are flavor 
preserving. Then, for simplicity, we describe this aspect in a supersymmetric 
model without additional flavor mixings, {\it i.e.} the minimal supersymmetric 
standard model (MSSM) with universal sfermion masses and with the trilinear 
SUSY-breaking scalar couplings proportional to the corresponding Yukawa 
couplings. (The generalization to the case of non-universal soft terms is 
straightforward.)  In such a constrained framework, there are four new phases 
beyond the two phases of the SM ($\delta_{\rm KM}$ and $\theta_{\rm QCD}$). One
arises in the bilinear $\mu$-term of the superpotential (\ref{superp}), while 
the other three arise in the soft supersymmetry breaking parameters of 
(\ref{Lsoft}): $\tilde m$ (the gaugino mass), $A$ (the trilinear scalar 
coupling) and $B$ (the bilinear scalar coupling). Only two combinations of the 
four phases are physical \cite{Dugan:1985qf,Dimopoulos:1996kn}:
\beq\label{phiAB}
\phi_A=\arg(A^* \tilde m),\ \ \ \phi_B=\arg(\tilde m\mu B^*).
\eeq
In the more general case of non-universal soft terms there is one independent 
phase $\phi_{A_{i}}$ for each quark and lepton flavor. Moreover, complex 
off-diagonal entries in the sfermion mass-squared matrices represent 
additional sources of CP violation.
 
The most significant effect of $\phi_A$ and $\phi_B$ is their contribution to 
electric dipole moments (EDMs). For example, the contribution from one-loop 
gluino diagrams to the down quark EDM is given by 
\cite{Buchmuller:1983ye,Polchinski:1983zd}:
\beq\label{ddsusy}
d_d=m_d{e\alpha_3\over 18\pi\tilde m^3}\left(
|A|\sin\phi_A+\tan\beta|\mu|\sin\phi_B\right),
\eeq
where we have taken $m^2_Q\sim m^2_D\sim m^2_{\tilde g}\sim\tilde m^2$, for 
left- and right-handed squark and gluino masses. We define, as usual,
$\tan\beta=\langle{H_u}\rangle/\langle{H_d}\rangle$. Similar one-loop diagrams 
give rise to chromoelectric dipole moments. The electric and chromoelectric 
dipole moments of the light quarks $(u,d,s)$ are the main source of $d_N$ (the 
EDM of the neutron), giving \cite{Fischler:1992ha}
\beq\label{dipole}
d_N\sim 2\, \left({100\, GeV\over \tilde m}\right )^2\sin \phi_{A,B}
\times10^{-23}\ e\, {\rm cm},
\eeq
where, as above, $\tilde m$ represents the overall SUSY scale. In a generic 
supersymmetric framework, we expect $\tilde m={\cal O}(m_Z)$ and 
$\sin\phi_{A,B}={\cal O}(1)$. Then the constraint (\ref{dnexp}) is generically 
violated by about two orders of magnitude. This is {\it the Supersymmetric CP 
Problem}.

Eq. (\ref{dipole}) shows two possible ways to solve the 
supersymmetric CP problem:

(i) Heavy squarks: $\tilde m\gsim1\ TeV$;

(ii) Approximate CP: $\sin\phi_{A,B}\ll1$.

\subsection{The Supersymmetric $\varepsilon_K$ problem}
The supersymmetric contribution to the $\varepsilon_K$ parameter is dominated 
by diagrams involving $Q$ and $\bar d$ squarks in the same loop. For $\tilde m=
m_{\tilde g}\simeq m_Q \simeq m_D$ (our results depend only weakly on this 
assumption) and focusing on the contribution from the first two squark 
families, one gets (see, for example, \cite{Gabbiani:1996hi}):
\beq\label{epsKSusy}
\varepsilon_K={5\ \alpha_3^2  \over 162\sqrt2}{f_K^2m_K\over\tilde
m^2\Delta m_K}\left [\left({m_K\over m_s+m_d}\right)^2+{3\over 25}\right]
\im{(\delta_{12}^d)_{LL}(\delta_{12}^d)_{RR}}.
\eeq
Here
\beqa\label{defdsusy}
(\delta_{12}^d)_{LL}&=&\left({m^2_{\tilde Q_2}-m^2_{\tilde Q_1}\over
m^2_{\tilde Q}}\right)\ K^{dL}_{12},\no\\
(\delta_{12}^d)_{RR}&=&\left({m^2_{\tilde D_2}-m^2_{\tilde D_1}\over
m^2_{\tilde D}}\right)\ K^{dR}_{12},
\eeqa
where $K^{dL}_{12}$ ($K^{dR}_{12}$) are the mixing angles in the gluino
couplings to left-handed (right-handed) down quarks and their scalar partners.
Note that CP would be violated even if there were two families only 
\cite{Nir:1986te}. Using the experimental value of $\varepsilon_K$, we get
\beq\label{epsKScon}
{(\Delta m_K\varepsilon_K)^{\rm SUSY}\over(\Delta m_K\varepsilon_K)^{\rm EXP}}
\sim10^7\left ({300 \ GeV\over\tilde m}\right)^2
\left({m^2_{\tilde Q_2}-m^2_{\tilde Q_1}\over m_{\tilde Q}^2}\right)
\left({m^2_{\tilde D_2}-m^2_{\tilde D_1}\over m_{\tilde D}^2}\right)
|K_{12}^{dL}K_{12}^{dR}|\sin\phi,
\eeq
where $\phi$ is the CP violating phase. In a generic supersymmetric framework, 
we expect $\tilde m={\cal O}(m_Z)$, $\delta m_{Q,D}^2/m_{Q,D}^2={\cal O}(1)$, 
$K_{ij}^{Q,D}={\cal O}(1)$ and $\sin\phi={\cal O}(1)$. Then the constraint 
(\ref{epsKScon}) is generically violated by about seven orders of magnitude. 

The $\Delta m_K$ constraint on $\re{(\delta_{12}^d)_{LL}
(\delta_{12}^d)_{RR}}$ is about two orders of magnitude weaker.
One can distinguish then three interesting regions for $\langle{\delta_{12}^d}
\rangle=\sqrt{(\delta_{12}^d)_{LL}(\delta_{12}^d)_{RR}}$\, :
\beq\label{ranmot}
\langle{\delta_{12}^d}\rangle\cases{
\gg0.003 & excluded; \cr
\in[0.0002,0.003]&viable with small phases;\cr
\ll 0.0002 & viable with ${\cal O}(1)$ phases.\cr}
\eeq
The first bound comes from the $\Delta m_K$ constraint (assuming that the
relevant phase is not particularly close to $\pi/2$). The bounds here apply to
squark masses of order 500~GeV and scale like $\tilde m$. There is also
dependence on $m_{\tilde g}/\tilde m$, which is here taken to be one.

Eq. (\ref{epsKScon}) also shows what are the possible ways to solve
the supersymmetric $\varepsilon_K$ problem:

(i) Heavy squarks: $\tilde m\gg300\ GeV$;

(ii) Universality: $(\Delta m_{Q,D}^2)_{21}\ll m_{Q,D}^2$;

(iii) Alignment: $|K_{12}^d|\ll1$;

(iv) Approximate CP: $\sin\phi\ll1$.

\subsection{More on supersymmetric flavor and CP violation}
The flavor and CP constraints on supersymmetric models apply to almost
all flavor changing couplings. The size of supersymmetric flavor violation
depends on the overall scale of the soft supersymmetry breaking terms,
on mass degeneracies between sfermion generations, and on the mixing
angles in gaugino couplings. One can choose a representative scale
(say, $\tilde m\sim300$ GeV) and then conveniently present the
constraints in terns of the $(\delta^q_{ij})_{MN}$ parameters [see
Eq.~(\ref{defdsusy})]. In a given supersymmetric flavor model, one can
find predictions for the $(\delta^q_{ij})_{MN}$ and test the model.  

A summary of upper bounds on the supersymmetric flavor changing
couplings is given in Table \ref{tab:del}. The bounds on the ${\cal
  I}m(\delta^d_{12})_{LR,RL}$ parameters are taken from
\cite{Eyal:1999gk}, on $\delta^d_{13}$ from \cite{Becirevic:2001jj}
and on $\delta^d_{23}$ from \cite{Ciuchini:2004ej,Silvestrini:2005zb}. The
bounds are expressed in powers of the Wolfenstein parameter $\lambda$,
which makes it easy to compare with model predictions. As an example,
we give the range of these parameters  that is expected in a large
class of viable models of alignment
\cite{Nir:1993mx,Leurer:1993gy,Nir:2002ah}.

\begin{table}[t]
\caption{Theoretical predictions for supersymmetric flavor changing
  couplings in 
  viable models of alignment, and the experimental constraints.}
\label{tab:del}
\begin{center}
  \begin{tabular}{c|c|c||c|c|c}
    \hline
$(\delta^q_{MN})_{ij}$  &  Prediction & Upper bound  & $(\delta^d_{MN})_{ij}$ &
Prediction & Upper bound \\ \hline\hline
$(\delta^d_{LL})_{12}$ & $\lambda^5-\lambda^{3}$ & $\lambda^3$
& $(\delta^d_{LR})_{12}$ & $\lambda^7(m_b/\tilde m)$ &
$\lambda^7({\cal I}m)$ \\
$(\delta^d_{RR})_{12}$ & $\lambda^7-\lambda^{3}$ &
$\lambda^{10}/(\delta^d_{LL})_{12}$
& $(\delta^d_{RL})_{12}$ & $\lambda^9(m_b/\tilde m)$ &
$\lambda^7({\cal I}m)$ \\ \hline
$(\delta^d_{LL})_{13}$ & $\lambda^3$ & $\lambda$
& $(\delta^d_{LR})_{13}$ & $\lambda^3(m_b/\tilde m)$ & $\lambda^2$ \\
$(\delta^d_{RR})_{13}$ & $\lambda^7-\lambda^3$ & $\lambda^4/(\delta^d_{LL})_{13}$
& $(\delta^d_{RL})_{13}$ & $\lambda^7(m_b/\tilde m)$ & $\lambda^2$ \\ \hline
$(\delta^d_{LL})_{23}$ & $\lambda^2$ & $\lambda^2({\cal R}e)-\lambda({\cal I}m)$
& $(\delta^d_{LR})_{23}$ & $\lambda^2(m_b/\tilde m)$ &
$\lambda^4({\cal R}e)-\lambda^3({\cal I}m)$ \\
$(\delta^d_{RR})_{23}$ & $\lambda^4-\lambda^2$ & $1$
& $(\delta^d_{RL})_{23}$ & $\lambda^4(m_b/\tilde m)$ & $\lambda^3$ \\ \hline
$(\delta^u_{LL})_{12}$ & $\lambda$ & $\lambda$ &&& \\
$(\delta^u_{RR})_{12}$ & $\lambda^4-\lambda^{2}$ &
$\lambda^4/(\delta^u_{LL})_{12}$ &&& \\ \hline
\end{tabular}
\end{center}
\end{table}

Until some time ago, the $\delta^d_{23}$ parameters have been only
weakly constrained (the improving accuracy of the measurements of
${\cal B}(B\to X\ell^+\ell^-)$ have strengthened the constraints
considerably). Furthermore, measurements of various CP asymmetries in
penguin dominated modes (particularly $S_{\phi K}$ and $S_{\eta^\prime
  K}$) gave central values that were far off the expected value $\sim
S_{\psi K}$ (at present the strongest discrepancy is down to the
$2\sigma$ level). One may still ask whether effects of order 0.1,
which is the order of the expected experimental accuracy and probably
above the theoretical error on $S_{\phi K}$ and $S_{\eta^\prime K}$,
are still possible within supersymmetric flavor models and, in
particular, alignment models.

To answer this question, we use the results of
ref. \cite{Ciuchini:2004ej}. From their Fig. 3, we make the following
estimates:
\beqa
\frac{\Delta S_{\phi K}}{\Delta {\cal
    I}m(\delta_{LL}^d)_{23}}&\sim&\frac{\Delta S_{\phi K}}{\Delta {\cal
    I}m(\delta_{RR}^d)_{23}}\sim0.3,\no\\
\frac{\Delta S_{\phi K}}{\Delta {\cal
    I}m(\delta_{LR}^d)_{23}}&\sim&\frac{\Delta S_{\phi K}}{\Delta {\cal
    I}m(\delta_{RL}^d)_{23}}\sim100.
\eeqa
Thus, for $S_{\phi K}$ to be shifted by ${\cal O}(0.1)$, we need at
least one of the following four options:
\beqa\label{obsbs}
{\cal I}m(\delta_{LL}^d)_{23}&\sim&\lambda,\ \ \
{\cal I}m(\delta_{RR}^d)_{23}\sim\lambda,\no\\
{\cal I}m(\delta_{LR}^d)_{23}&\sim&\lambda^4,\ \ \
{\cal I}m(\delta_{RL}^d)_{23}\sim\lambda^4.
\eeqa
Examining Table \ref{tab:del}, we learn that in alignment models
${\cal I}m(\delta_{LR}^d)_{23}\sim 7\times10^{-4}(350\ GeV/\tilde m)$
is the closest to satisfying the condition in Eq.~(\ref{obsbs}),
though the unknown numbers of order one should be on the large side to
give an observable effect.

\subsection{Discussion}
We define two scales that play an important role in supersymmetry:
$\Lambda_S$, where the soft supersymmetry 
breaking terms are generated, and $\Lambda_F$, where flavor dynamics takes 
place. When $\Lambda_F\gg\Lambda_S$, it is possible that there are no genuinely
new sources of flavor and CP violation. This class of models, where
the Yukawa couplings (or, in the mass basis, the CKM matrix) are the
only source of flavor and CP breaking, are often called `minimal
flavor violation.' The most important features of the supersymmetry
breaking terms are universality of the scalar masses-squared and
proportionality of the $A$-terms. When $\Lambda_F\lsim\Lambda_S$, we do
not expect, in general,  
that flavor and CP violation are limited to the Yukawa matrices. One way to 
suppress CP violation would be to assume that, similarly to the Standard Model,
CP violating phases are large, but their effects are screened, possibly by the 
same physics that explains the various flavor puzzles, such as models with 
Abelian or non-Abelian horizontal symmetries. It is also possible that CP 
violating effects are suppressed because squarks are heavy. Another option,
which is now excluded, was to assume that CP is an approximate symmetry 
of the full theory (namely, CP violating phases are all small). 

We would like to emphasize the following points:

(i) For supersymmetry to be established, a direct observation of supersymmetric
particles is necessary. Once it is discovered, then measurements of CP 
violating observables will be a very sensitive probe of its flavor structure 
and, consequently, of the mechanism of dynamical supersymmetry breaking.

(ii) It seems possible to distinguish between models of exact universality and 
models with genuine supersymmetric flavor and CP violation. The former tend to
give $d_N\lsim10^{-31}$ e cm while the latter usually predict 
$d_N\gsim10^{-28}$ e cm.

(iii) The proximity of $S_{\psi K_S}$ to the SM predictions is obviously
consistent with models of exact universality. It disfavors models
of heavy squarks such as that of ref. \cite{Cohen:1997sq}. Models of flavor
symmetries allow deviations of order 20\% (or smaller) from the SM predictions.
To be convincingly signalled, an improvement in the theoretical calculations 
that lead to the SM predictions for $S_{\psi K_S}$ will be required 
\cite{Eyal:2000ys}.

(iv) Alternatively, the fact that $K\to\pi\nu\bar\nu$ decays are not affected 
by most supersymmetric flavor models 
\cite{Nir:1998tf,Buras:1998ij,Colangelo:1998pm}
is an advantage here. The Standard Model correlation between 
$a_{\pi\nu\bar\nu}$ and $S_{\psi K_S}$ is a much cleaner test than 
a comparison of $S_{\psi K_S}$ to the CKM constraints.

(v) The neutral $D$ system provides a stringent test of alignment. 
Observation of CP violation in the $D\to K\pi$ decays will make a convincing
case for new physics.

(vi) CP violation in $b\to s$ transition remains an interesting probe
of supersymmetry. Deviations of order $0.1$ from the SM predictions
are possible if at least one of the conditions in Eq.~(\ref{obsbs}) is
satisfied. 

\section{Lessons from the B Factories}
Let us summarize the main lessons that have been
learned from the measurements of CP violation in B decays:
\begin{itemize}
  \item The KM phase is different from zero, that is, the SM violates CP.
  \item The KM mechanism is the dominant source of CP
    violation in meson decays.
    \item The size and the phase of new physics contributions to $b\to
      d$ transitions ($B^0-\overline{B}^0$ mixing) is severely
      constrained ($\leq{\cal O}(0.2)$).
      \item Complete alternatives to the KM mechanism (the superweak
        mechanism and approximate CP) are excluded.
        \item Corrections to the KM mechanism are possible,
          particularly for $b\to s$ transitions, but there is no
          evidence at present for such corrections.
        \item There is still a lot to be learned from future
          measurements.
          \end{itemize}
        
\acknowledgments
I am grateful to Andreas H\"ocker, Sandrine Laplace and, in
particular, Stephane T'Jampens for providing me with beautiful
plots of CKM constraints. Their work has helped me to understand and,
hopefully, to explain the significance of the B-factory measurements
of CP violating asymmetries to our understanding of flavor and CP
violation. I am grateful to Guy Raz and to Zoltan Ligeti for their
contributions to the basic ideas and to the details of this review. I
thank David Kirkby for collaboration on the PDG review on CP violation
in meson decays \cite{kirkbynir} which is the basis of some sections
in these lecture notes.  
This work was supported by a grant from the G.I.F., the
German--Israeli Foundation for Scientific Research and Development,
by the Israel Science Foundation
founded by the Israel Academy of Sciences and Humanities, by EEC RTN
contract HPRN-CT-00292-2002, by the Minerva Foundation (M\"unchen),
and by the United States-Israel Binational Science Foundation (BSF),
Jerusalem, Israel. 


\tighten

\end{document}